\RequirePackage{silence}
\documentclass[10pt,journal,compsoc]{sty/IEEEtran/IEEEtran}

\newcommand{\ourTitle}{Space Partitioning Schemes and Algorithms for Generating Regular and Spiral Treemaps}
\newcommand{\ourAuthors}{Reyhaneh Mohammadi, Mehdi Behroozi, and Cody Dunne}
\newcommand{\ourKeywords}{Regular and Spiral Treemaps, Space Partitioning Optimization, Data Visualization, Computational Geometry}
\newcommand{\shortitle}{Behroozi \etal: Space Partitioning for Treemaps}

\newcommand{\mailtohref}[1]{\href{mailto:mohammadi.re@northeastern.edu
;m.behroozi@northeastern.edu;c.dunne@northeastern.edu}{\color{black}#1}}

\newcommand{\ourAuthorsFormatted}{%
    \author{
            \orcidiconlink{Mehdi Behroozi}{0000-0002-8295-3248}, 
        \orcidiconlink{Reyhaneh Mohammadi}{0000-0002-7358-8023}, and
        \orcidiconlink{Cody Dunne}{0000-0002-1609-9776}%
        \thanks{Mehdi Behroozi is the corresponding author. All authors are with Northeastern University, Boston, MA, USA. E-mails: \mailtohref{%
            mohammadi.re@northeastern.edu;
            m.behroozi@northeastern.edu;
            c.dunne@northeastern.edu}
        }%
        \thanks{ A full version of this paper with all appendices and data sets is available at \anonymizeOSF{\OSFSupplementText}.%
        }
    }%
}

\usepackage[nocompress]{cite}

\usepackage{graphicx}
\graphicspath{{./figs/}}

\usepackage{url}
\usepackage[bookmarks=true, pdfpagelabels=true, pdftex]{hyperref}
\hypersetup{
    bookmarksnumbered=true,
    pdfpagemode={UseOutlines},
    plainpages=false,
    colorlinks=true,
    linkcolor={black},
    citecolor={black},
    urlcolor={black},
    pdftitle={\ourTitle},
    pdfsubject={Typesetting},
    pdfauthor={\ourAuthors},
    pdfkeywords={\ourKeywords},
}
\usepackage[TVCG-2021]{sty/dunnemacros}

\begin{document}

\title{\ourTitle}
\ourAuthorsFormatted
\markboth{\shortitle}%
{fish}

\IEEEtitleabstractindextext{%
\begin{abstract}
    Treemaps have been widely applied to the visualization of hierarchical data.
    A treemap takes a weighted tree and visualizes its leaves in a nested planar geometric shape, with sub-regions partitioned such that each sub-region has an area proportional to the weight of its associated leaf nodes.
    Efficiently generating visually appealing treemaps that also satisfy other quality criteria is an interesting problem that has been tackled from many directions.
    We present an optimization model and five new algorithms for this problem, including two divide and conquer approaches and three spiral treemap algorithms.
    Our optimization model is able to generate superior treemaps that could serve as a benchmark for comparing the quality of more computationally efficient algorithms.
    Our divide and conquer and spiral algorithms either improve the performance of their existing counterparts with respect to aspect ratio and stability or perform competitively. Our spiral algorithms also expand their applicability to a wider range of input scenarios.
    Four of these algorithms are computationally efficient as well with quasilinear running times and the last algorithm achieves a cubic running time. A full version of this paper with all appendices, data, and source codes is available at \anonymizeOSF{\OSFSupplementText}.%
\end{abstract}

}

\maketitle

\IEEEraisesectionheading{\section{Introduction}\label{sec:introduction}}

\IEEEPARstart{T}{reemaps} are well-known planar structures for visualization of tree-structured hierarchical data with a space-filling approach.
A nested treemap maps the nodes of a tree to a nested non-overlapping structure made up with simply connected planar figures, usually rectangles or convex polygons, where the areas of these figures are proportionate to some attribute of the nodes of the tree.
Another way of looking at treemap layouts is that they partition a given planar region into a number of sub-regions with given areas.

Treemaps were introduced by Shneiderman in 1991 as a space-filling technique to visualize space usage within computer directory structures \cite{shneiderman1991tree,shneiderman1992tree}.
\Cref{fig:TreemapExample} shows an example of tree-structured data and an associated treemap generated with the original \modelSliceDice algorithm.
Shneiderman's approach was to make a rectangular treemap, in which a rectangle was recursively divided with alternating horizontal and vertical cuts into rectangular sub-regions.
\modelSliceDice is simple to implement, but suffers some issues particularly with creating long and skinny rectangles.
The poor aspect ratio raises issues with interpreting any associated visual encodings \cite{kong2010perceptual}.

Over the years several treemap variants have been proposed.
These include polygonal treemaps where the sub-regions are simple polygons \cite{liang2015divide},
convex treemaps where the sub-regions are convex polygons \cite{onak2008circular}, 
Voronoi treemaps where the sub-regions are Voronoi cells \cite{balzer2005voronoiA,balzer2005voronoiB}, 
cascaded treemaps where the sub-regions are cascaded as opposed to be nested \cite{lu2008cascaded}, 
spiral treemaps where layout presents a spiral structure \cite{tu2007visualizing,Chaturvedi14GroupinBox}, and 
treemaps with space-filling fractal-like curves such as Gosper \cite{auber2013gospermap} and Hilbert curves \cite{tak2012enhanced}. 

Treemaps are used in a wide range of application domains including software visualization; file/directory structures; news and multimedia visualization; and visualization of financial data such as stocks, budgets, and import/export trades.
Its associated spatial partitioning problem also has a wide range of applications such as designing service districts, land allocation in farming reforms, and matrix multiplication algorithms in parallel computing.
Beyond area encoding, treemaps can be enhanced to encode additional attributes using color, shade, filling pattern, label, and nesting/cascading structure.
Creative and new applications of treemaps for different visualization purposes are continuously being proposed using a variety of treemap algorithms (see \cite{scheibel2020survey} for a survey on treemap algorithms).

A regular treemapping problem with rectangular sub-regions could be defined as follows. Given a rectangle $R$ with $\Area(R)=A$ and a hierarchical list of areas $A_1,...,A_n$ with $\sum_{i=1}^n A_i = A$ arranged in a tree data structure $T$ we want to layout $n$ non-overlapping rectangles with areas $A_1,...,A_n$ inside $R$ in a nested way that follows the structure of $T$ and the resulting rectangles are as square as possible with $\cup_{i=1}^n R_i =R$. The squareness of the rectangles is important practically and aesthetically in data visualization. In this paper, we also study another variant of this problem, where we relax the partitioning condition ($\cup_{i=1}^n R_i =R$) and instead require the layout to follow certain \emph{spiral} patterns. In other words, for regular treemaps we are given an input region while for the spiral treemaps, there is no input region and instead we aim to form a spiral structure with rectangular sub-regions of given areas. For regular treemaps we propose an optimization model, one divide \& conquer algorithm, and one dynamic programming algorithm and for the spiral treemaps we propose three constructive algorithms that mimic the ``golden spiral'' structure in different ways. Our divide and conquer and dynamic programming algorithms take a \emph{subdivision} approach and our spiral algorithms take a \emph{packing} approach, while all proposed algorithms fall under \emph{space-filling} treemaps according to the existing design space categories \cite{schulz2010design,scheibel2020taxonomy,scheibel2020survey}.

\subsection*{Contributions, Outline, and Supplemental Materials}

In this paper, we consider nested rectangular, convex, and spiral treemaps.
Our main contributions, listed in presentation order, are:
\begin{enumerate}
    \item \textbf{A treemap optimization model} that minimizes the total perimeter of the component rectangles, thus favoring square sub-regions (\cref{sec:optModel}). To the best of our knowledge, this is the most comprehensive model in the literature of the rectangular treemapping problem, incorporating several variants of the problem.
    
    \item \textbf{Two subdivision treemap algorithms} based on divide \& conquer and dynamic programming that resolve the shortcoming of the prior divide \& conquer approach when faced with uneven area distributions (\cref{sec:SubdivisionAlgs}).  
    
    \item \textbf{Three spiral treemap algorithms} that mimic the Symmetric Spirals that are ubiquitous in nature and famous for their visual attractiveness (\cref{sec:SpiralAlgorithms}). Spiral algorithm \emph{construct} treemaps independent of the input region. Removing this restriction can lead to both significant improvements or deterioration depending on the problem instance. 
    
    \item \textbf{The usage of Hausdorff distance} as a metric for stability is established for the first time. 
    
    \item \textbf{A performance comparison} of treemap algorithms based on aspect ratio and stability (\cref{sec:ComputationalResults}).
    We demonstrate that our optimization model generates superior treemaps and could serve as a benchmark for assessing more computationally efficient approaches.
    Also, our other treemap algorithms offer improved aspect ratio vs. prior methods, while staying competitive on stability.
    Finally, usage examples on several diverse datasets (\cref{sec:UsageExamples}) are also provided for further comparison of output visualizations.
\end{enumerate}

A full version of this paper with all appendices, data, and source code is available at \anonymizeOSF{\OSFSupplementText}.

\section{Background}
\label{sec:background}

\subsection{Notational Conventions}

The following notational conventions are assumed throughout this paper: 
By bounding box (rectangle) of a region $C$, we mean minimum-area axis-aligned bounding box (rectangle) of $C$ and we show it with $\Box(C)$.
The width and height of a region $C$ are defined as the width and height of $\Box(C)$ and are denoted by $\width(C)$ and $\height(C)$.
$\Area(C)$ denotes the area of a region $C$ and $\Perim(C)$ shows its perimeter. We define the perimeter of a rectangle $R$ by $\Perim(R)=\width(R)+\height(R)$.
We also define the aspect ratio of a rectangle $R$ as
\[
\AR(R)=\max\left\{ \frac{\width(R)}{\height(R)}, \frac{\height(R)}{\width(R)}\right\},
\]
 i.e., $\AR\geq 1$ and the aspect ratio of a square is one.
 We define the aspect ratio of a non-rectangular polygon $P$ as the aspect ratio of its bounding rectangle $\Box(P)$.
Finally, $\left\vert S \right\vert$ shows the cardinality (size) of set $S$.

\begin{figure}[tbp]%
    \centering%
    \begin{subfigure}[b]{\textwidth}%
    \begin{center}
        \includegraphics[width=0.9\textwidth]{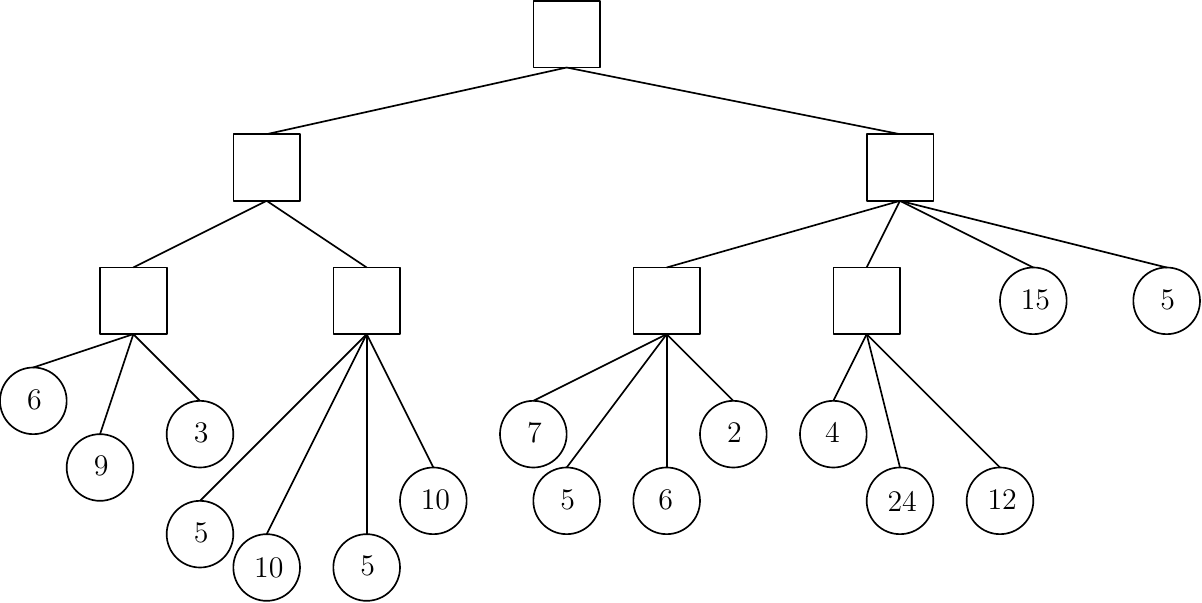}%
        \caption{}%
        \label{fig:Tree}%
    \end{center}
    \end{subfigure}%
    \\%
    \begin{subfigure}[b]{\textwidth}%
    \begin{center}
        \includegraphics[width=0.9\textwidth]{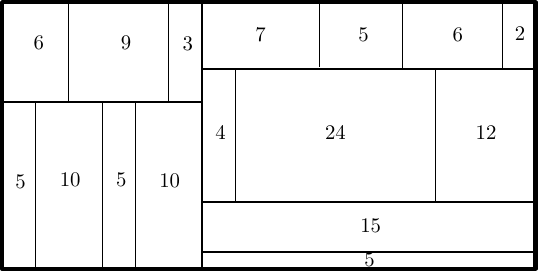}%
        \caption{}%
        \label{fig:SliceDice}%
     \end{center}
    \end{subfigure}%
    \subfigsCaption{\protect
        Hierarchical data \subref{fig:Tree} modeled as a height-3 weighted tree \& \subref{fig:SliceDice} the associated treemap visualization produced by Shneiderman's original \modelSliceDice algorithm \cite{shneiderman1991tree,shneiderman1992tree}.%
    }%
    \label{fig:TreemapExample}%
\end{figure}

\subsection{Evaluation Criteria}
\label{subsec:criteria}

Depending on the application area, the goal in generating a treemap and thus the evaluation metrics for the results could be different.
Generally, the utility of a treemap-generating algorithm for decision-makers could be evaluated according to the following metrics: 

\textbf{Space-filling}---The main goal of most treemap algorithms is to present information in a 2D space in an efficient, space-filling way and minimize unused space \cite{johnson1991tree}.
This enables the presentation of a lot of information in a concise format.
The goal in some other treemap algorithms is to show \emph{containment} between parents and children nodes where unused spaces can be present. Here, we only consider space-filling treemaps.

\textbf{Aspect ratio}---Treemap algorithms should avoid generating thin and elongated sub-regions as much as possible.
A planar geometric shape is called ``\emph{fat}'' if the aspect ratio of its bounding rectangle is one or close to one (almost square).
Using convex or rectangular fat shapes for a treemap collectively forms a visually appealing pattern, makes mouse selection easier \cite{bruls2000squarified,MacKenzie1992FittsLaw}, allows larger labels, and can make area comparison tasks more accurate \cite{kong2010perceptual}.
Treemap algorithms have often been designed or evaluated in terms of their ability to minimize both maximum and average aspect ratio ($\AR$) of the sub-regions, e.g., \modelSquarified by Bruls \etal \cite{bruls2000squarified}.
In \cref{sec:SpiralAlgorithms} we further discuss the experiments of Kong \etal \cite{kong2010perceptual}---including their finding that both perfect squares and rectangles with extreme aspect ratios can create misjudgment in area comparison tasks.

\textbf{Stable}---In many applications, the data of interest may change frequently or on a dynamic basis.
Thus stability in the face of dynamically changing data is desired so that changes to the layout are minimal if the data is just slightly perturbed \cite{shneiderman2001ordered,sud2010fast}.
This helps the user to track what has changed without completely distorting their mental map, as defined by Misue \etal \cite{Misue1995MentalMap}.
To measure stability we need a distance measure for corresponding pairs of sub-regions between two layouts to measure the change in location of a sub-region in different treemap layouts, generated after perturbation of areas.
Several such measures have been used for treemaps that can be categorized as: (1) \emph{absolute} metrics that measure how much individual rectangles move/change (e.g., \emph{average Euclidean distance change} (between corners) \cite{shneiderman2001ordered,wattenberg1999visualizing} \emph{location drift} \cite{tak2012enhanced}, \emph{corner travel distance} \cite{vernier2020quantitativeJ}),  and (2) \emph{relative} metrics that measure how much positions of pairs of rectangles change relative to each other, usually by comparing their centers or aspect ratios (e.g., \emph{average Euclidean distance change} (between centroids) \cite{wood2008spatially,hahn2014visualization},  \emph{average angular displacement} \cite{wood2008spatially}, \emph{relative direction change} \cite{hahn2017relative}, \emph{relative position change} \cite{sondag2018stable}, and \emph{average aspect ratio change} and \emph{relative parent change} \cite{scheibel2018evocells}). Some of these measures fail or perform poorly for certain cases. For example, two skinny rectangles can form a cross (which shows a significant move of the corners and edges), while having the same centers. 

\begin{figure}[tbp]%
    \centering%
    \includegraphics[width=.65\textwidth]{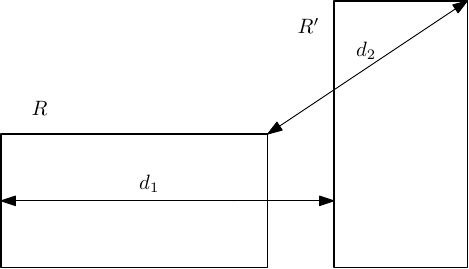}
    \caption{
        The Hausdorff distance between rectangles $R$ \& $R^{\prime}$ is defined as $\HD(R,R^{\prime}) = \max(d_1,d_2)$, where $d_1$ is the largest Euclidean distance of a point in $R$ to its closest point in $R^{\prime}$ and $d_2$ is the largest Euclidean distance of a point in $R^{\prime}$ to its closest point in $R$.%
    }%
    \label{fig:Hausdorff}%
\end{figure}

Here we establish the maximum and average \emph{Hausdorff distance} ($\HD$) as our metrics for stability.  We define the \emph{Hausdorff distance} between two compact sets $S_1$ and $S_2$, both in the same metric space, as
\[
\HD(S_1,S_2) = \max \left\{\,\sup _{x\in S_1}\inf _{y\in S_2} \|x-y\|,\,\sup _{y\in S_2}\inf _{x\in S_1}\|x-y\|\,\right\},
\]
where $\|\cdot\|$ is the Euclidean norm, $\sup$ is the supremum and $\inf$ the infimum. This is illustrated in \cref{fig:Hausdorff}.
HD provides a measure of distance between two corresponding sub-regions in two layouts. It shows the maximum of the maximum distances of each sub-region in one layout to the nearest point of its corresponding sub-region in the other layout. This is the first use of this stability measure for treemaps and it is an absolute metric in the same spirit of the Euclidean distance change metric proposed by Shneiderman \& Wattenberg \cite{shneiderman2001ordered}.

\textbf{Ordered}---Preserving an input order in the output can help with recognizing the tree structure and any hierarchy more quickly, as well as increasing the readability and scannablity of individual items \cite{shneiderman2001ordered}.
A treemap is ``\emph{ordered}'' if its sub-regions are scannable in close to the same order as was given.
The essentiality of preserving order as a desired property depends on the application domain.
For example, it is faster to find a node with a given label if the nodes are ordered alphabetically, or faster to find bonds that are paying out soon if they are ordered by maturity date. We leave generating ordered treemaps as a future research direction for the algorithms developed in this paper.

\textbf{Encoding Multivariate Data}---Conveying maximum information through size, color, shade, placement location, adjacency relations, label, filling pattern, and nesting structure of the layout can enhance the utilization of treemaps \cite{yang2015cabinet}.
Most algorithms could be easily enhanced with respect to this metric with some simple modifications.
But these are outside of the focus of this paper.

Here we focus solely on the first three metrics for treemaps: space-filling, low aspect ratio, and stable. For a survey on user studies on the effectiveness of treemaps see \cite{fiedler2020survey}.

\subsection{Treemap Algorithms}
\label{subsec:literature}

The first treemap algorithm (and the visualization approach in general) was developed by Shneiderman \cite{shneiderman1991tree,shneiderman1992tree}.
His \modelSliceDice algorithm performs an alternating sequence of vertical and horizontal cuts.
Recursively, the direction of the layout for each level of the tree is reversed with cuts vertically at even levels and horizontally at odd levels.
If the tree has only one level other than the root then all cuts will be vertical.
This creates the so-called slices and dices.
An example of \modelSliceDice is shown in \cref{fig:SliceDice}.
However, \modelSliceDice is prone to generating thin, ``\textit{skinny}'' rectangles which do poorly on our aspect ratio criteria.

\subsubsection{Aspect Ratio Optimization}

Wattenberg's \modelCluster algorithm \cite{wattenberg1999visualizing} was the first to address aspect ratio.
\modelCluster was presented as a modification of \modelSliceDice that avoids high-aspect-ratio rectangles by employing both vertical and horizontal partitions at each level of hierarchy and clustering similar rectangles adjacent or close to each other.
Wattenberg used \modelCluster for what was likely the first stock market treemap \cite{Wattenberg1998MarketMap}.

The \modelSquarified algorithm by Bruls \etal \cite{bruls2000squarified} similarly focused on generating sub-regions that are as close as possible to squares.
It takes an input rectangle and the desired areas of sub-rectangles.
First, the desired areas are sorted in descending order.
Then, in each iteration 
the algorithm places rectangles either vertically or horizontally according to a subset of areas from the sorted list.
The subset is selected depending on the width and height of the remaining area of the input rectangle and such that the maximum aspect ratio of rectangles in this subset is minimized. Nguyen and Huang \cite{nguyen2005enccon} took a similar approach with a flexible acceptable maximum aspect ratio in each mentioned subset and some adjustments to the weights of sub-regions.

A third approach to this problem was developed by Vernier \& Nigay \cite{vernier2000modifiable}.
Their \modelModifiable treemaps algorithm allocates to each node of the tree a bounding box with fixed aspect ratio, which can be modified by the user.
It then creates a new set of recursive steps between two recursive steps of \modelSliceDice \cite{shneiderman1991tree,shneiderman1992tree}.
At each iteration, \modelModifiable generates several options from the current set of sub-rectangles and picks the one with the minimum sum of differences between the fixed aspect ratio and the aspect ratio of each contained rectangle.
The chosen option is then recursively partitioned.
If the fixed aspect ratios are set to 1 then the algorithm generates square-like rectangles. 

\subsubsection{Ordering \& Stability}

While \modelSliceDice \cite{shneiderman1991tree,shneiderman1992tree} preserves the original order of the leaves in the tree structure, \modelCluster \cite{wattenberg1999visualizing}, \modelSquarified \cite{bruls2000squarified}, and \modelModifiable \cite{vernier2000modifiable} do not---violating the ordered criteria.
Moreover, they violate the stable criteria as they are not stable to changes in the input data.

Engdahl was able to preserve order with \modelSplit \cite{engdahl2005ordered}, while still offering improved aspect ratios over \modelSliceDice  \cite{shneiderman1991tree,shneiderman1992tree}.
\modelSplit takes an ordered list of desired sub-rectangle areas and splits them into two lists, without changing the order, such that the sum of areas is as close as possible across lists.
\modelSplit then divides the input rectangle with a vertical or horizontal cut, depending on whether the width is larger than the height, then recurses.

Introducing the stability criteria, Shneiderman \& Wattenberg's \modelPivot \cite{shneiderman2001ordered} and the Bederson \etal's \modelStrip \cite{bederson2002ordered} algorithms take compromise approaches that still preserve ordering and good aspect ratios.
\modelPivot selects one of the input areas as the pivot element, places it as a square-like sub-rectangle to partition the input rectangle $R$ into three areas, and then recursively lays out rectangles from three lists $L_1$--$L_3$ into each of the areas.
$L_1$ includes all sub-rectangles whose index in the input order is less than the index of the pivot.
$L_2$ and $L_3$ are chosen such that the indices of sub-rectangles in $L_2$ is less than the indices of sub-rectangles in $L_3$ and the aspect ratio of the pivot is as close to 1 as possible.

\modelStrip adds rectangles one-by-one to strips, just like \modelSquarified \cite{bruls2000squarified}, but fixes the orientation of strips throughout and omits the initial sorting step to preserve order. Bederson \etal also introduce \modelQuantum \cite{bederson2002ordered}, which is specially suited for cases where all leaf nodes should have the same size (e.g., for photo browsers). For preserving order, Wood and Dykes took another approach to modify \modelSquarified by assigning to each node a two dimensional location and proposed \modelOrderedSquarified \cite{wood2008spatially}. In a similar approach, Tu \& Shen \cite{tu2007visualizing} introduce \modelSpiral treemaps that tries to preserve order and continuity while minding the aspect ratios as the rectangles are being added.
The algorithm works just like the \modelStrip algorithm except that starting from the top left corner of $R$ the direction of the strip alternates in an east-south-west-north manner. 
Other space-filling curves are also considered for generating rectangular treemaps. Tak and Cockburn proposed \modelHilbert and \modelMoore treemaps \cite{tak2012enhanced} based on Hilbert and Moore fractal-like curves and showed that they perform well with respect to stability measures. Fractal-like treemaps is also used by D'Ambros \etal \cite{dAmbros2005fractal} to visualize the evolution of software systems which can be used for over preservation.
The main goal in these space-filling curve-based treemaps is to preserve order and maximize stability and \emph{readability}, where the latter is a measure to show the benefit of ordered layouts and quantifies the simplicity of visually scanning a treemap layout by counting e.g. the number of changes in direction a viewer's eye must make when scanning the rectangles in order \cite{bederson2002ordered}. In contrast to these, our spiral algorithms do not aim to preserve order and instead try to construct a balance between stability and aspect ratios. For this reason and the fact that they generally perform poorly with respect to aspect ratio, as shown in \cite{tu2007visualizing,tak2012enhanced,sondag2018stable,vernier2020quantitativeJ}, we skip \modelSpiral, \modelHilbert, and \modelMoore in our computational comparisons.

Vernier \etal developed \modelGreedyInsert \cite{vernier2018stable}, aiming to balance aspect ratio and stability criteria, although the results suggests that the algorithm performs well on stability while scoring poorly on aspect ratio.
It inserts sub-rectangles into the given rectangle one-by-one greedily and in the given order. Sondag \etal proposed \modelIncremental \cite{sondag2018stable} that tries to achieve stability through local moves when an initial layout is given. 
For a thorough comparative evaluation of dynamic treemap algorithms, we refer the reader to Vernier \etal \cite{vernier2020quantitativeJ}. This study runs an extensive computational experiments over more than 2000 data sets and compares 14 well-known treemap algorithms, summarized in Fig. 8 -- Fig. 11 in \cite{vernier2020quantitativeJ}, and suggests that \modelSquarified \cite{bruls2000squarified} almost consistently outperforms all other well-known algorithms with respect to aspect ratio (our main metric in this paper). This is also demonstrated by the computational results over two large data sets summarized in Fig. 21 -- Fig. 26 in the paper \cite{sondag2018stable}. The latter paper also shows that it performs reasonably well with respect to stability on one of the large data sets and performs weaker on the other one. For this reason we compare our results with those of \modelSquarified. For the same reason

\subsubsection{Other Variants}

There have also been a number of modifications to these algorithms to increase aesthetics and readability.

Baker \& Eick \cite{baker1995space} adapted \modelSliceDice \cite{shneiderman1991tree,shneiderman1992tree} by adding extra layers of information for visualization of software subsystems and their associated statistics.
Likewise, van Wijk \& van de Wetering \cite{van1999cushion} added a shaded surface to each rectangle to highlight the hierarchical structure.

To have treemaps simulate standard chart types, Vliegen \etal \cite{vliegen2006visualizing} created several treemap approaches and pixel transformations.
As part of this, they developed a \modelMixed algorithm that partitions higher-level nodes according to \modelSliceDice  \cite{shneiderman1991tree,shneiderman1992tree} and places leaf nodes using \modelSquarified \cite{bruls2000squarified}.
They also let users manually set the direction of cut in \modelSliceDice at each iteration, which would, e.g., let the user compare different categories of data side-by-side.
Another of their variations was to change \modelSquarified so that the vanishing point of small rectangles could be set to center instead of a default corner.
Likewise, Yang \etal \cite{yang2015cabinet} presented a variant of \modelSliceDice intended to better highlight the hierarchical structure in which the layout resembles objects stacked in a cabinet.

To have treemaps better show spatial ordering, e.g., for a cartogram, Wood \& Dykes \cite{wood2008spatially} modified \modelSquarified \cite{bruls2000squarified} so that it both follows the given order of the nodes and at the same time the Euclidean distance between the first placed rectangle (the top left corner) and all other rectangles increases to the last placed rectangle in a more smooth spatial transition.
The algorithm modifies the ordered list by associating a location dimension to each node and then sorting the list again with regards to the ascending order of the spatial distance between the child  nodes (rectangles) and their parent nodes.
It then follows the recursive steps of \modelSquarified but instead of selecting nodes from a sorted list of areas, it selects the node closest to the current position in the layout rectangle.

Duarte \etal \cite{duarte2014nmap} generalized the concept of order preservation to neighborhood preservation. They developed the \modelNmap algorithm that uses a slice and scale strategy to take a spatial data set in which each data point is assigned a location $(x,y)\in\mathbb{R}^2$ and a weight $p\in \mathbb{R}$ and present it as a treemap where the closeness of rectangles tends to follow the closeness of the data points in $\mathbb{R}^2$, while maintaining the visual quality.

Unlike most treemap algorithms which are \textit{constructive}, the \modelIncremental by Sondag \etal \cite{sondag2018stable} uses a local search.
It can take an initial layout generated by any of the constructive algorithms and then tries to improve through a series of local moves like stretching and flipping, followed by the readjustment of areas.
As a result, it can potentially generate all possible layouts---including \emph{non-sliceable} layouts, which are those that cannot be recursively sliced into two parts having at least one rectangle as a center of a windmill pattern.
The algorithm generates layouts that score well both in aspect ratio and stability.
However, the output quality and computational time greatly depends on the quality of the initial constructed layout as it takes $\mathcal{O}(n^2)$ to move from one layout to another---in addition to solving several linear equality systems for readjusting areas in each iteration.

Non-rectangular treemaps have also been very popular.
Balzer \etal \cite{balzer2005voronoiA,balzer2005voronoiB} developed an algorithm for generating Voronoi treemaps, where the planar figures representing the leaf nodes are the cells of a Voronoi diagram instead of rectangles.
The algorithm is based on computing a weighted centroidal Voronoi tessellation (CVT) in which the weight of each sub-region is adjusted so the area of each cell is within a threshold $\epsilon$ of the input (desired) areas.
They provide an variant using additively-weighted Voronoi diagrams and power diagrams.
Voronoi treemaps enable layouts within areas of arbitrary shape, such as triangles and circles.
However, computation of CVTs is expensive and not suitable for dynamic updates in the input data.

Others tried to improve the computational efficiency and other aspects of Voronoi treemaps \cite{reitsma2007weight,sud2010fast,nocaj2012computing,nocaj2012organizing,tua2021voronoi}. Wang \etal \cite{wang2020generating} developed an orthogonal Voronoi treemap in which borders of cells are axis-aligned. Similarly, Wang \etal \cite{wang2023voronoi} proposed algorithms for generating Voronoi treemaps based on Manhattan ($\ell_1$) and Chebychev ($\ell_{\infty}$) distances. 
Liang \etal \cite{liang2015divide} extend Engdahl's \modelSplit \cite{engdahl2005ordered} and the polygonal division idea of Nguyen \& Huang \cite{nguyen2003space} and use \modelDC to create polygonal, angular, and---relevant to this paper---rectangular treemaps that could be laid out in almost any container shape.

The use of space-filling fractal-like curves may also produce non-rectangular layouts such as \modeJigsaw treemaps \cite{wattenberg2005note} and \modelGosper treemaps \cite{auber2013gospermap}.
Later, Chaturvedi \etal \cite{Chaturvedi14GroupinBox} proposed two heuristic-based treemap layouts which did not have any space-filling guarantees. Here, we only consider deterministic data, but rectangular treemaps are also produced in presence of uncertainty \cite{sondag2020uncertainty}, in which case the layout structure could be overlapping. For a review on different variants of treemaps see \cite{scheibel2020taxonomy}.

\subsubsection{Performance Considerations}

Despite the wide variety of rectangular and non-rectangular treemap algorithms, less attention has been paid to performance guarantees.

Nagamochi \& Abe \cite{nagamochi2007approximation} consider the problem of partitioning a rectangle $R$ into $n$ rectangles with specified areas $A_1,...,A_n$ with objective functions such as the sum of the perimeters of the sub-rectangles (PERI-SUM), the maximum perimeter of the sub-rectangles (PERI-MAX), and the maximum aspect ratio of the sub-rectangles.
They then present an $\mathcal{O}(n\log n)$ algorithm with a 1.25 approximation factor for PERI-SUM,  a $2/\sqrt{3}$ approximation factor for PERI-MAX, and aspect ratio of at most $\max\{\AR(R), 3, 1+\max_{i=1,...,n-1} \frac{A_{i+1}}{A_i}\}$.
The algorithm recursively partitions $R$ into two rectangles $R_1$ and $R_2$ such that $\Area(R_1)\geq \frac{1}{3}\Area(R)$ and $\Area(R_2)\geq \frac{1}{3}\Area(R)$---except when the maximum input area is greater than $ \frac{1}{3}\Area(R)$, in which case $\Area(R_1)$ may not meet this condition.
If the input rectangle is a square, this improves the 1.75 approximation factor and the computational complexity of Beaumont \etal's algorithm for PERI-SUM \cite{beaumont2002partitioning}, while extending the results for both PERI-SUM and PERI-MAX to the case where the input rectangle is not necessarily a square.
Bui \etal \cite{bui2019three} study the same problems under a strict rule that all sub-rectangles must be constructed by two-stage guillotine cuts, first with cuts parallel to the longer edge of $R$ and then with cuts perpendicular to the first layer of cuts.

Onak \& Sidiropoulos \cite{onak2008circular} present an algorithm using convex polygons as the sub-regions, where the aspect ratio of each polygon, defined as $\AR(P)=\frac{\diam(P)^2}{\vol(P)}$, is small.
For a tree with $n$ leaf nodes and depth $d$, this aspect ratio is of $\mathcal{O}\left((d\cdot\log n)^{17}\right)$.
This weak bound was later improved to $\mathcal{O}(d+\log n)$ in \cite{deBerg2010fat,de2013fat} and to $\mathcal{O}(d)$ in \cite{deBerg2014Treemaps}.
They further prove that orthoconvex treemaps; where the sub-regions representing leaf node are rectangles, L-, and S-shapes, and sub-regions representing internal nodes are orthoconvex polygons; can be constructed with constant aspect ratio.

For this paper, note the time complexities of \modelSquarified \cite{bruls2000squarified}---and \modelDC \cite{liang2015divide}---$\mathcal{O}(n\log n)$.

\section{Rectangular Treemapping as an Optimization Model}
\label{sec:optModel}

Treemapping is essentially a geometric optimization problem.
There are many closely-related geometric and space partitioning optimization problems that include packing, covering, and tiling---generally focused on minimizing wasted space or optimally allocating geographical resources.
Related problems include: cutting stock; knapsack; bin packing; guillotine; disk covering; polygon covering; kissing number; strip packing; square packing; squaring the square; squaring the plane; and, in 3D space, cubing the cube and tetrahedron packing.

The treemapping problem---specifically with the goal of minimizing the maximum aspect ratio of all sub-rectangles---was noted as NP-hard by Bruls \etal \cite{bruls2000squarified}.
de Berg \etal later proved the problem is strongly NP-hard with a reduction from the square packing problem \cite{deBerg2014Treemaps}.
The related treemapping problem of minimizing the total perimeter of all sub-rectangles was proved by Beaumont \etal \cite{beaumont2001matrix} to be NP-hard, using a reduction from the problem of partitioning a set of integers into two subsets of equal sum.
Given this computational complexity, several heuristics have been developed for generating treemaps efficiently, as reviewed in the last section. In this paper, we will also introduce several new heuristic algorithms.

Optimization approaches is understudied in the literature of the treemapping problem. Zhao and Lu \cite{zhao2015variational} modeled circular treemapping problem as an optimization problem and developed a variational layout heuristic algorithm based on power diagram to solve it. Carrizosa \etal \cite{carrizosa2017visualizing} proposed a specific type of treemap called space-filling box-connected map (SBM), with orthogonal sub-regions that are made of grid cells and have to satisfy the so-called \emph{box connectivity constraints}, and modeled it as a mixed integer nonlinear program (MINLP) in which the objective is to minimize total dissimilarities between sub-regions as measured by various distance functions defined specifically for such sub-regions. They then solved it heuristically using a Large neighborhood Search algorithm. Fried \etal \cite{fried2015isomatch} proposed a $p$-norm (energy function) minimization model for the problem of assigning a set of visual objects to a set spatial locations in form of a 2D grid. The pairwise distances between objects are given and the goal is to minimize the discrepancies between the given distances and the mapped distances measured by the Euclidean distances between the grid cells. They then solve the problem using a heuristic called \emph{IsoMatch}.

F{\"u}genschuh and F{\"u}genschuh \cite{fugenschuh2008integer} modeled a metal sheet product design problem as a 2D grid slicing problem as a binary integer program that in essence has some similarities with the border length minimization in a grid-based orthogonal treemapping problem. 
F{\"u}genschuh \etal \cite{fugenschuh2014exact} considered the problem of minimizing total border length in partitioning a rectangle to a set of sub-rectanlged with given areas. They reiterated an MINLP model previously presented and linearized as a mixed integer linear program in \cite{fugenschuh2008verfeinerte}, modified the approximation algorithm of \cite{nagamochi2007approximation}, and used the solutions provided by that as advanced starters and showed how this helps to speed-up the optimization process. We noticed this work after finalizing this research and it seems to be the closest in the literature to our model, which is more detailed, explicit, and comprehensive. Our model incorporates several variants of rectangular treemapping and considers a much bigger feasibility region. This gives the user much more flexibility in generating treemaps with specifics designs or characteristic requirements.

In order to provide a baseline for comparison, we propose an integer nonlinear \modelOptimization that minimizes the total perimeter of all sub-rectangles.
This is motivated by the fact that the perimeter of a rectangle is minimized when it is a square. 
Note that this is a proxy for minimizing the average aspect ratio and the optimal solution will consist of square-like rectangles.
One could consider minimizing the maximum perimeter or the maximum aspect ratio as well.
Nevertheless, the solution to these problems may not be the same. To see this, consider an example with four rectangles and two solutions with aspect ratios $\{1,1,1,5\}$ and $\{1,2,3,4\}$. The former is a better solution concerning the average aspect ratio, while the latter is better regarding the maximum aspect ratio.

To formally define our problem, suppose we are given a rectangle $R$ with $\width(R)=w$ and $\height(R)=h$ aligned orthogonally to the $x$ and $y$ axes with its lower-left corner at the origin.
We are also given numeric areas $A_1,...,A_n$ s.t. $\sum_{i=1}^n A_i = \Area(R)$.
We want to partition $R$ into $n$ non-overlapping rectangular sub-regions $R_1,...,R_n$ with areas $A_1,...,A_n$ in a way that the total perimeter of all rectangles (cut length) is minimized.

One of the advantages of our \modelOptimization is that it can be easily adjusted to incorporate specific constraints or other evaluation metrics. One could change the objective function from total perimeter to maximum aspect ratio or area weighted average aspect ratio or any other metric that could be constructed as a function of dimensions or coordinates of the sub-regions. 

Here, we consider a parametrized objective function to take advantage of this flexibility. We consider an area-weighted total perimeter. We can switch between the total perimeter and area-weighted version using a binary parameter $\alpha \in \{0,1\}$.
We also incorporate another degree of flexibility in our objective function which may be very useful in practice. Often we may want to compromise the aspect ratio slightly and in return gain rectangles that are more horizontal, suitable for longer labels, or more aligned with each other, making it easier to compare the areas. In order to do that, we define our objective function as a weighted-average of total area-weighted perimeter and the number of rectangles having greater width than height. 
Greater values for the parameter $\beta \geq 0$ puts more weight on creating more horizontal rectangles. 

Similarly, we might be interested to modify the objective function to incorporate our stability metric as well to make a trade-off between perimeter (aspect ratio) and stability. We must point out that since all of sub-regions in the generated treemaps are rectangles and thus convex, for any pair $R, R^{\prime}$, the function $\HD(R,R^{\prime})$ is convex. However, since $R^{\prime}$ is the corresponding sub-rectangle to $R$ in the updated layout after perturbing areas (i.e., areas changing over time), the exact formula of this metric and the way it should be added to the objective function and then the solution analysis of that model is beyond the scope of this paper. However, we can easily incorporate some design preferences such as closeness of some sub-regions. Depending on the application, we may be interested in having certain sub-rectangles close to each other in the final treemap. Let, $\bm{v}^i$ denote the lower-left vertex of $R_i$. We can measure the closeness between $R_i$ and $R_j$ by the Euclidean distance between their lower-left corners, i.e., $\|\bm{v}^i - \bm{v}^j \|$. 

We can also consider additional positioning and shape constraints. For example, here we allow the user to opt, via parameter $\delta_i \in \{0,1\},\; \forall i$, to fix the position of one sub-rectangle to one of the corners of the input rectangle. We also allow the user to have two particular sub-rectangles to be adjacent, i.e., sharing a corner, an edge (two corners), or part of an edge, by adjusting parameters $\eta_{ij}, \theta_{ij} \in \{0,1\},\; \forall i,j$.
Note that one could do this for multiple rectangles too, however it may increase the risk of infeasibility.

In the following we present an MINLP model for the considered problem. The parameters for any rectangles $R_i, R_j$ with $i,j\in\{1,...,n\}$ are:
\begin{IEEEeqnarray*}{lCl}
    \alpha  &=& \left\{ \begin{array}{ll}
                1 & \text{area-weighted objective};\\
                0 & \mbox{otherwise}.\end{array} \right. \\
    \beta    &:& \text{a weight parameter for horizontality} \\
    \gamma_{ij}  &=& \left\{ \begin{array}{ll}
                1 & \text{if } R_i \text{ is preferred to be close to } R_j;\\
                0 & \mbox{otherwise}.\end{array} \right. \\
    \delta_i  &=& \left\{ \begin{array}{ll}
                1 & \text{if } R_i \text{ has to be on the lower left corner};\\
                0 & \mbox{otherwise}.\end{array} \right. \\
                & & \text{We must have } \delta_1+ \cdots +\delta_n \leq 1 \\ 
    \eta_{ij}  &=& \left\{ \begin{array}{ll}
                1 & \text{if } R_j \text{ must be adjacent to and to the right of } R_i;\\
                0 & \mbox{otherwise}.\end{array} \right. \\
                & & \text{We must have } \eta_{ij} + \eta_{ji} \leq 1. \\
    \theta_{ij}  &=& \left\{ \begin{array}{ll}
                1 & \text{if } R_j \text{ must be adjacent to and above } R_i;\\
                0 & \mbox{otherwise}.\end{array} \right. \\
                & & \text{We must have } \theta_{ij} + \theta_{ji} \leq 1.                 
\end{IEEEeqnarray*}

The decision variables, for any rectangles $R_i, R_j$ with $i,j\in\{1,...,n\}$ are:
\begin{IEEEeqnarray*}{lCl}
    w_i     &:& \text{width of~} R_i\\
    h_i     &:& \text{height of~} R_i\\
    \bm{v}^i & : & \text{ lower-left corner of~} R_i \text{ with }  \bm{v}^i = (v^i_{x},v^i_{y}) \\
    x_{ij}  &=& \left\{ \begin{array}{ll}
                1 & \mbox{if $v^{j}_{x}<v^{i}_{x}+w_i$};\\
                0 & \mbox{otherwise}.\end{array} \right.\\
    y_{ij}  &=& \left\{ \begin{array}{ll}
                1 & \mbox{if $v^{j}_{y}<v^{i}_{y}+h_i$};\\
                0 & \mbox{otherwise}.\end{array} \right. \\
    z_{i}  &=& \left\{ \begin{array}{ll}
                1 & \mbox{if $h_i \leq w_i$};\\
                0 & \mbox{otherwise}.\end{array} \right.
\end{IEEEeqnarray*}

We can write our optimization problem as:
\begin{IEEEeqnarray}{rCl}
    \IEEEeqnarraymulticol{3}{l}{
    \minimize \sum_{i=1}^{n} \left( (1-\alpha +\alpha A_i) ((w_i+h_i) - \beta z_i) \right) }  \nonumber \\* 
     \IEEEeqnarraymulticol{3}{l}{ \qquad \qquad \qquad +  \sum_{i,j=1}^n \gamma_{ij} \|\bm{v}^i - \bm{v}^j \|     \qquad \qquad \qquad \sut } \nonumber  \\*
    \log A_i - \log w_i - \log h_i     & \leq  & 0, \qquad \qquad \qquad \forall i                      \label{eq:modelArea}                   \\
    v^{i}_{x}+w_i                           & \leq  & w, \qquad \qquad \qquad \forall i                      \label{eq:modelWidth}                    \\
    v^{i}_{y}+h_i                           & \leq  & h, \qquad \qquad \qquad \forall i                      \label{eq:modelHeight}                     \\
    v^{i}_{x}+v^{i}_{y}                     & \leq  & (w+h)(1-\delta_i), \qquad \forall i               \label{eq:LLCornerRect}     \\
    v^{i}_{x}-v^{j}_{x}+w_i           & \leq  & wx_{ij},  \qquad \qquad \forall i,j  \label{eq:modelNonOverlappingX}   \\
    v^{i}_{x}-v^{j}_{x}+w_i           & \geq  & (\epsilon-w(1-x_{ij}))(1-\eta_{ij}), \, \forall i,j  \; \label{eq:modelOverlappingX}     \\
    x_{ij}				& \leq & 1-\eta{ij} \qquad \qquad \forall i,j  \label{eq:modelAdjacentX}   \\
    v^{i}_{y}-v^{j}_{y}+h_i          & \leq  & hy_{ij},  \qquad \qquad \forall i,j   \label{eq:modelNonOverlappingY}   \\
    v^{i}_{y}-v^{j}_{y}+h_i          & \geq  & (\epsilon-h(1-y_{ij}))(1-\theta_{ij}), \: \forall i,j  \label{eq:modelOverlappingY}     \\
    y_{ij}				& \leq & 1-\theta_{ij} \qquad \qquad \forall i,j  \label{eq:modelAdjacentY}   \\
    x_{ij} + x_{ji} + y_{ij} + y_{ji}  & \leq  & 3, \quad \forall i,j    \label{eq:modelNonOverlapping}    \\
    w_i                           & \geq  & h_i-h(1-z_{i}), \qquad \forall i 			\label{eq:horizontalW}  \\
    h_i                           & \geq  & w_i-wz_{i}, \qquad \qquad \forall i 			\label{eq:horizontalH}  \\
    v^{i}_{x}, \; v^{i}_{y} , \; w_i, \; h_i  & \geq & 0, \qquad \qquad \forall i \nonumber \\
    x_{ij}, \; y_{ij}                  & \in   & \{0,1\}, \qquad \forall i,j \nonumber \\
        z_{i}                  & \in   & \{0,1\}, \qquad \forall i \nonumber 
\end{IEEEeqnarray}
where $\epsilon>0$ is a very small real number.
Constraint \eqref{eq:modelArea}, written in its convex form, is to ensure that each rectangle $R_i$ with width $w_i$ and height $h_i$ will have its associated area as $A_i = w_i \times h_i$. Note that this constraint will be active at optimality.
Constraints \eqref{eq:modelWidth} and \eqref{eq:modelHeight} guarantee that the ending point of any rectangle on both axes should not violate the width and height of $R$.
Constraint \eqref{eq:LLCornerRect} allows the solver to fix one rectangle's position to have its lower left corner on the origin.
Constraints \eqref{eq:modelNonOverlappingX}--\eqref{eq:modelNonOverlapping} are added to avoid overlapping among rectangles as well as to enforce adjacencies if required. Finally, constraints \eqref{eq:horizontalW} and \eqref{eq:horizontalH} determine weather a sub-rectangle is horizontal.
Fig. \ref{fig:optModelHorizontal} shows an implementation of this new model over random examples in layout boxes with aspect ratios 1,2, and 3, and compares the sensitivity of the results to one of the parameters, i.e., different values of $\beta$.
\newcommand\tabCellImage[3]{
    \begin{subfigure}[c]{.2\textwidth}%
        \includegraphics[sqrtofarea=.85\textwidth]{#2}%
        \caption{Perimeter = #1}%
        \label{#3}%
    \end{subfigure}%
}
\begin{figure*}[tbp]
   \centering
    \hspace{-0.8cm}
    \begin{tabular}{ccc}
        	 \hspace{0.3cm} 		$\beta$=0 & \hspace{0.6cm} $\beta$=0.05  &  \hspace{1cm}   $\beta$=0.10 \\
         \hspace{0.6cm} \tabCellImage{\best{10.3906}}  {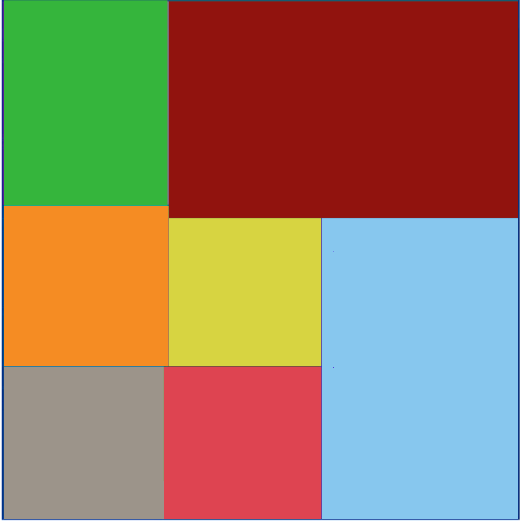}  {fig:optimal_AR=1}
        &\hspace{1.2cm} \tabCellImage       {\rnup{10.4856}}   {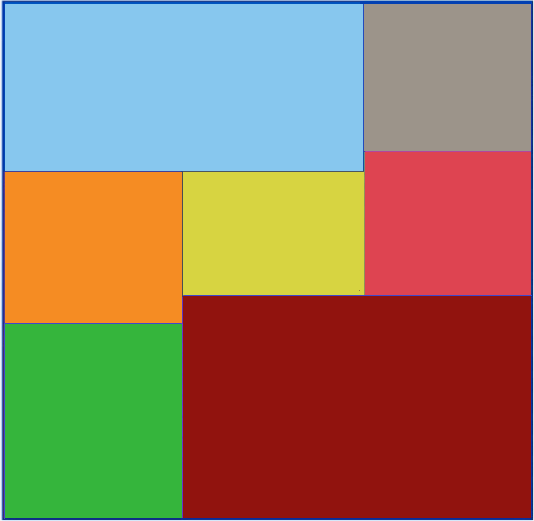}  {fig:beta0.05_AR=1}
        &  \hspace{1.5cm} \tabCellImage      {10.6854}  {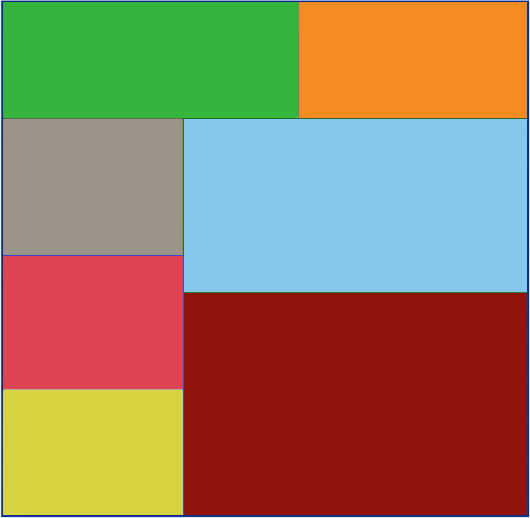} {fig:beta0.10_AR=1}
        \\
         \hspace{-0.8cm} \tabCellImage{\best {11.2766}}  {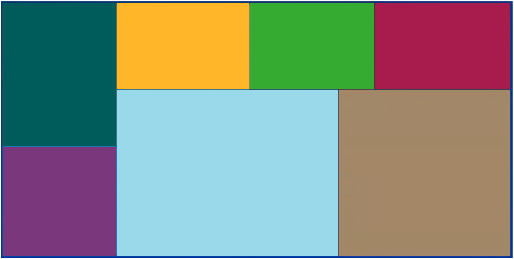}     {fig:optimal_AR=2} 
       & \hspace{0.3cm} \tabCellImage {\best{11.2766}}   {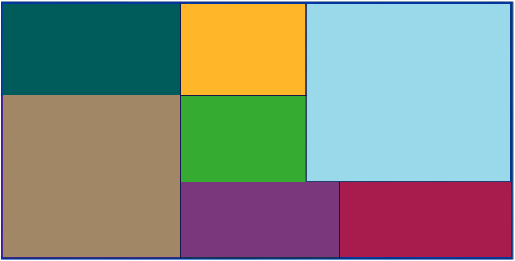}   {fig:beta0.05_AR=2} 
        & \hspace{0.5cm} \tabCellImage       {11.4870}    {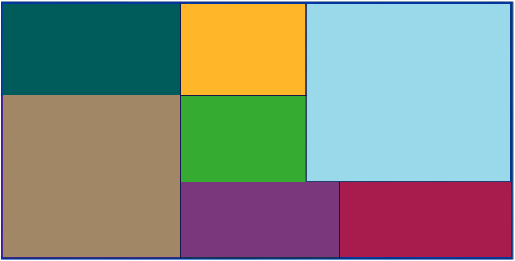}      {fig:beta0.10_AR=2} 
        \\
         \hspace{-1.5cm} \tabCellImage{\best {10.3842}}     {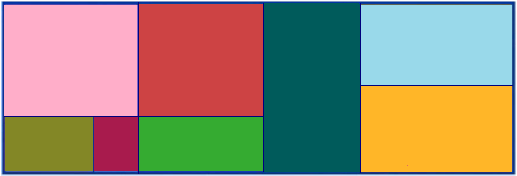}     {fig:optimal_AR=3} 
      	 \hspace{0.2cm} &    \tabCellImage {\best{10.3842}}    {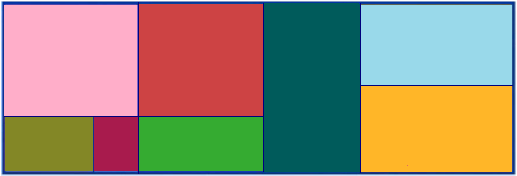}   {fig:beta0.05_AR=3} 
	  \hspace{0.4cm} &   \hspace{-0.45cm}      \tabCellImage       {11.7842}    {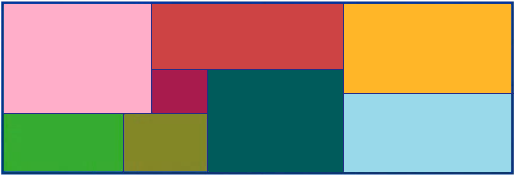}        {fig:beta0.10_AR=3} 

        \\

     \end{tabular}
    \caption{\protect
    An illustration of treemaps generated by our \modelOptimization for three different wight factors $\beta=0,\;0.05,\; 0.1$, while all other parameters are set to zero. Larger values of $\beta$ generate more horizontal rectangles to fit longer labels. Note the \emph{non-sliceable} layouts in cases: \subref{fig:optimal_AR=1}, \subref{fig:beta0.05_AR=1}, and \subref{fig:beta0.10_AR=3}. For each treemap, the total perimeter of the component subrectangles is shown with the \bestName (shortest) and \rnupName values shown using color.
    }
    \label{fig:optModelHorizontal}
\end{figure*}

A major advantage of this \modelOptimization approach is that it focuses on the optimal placement of each sub-region as opposed to dividing the region by guillotine cuts.
Therefore, it explores all possible layouts including the \emph{non-sliceable} layouts (see \cref{fig:Random_Optimal,fig:Midwest_Unitbox_Optimal}).
This problem is NP-hard, as mentioned above and proved in \cite{beaumont2001matrix}.

In the rest of this paper, we provide several suboptimal but efficient algorithms for this optimization problem.
In \cref{sec:SubdivisionAlgs} we will present our subdivision-based algorithms and in \cref{sec:SpiralAlgorithms} we will develop our packing-based algorithms. In these algorithms, similar to our \modelOptimization, minimizing the total perimeter (and thus the relevant aspect ratio measures) is our main goal and stability is more of a secondary metric for comparison and is not directly targeted. However, by construction, stability has a higher weight in our spiral algorithms. 
We compare these to \modelOptimization which gives the absolute best treemaps concerning aspect ratios (when the input region is given), \modelSquarified \cite{bruls2000squarified} which has one of the best performance in the literature with respect to creating square-like rectangles, and \modelDC \cite{liang2015divide} due to its similar approach and that it also performs well and can lay out rectangles in any container shape.

\section{Subdivision Algorithms}
\label{sec:SubdivisionAlgs}

\subsection{A Modified Divide and Conquer Algorithm}
\label{sec:ModifiedDC} 
         
Given the structure of the problem, which is based on a tree, it is natural to take a divide and conquer approach.
Our divide and conquer algorithm is in the same spirit of Liang \etal's \modelDC \cite{liang2015divide} algorithm, which appears to be an extension of Engdahl's \modelSplit \cite{engdahl2005ordered} and the best already existing divide and conquer approach to treemapping.
\modelDC tries to divide the input areas into two lists of equal weights in each iteration and recursively continues that until each sublist has only one area.

However, this is sensitive to the input parameters and works well only when there is no extreme area.
E.g., take a $8\times 4$ rectangle to be partitioned into 18 sub-rectangles with areas $\{15,1,1,1,...,1\}$.
The algorithm sets $S_1=\{R_1,R_2\}$ and $S_2=\{R_3,...,R_{18}\}$ with $\sum_{i: R_i\in S_1} A_i = \sum_{i: R_i\in S_2} A_i = 16$.
This will generate a very thin rectangle $R_2$.
In the rectangular version of the algorithm we would have $\Perim(R_2)=2\times(4+1/4)=8.5$ and $\AR(R_2) = 16$.
It is clear that setting $S_1=\{R_1\}$ and $S_2=\{R_2,R_3,...,R_{18}\}$ could generate a much better result with the maximum aspect ratio very close to 1.

An alternative approach is to divide the list of areas in half according to the indices, i.e., $S_1=\{R_1,...,R_8\}$ and $S_2=\{R_9,...,R_{18}\}$ for our example above, and then recur on each sub-list.
But this suffers from the same problem as \modelDC \cite{liang2015divide} with extreme areas.
However, we propose a remedy.

In our new \modelModifiedDC algorithm we first divide the input set of areas into two equally weighted lists $S_1=\{A_1,...,A_{k_1},A_k\}$ and $S_2=\{A_{k+1},A_{k+2},...,A_n\}$. We then check an additional condition that could change the list that $A_k$ or $A_{k+1}$ belong to. For some real $c > 0$, if $|A_{\max\{1,k-1\}} - A_k | > c * |A_k - A_{\min\{k+1,n\}}|$, we pick the best out of adding $A_k$ to $S_2$ or $A_{k+1}$ to $S_1$ by running both. Otherwise, we make no change and continue the divide and conquer as usual. Parameter $c$ can be adjusted by the user and gives additional flexibility to the algorithm (in our experiments we use $c=2$). Intuitively, this algorithm tries to avoid generating long and skinny rectangles at the division points. So, a significant improvement in the maximum aspect ratio metric is expected. 

The pseudocode for \modelModifiedDC, for the case where cuts are either vertical or horizontal (depending on the width \& height of the remaining segment), is shown in \cref{alg:ModifiedDC}. As with \modelDC, it can be easily modified and generalized to handle polygonal and angular cuts and to have no restriction on input layout container shape.
This \modelModifiedDC algorithm remedies the issue in \modelDC in the same computational time of $\mathcal{O}(n \log n)$.

\begin{algorithm}[tbp]
    \SetAlgoLined
    \BlankLine   
    \SetKwFunction{main}{main}
    \SetKwFunction{proc}{Partition}
    \SetKwProg{Fmain}{Function}{}{}
    \SetKwProg{Fproc}{Procedure}{}{}
     
    \KwIn{Rectangle $R$, a list of $n$ areas $L=\{A_1,A_2,...,A_n\}$ with $\sum_{i=1}^n A_i = \Area (R)$, and a real constant $c$.}
    \KwOut{Partition $R_1,R_2,...,R_n$ with areas $A_1,A_2,...,A_n$.} 
    \BlankLine
    \tcc{***********************************************}
     \Fmain{\main{R,L}}{
    	Let $w$ denote the width of $R$ and $h$ be the height\;
        \eIf{$\left\vert L \right\vert =1$}{
            \Return{$(w+h,R)$}\;
        }{ 
        Sort $L$ in a non-increasing order and reindex the sorted areas as $A_1,A_2,...,A_n$\; 
         \KwRet{${\sf Partition}(R,L,1,n,c)$}\;
         }
     }
    \setcounter{AlgoLine}{0}
    \Fproc{\proc{$Q, L,\mbox{start, stop},c$}}{
        Let $w$ denote the width of $Q$ and $h$ be the height\;
        Set $S_1=0$ and $S_2=\sum_{i=start}^{stop}{A_i}$\;
        \For{$k=\textrm{start : stop}-1$}{
              Set $S_1 = S_1 + A_k$\;
              Set $S_2 = S_2 - A_k$\;
              \If{$|S_1-S_2| < |(S_1+A_{k+1}) - (S_2-A_{k+1})|$}{
                	 break\;
	      }
        }
        \eIf{$A_{\max(start,k-1)}-A_k > c * (A_k-A_{\min(k+1,stop)})$}{
            Set $S_1^{\prime} = S_1 - A_k$ and $S_2^{\prime} = S_2 + A_k$\; 
            Set $S_1^{\prime\prime} = S_1 + A_{k+1}$ and $S_2^{\prime\prime} = S_2 - A_{k+1}$\;
            \tcc{Set $L_1^{\prime}=\{A_{start},...,A_{k-1}\}$ and $L_2^{\prime} = \{A_{k},...,A_{stop}\}$;}
            \tcc{Set $L_1^{\prime\prime}=\{A_{start},...,A_{k+1}\}$ and $L_2^{\prime\prime} = \{A_{k+2},...,A_{stop}\}$;}
            \eIf{$w > h$}{
              Divide $Q$ vertically into two pieces $Q_1^{\prime}$ with $w_1^{\prime}=S_1^{\prime}/h$ and $h_1^{\prime}=h$ on the left and $Q_2^{\prime}$ with $w_2^{\prime}=w-S_2^{\prime}/h$ and  $h_2^{\prime}=h$ on the right\;
              Divide $Q$ vertically into two pieces $Q_1^{\prime\prime}$ with $w_1^{\prime\prime}=S_1^{\prime\prime}/h$ and $h_1^{\prime\prime}=h$ on the left and $Q_2^{\prime\prime}$ with $w_2^{\prime\prime}=w-S_2^{\prime\prime}/h$ and  $h_2^{\prime\prime}=h$ on the right\;
              }{
              Divide $Q$ horizontally into two pieces $Q_1^{\prime}$ with $w_1^{\prime}=w$ and $h_1=S_1^{\prime}/w$ on top and $Q_2^{\prime}$ with $w_2^{\prime}=w$ and $h_2=h-S_2^{\prime}/w$ in bottom\;
               Divide $Q$ horizontally into two pieces $Q_1^{\prime\prime}$ with $w_1^{\prime\prime}=w$ and $h_1^{\prime\prime}=S_1^{\prime\prime}/w$ on top and $Q_2^{\prime\prime}$ with $w_2^{\prime\prime}=w$ and $h_2^{\prime\prime}=h-S_2^{\prime\prime}/w$ in bottom\;
        } 
        \KwRet{the best out of \\ $\proc(Q_{1}^{\prime},L,start,k-1,c) \cup \proc(Q_{2}^{\prime},L,k,stop,c)$ \\ and \\
        $\proc(Q_{1}^{\prime\prime},L,start,k+1,c) \cup \proc (Q_{2}^{\prime\prime},L,k+2,stop,c)$}\; 
        }{
           \tcc{Set $L_1=\{A_{start},...,A_k\}$ and $L_2 = \{A_{k+1},...,A_{stop}\}$; }
           \eIf{$w > h$}{
              Divide $Q$ vertically into two pieces $Q_1$ with $w_1=S_1/h$ and $h_1=h$ on the left and $Q_2$ with $w_2=w-S_2/h$ and  $h_2=h$ on the right\;
              }{
              Divide $Q$ horizontally into two pieces $Q_1$ with $w_1=w$ and $h_1=S_1/w$ on top and $Q_2$ with $w_2=w$ and $h_2=h-S_2/w$ in bottom\;
        } 
        \KwRet{$\proc(Q_{1},L,start,k,c) \cup \proc(Q_{2},L,k+1,stop,c)$}\;
        }         
        }
    \caption{
        \modelModifiedDC $(R,L,c)$ 
        Generates a rectangular treemap with the given areas and bounding rectangle according to a modified divide and conquer approach.
    }
    \label{alg:ModifiedDC} 
\end{algorithm}

\subsection{A Dynamic Programming Approach}
\label{sec:DP}
We can improve the quality even further. Here, we present a new \modelDynamicProg algorithm that dynamically picks the best dividing point to minimize the objective function.
First, we sort the list of areas in non-ascending order.
Then, in each step, we divide the areas into two sub-lists.
With these, we dissect $R$ into two sections by a guillotine cut so that each has an area equal to the total area of its associated list.
We find the dividing point in the list of areas in each iteration using the recursive equation:
\begin{IEEEeqnarray}{lCl}
    P_{(i)}     &:& \text{The perimeter of sub-rectangle~} R_i	\nonumber \\
    P_{(i,j)}   &:& \text{The total perimeter of sub-rectangles~} R_{i \ldots j} 		\nonumber \\
    P_{(1,n)}   &=& \min_{1\leq k\leq n} \{P_{(1,k)}+P_{(k+1,n)}\} 		\nonumber\\
    P_{(i,j)}   &=& \left\{ \begin{array}{ll} %
            P_{(i)}                                         & \mbox{if $i=j$}\\ ~\\
            \min_{i\leq k < j} \{P_{(i,k)}+P_{(k+1,j)}\}  & \mbox{if $i<j$}
    \end{array} \right. 
    \label{eq:dynamicRecursion}
\end{IEEEeqnarray}

This approach of course leads to a high quality solution due to considering a much larger feasible space, but also leads to an exponential running time of $\mathcal{O}(3^n)$. The reason is unlike the partitioning of a discrete set of numbers {areas}, each subset of areas $\{A_i,...,A_j\}$ could represent several different subproblems depending on the the location of those sub-rectangles and the sequence of vertical and horizontal cuts that lead us there. This makes it impossible to cache the best solutions for each subproblem to be used in future calls which leads to solving each subproblem from scratch.

\begin{algorithm}[tbp]
    \SetAlgoLined
    \BlankLine
    \SetKwFunction{main}{main}
    \SetKwFunction{proc}{Partition}
    \SetKwProg{Fmain}{Function}{}{}
    \SetKwProg{Fproc}{Procedure}{}{}
    
    \KwIn{Rectangle $R$ 
    and a list of $n$ areas $L=\{A_1,A_2,...,A_n\}$ with $\sum_{i=1}^n A_i = \Area (R)$.}
    \KwOut{Partition $R_1,R_2,...,R_n$ with areas $A_1,A_2,...,A_n$.} 
    \BlankLine
    \tcc{***********************************************}
    \Fmain{\main{R,L}}{
    	Let $w$ denote the width of $R$ and $h$ be the height\;
        \eIf{$n=1$}{
            \Return{$(w+h,R)$}\;
        }{ 
            Sort $L$ in a non-increasing order and reindex the sorted areas as $A_1,A_2,...,A_n$\; 
             \tcc{pBest is a structured array to store best perimeter and best partition for a block $A_i,...,A_j$ in a rectangle with width $w$ and height $h$ and its lower left corner on the origin.}
            Let $\mbox{pBest}=\{\;\}$\;
           \KwRet{${\sf Partition}(R,L,1,n)$}\;
        }
    }
    \setcounter{AlgoLine}{0}
    \Fproc{\proc{$Q, L,\mbox{start, stop}$}}{
    Let $w$ denote the width of $Q$ and $h$ be the height\;
    Let $\bm{q}=(q_x,q_y)$ be the lower left corner of $Q$ and $\bm{q_0}=(0,0)$\;
    \tcc{$\mbox{pBest}$ stores the best partition of a box $B$ with $\bm{q}_B=\bm{q_0}$ and area $\sum_{i=\mbox{\tiny start}}^{\mbox{\tiny stop}} A_i$ into sub-rectangles $R_{\mbox{start}},...,R_{\mbox{stop}}$}
    \eIf{\textrm{the 4-tuple} $\mbox{(start, stop,}w,h)$ \textrm{exists in pBest.ID}}{ 
        Let $k$ be the index in pBest.ID that stores $\mbox{(start, stop,}w,h)$\;
    	\KwRet{$(\mbox{pBest.Perim(k)},(\mbox{pBest.Par(k)} + \vec{\bm{q}}))$}\;
	}{
        Set $\mbox{bestPrm} =$ An arbitrary large number\;
        \eIf{start = stop}{
            Set $\mbox{bestPrm}=w+h$\;
            Set $\mbox{bestPar} = Q$\;
        }{
            \For{$k=\textrm{start : stop}-1$}{
                Set $S = \sum_{i=\mbox{\tiny start}}^{k}{A_i}$\;
                    Divide $Q$ vertically into two pieces $Q_1$ with $w_1=S/h$ and $h_1=h$ on the left and $Q_2$ with $w_2=w-S/h$ and  $h_2=h$ on the right\;
                    Divide $Q$ horizontally into two pieces $Q_3$ with $w_3=w$ and $h_3=S/w$ on top and $Q_4$ with $w_4=w$ and $h_4=h-S/w$ in bottom\;
                Let $[\mbox{Prm}_1, \mbox{Par}_1] = {\sf Partition}(Q_1,L,\mbox{start},k)$\;
                Let $[\mbox{Prm}_2, \mbox{Par}_2] = {\sf Partition}(Q_2,L,k+1,\mbox{stop})$\;
                Let $[\mbox{Prm}_3, \mbox{Par}_3] = {\sf Partition}(Q_3,L,\mbox{start},k)$\;
                Let $[\mbox{Prm}_4, \mbox{Par}_4] = {\sf Partition}(Q_4,L,k+1,\mbox{stop})$\;
                \eIf{$(\mbox{Prm}_1 + \mbox{Prm}_2) < (\mbox{Prm}_3 + \mbox{Prm}_4)$}{
                		Set $\mbox{tmpPrm} = \mbox{Prm}_1 + \mbox{Prm}_2$\; 
                		Set $\mbox{tmpPar} = \mbox{Par}_1 \cup \mbox{Par}_2$\;
                }{
                		Set $\mbox{tmpPrm} = \mbox{Prm}_3 + \mbox{Prm}_4$\; 
                		Set $\mbox{tmpPar} = \mbox{Par}_3 \cup \mbox{Par}_4$\;
		}
                \If{$\mbox{tmpPrm} < \mbox{bestPrm}$}{
                    Set $\mbox{bestPrm} = \mbox{tmpPrm}$\;
                    Set $\mbox{bestPar} = \mbox{tmpPar}$\;
                }
              }{
            }
        }
        Set pBest.ID = [pBest.ID, ($\mbox{start,stop,}w,h$)]\;
        Set pBest.Perim = [pBest.Perim, bestPrm]\;
        Set pBest.Par = [ pBest.Par, ($bestPar-\vec{\bm{q}}$)]\;
        }
        \KwRet{$(\mbox{bestPrm}, \mbox{bestPar})$}\;
    }
    \caption{
        \modelDynamicProg $(R,L)$ 
        Generates a rectangular treemap with the given areas and bounding rectangle according to a divide and conquer approach with optimized division process.
    }
    \label{alg:DynamicProg} 
\end{algorithm}
However, we can remedy this with some modifications. By adding the width $w_b$ and height $h_b$ of the box $B$ which we want to partition it to a block of areas $A_i,...,A_j$ to the sequence $\{i,...,j\}$ as the identifier of a subproblem we can save solved subproblems and potentially reuse them in future calls. Regardless of the location of $B$ inside the input rectangle $R$,  \eqref{eq:dynamicRecursion} will always find the same value, as long as the shape (width and height) of $B$ does not change. Hence, we define a subproblem by $(i,j,w_b,h_b)$. Using this identifier, we will save the best solution for each subproblem to avoid solving these subproblems from scratch each time they occur. Besides the perimeter we also need to save the partitioning structure. This creates a challenge because rectangles of dimension $w_b\times h_b$ can have their lower left corner almost anywhere inside $R$. To do this, we assume the left corner of $B$ in the saved subproblem to be the origin and then we define $T: \mathcal{P}+t\vec{\bm{v}}$ with $t \in \mathbb{R}$ as a translation operator on the set $\mathcal{P}$ that in the direction $\vec{\bm{v}}$, in which the shift is applied to every point $\bm{x}\in \mathcal{P}$. This operator handles the necessary shifts between the boxes. This modification improves the running time down to $\mathcal{O}(n^3)$. 

\begin{figure*}[tbp]%
    \centering%
    \begin{subfigure}[b]{.35\textwidth}%
        \includegraphics[width=\textwidth]{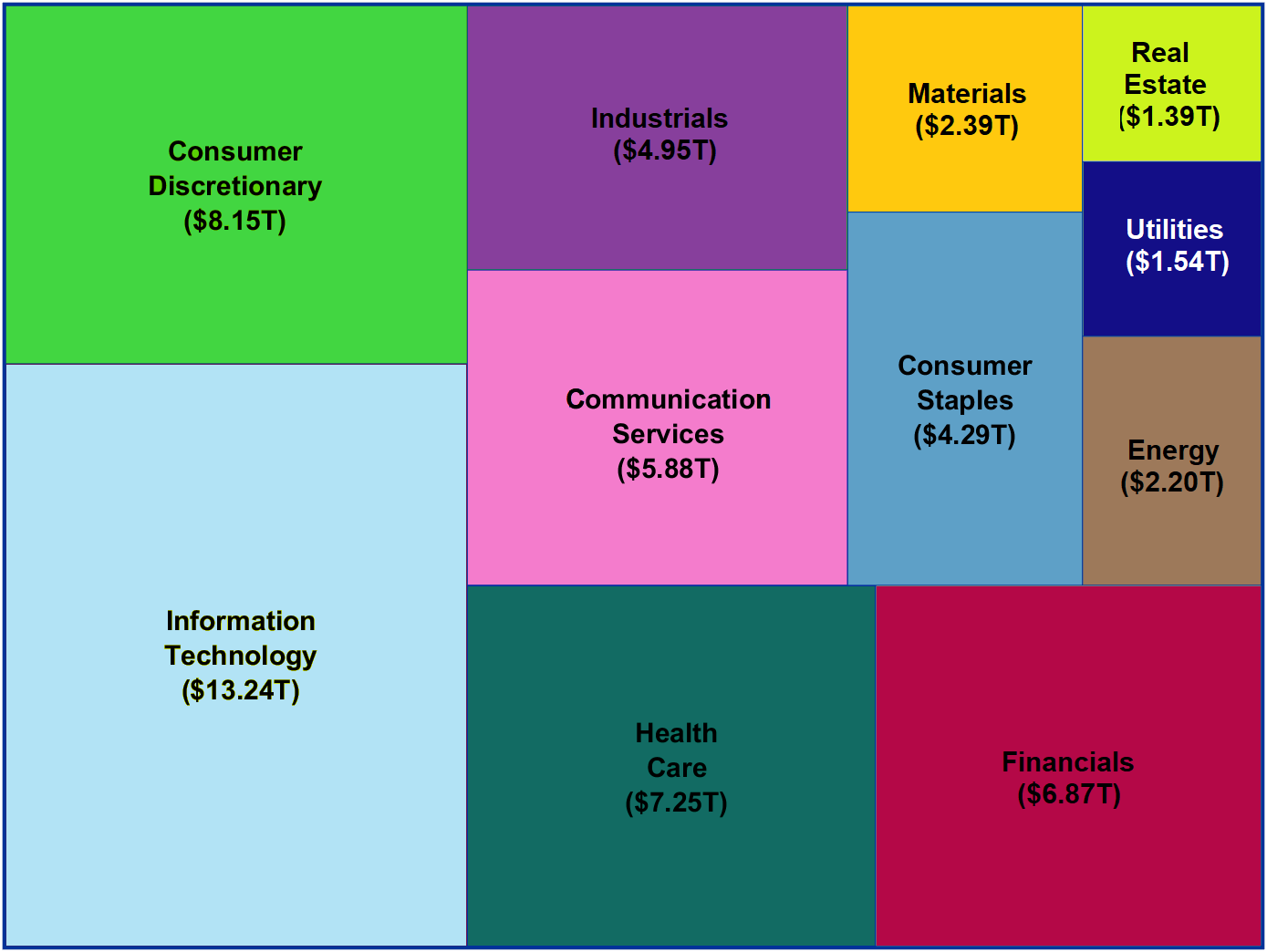}%
        \caption{}%
        \label{fig:StockMarketRect}%
    \end{subfigure}%
    %
    \hspace{0.05\textwidth}
    \begin{subfigure}[b]{.3\textwidth}%
        \includegraphics[width=\textwidth]{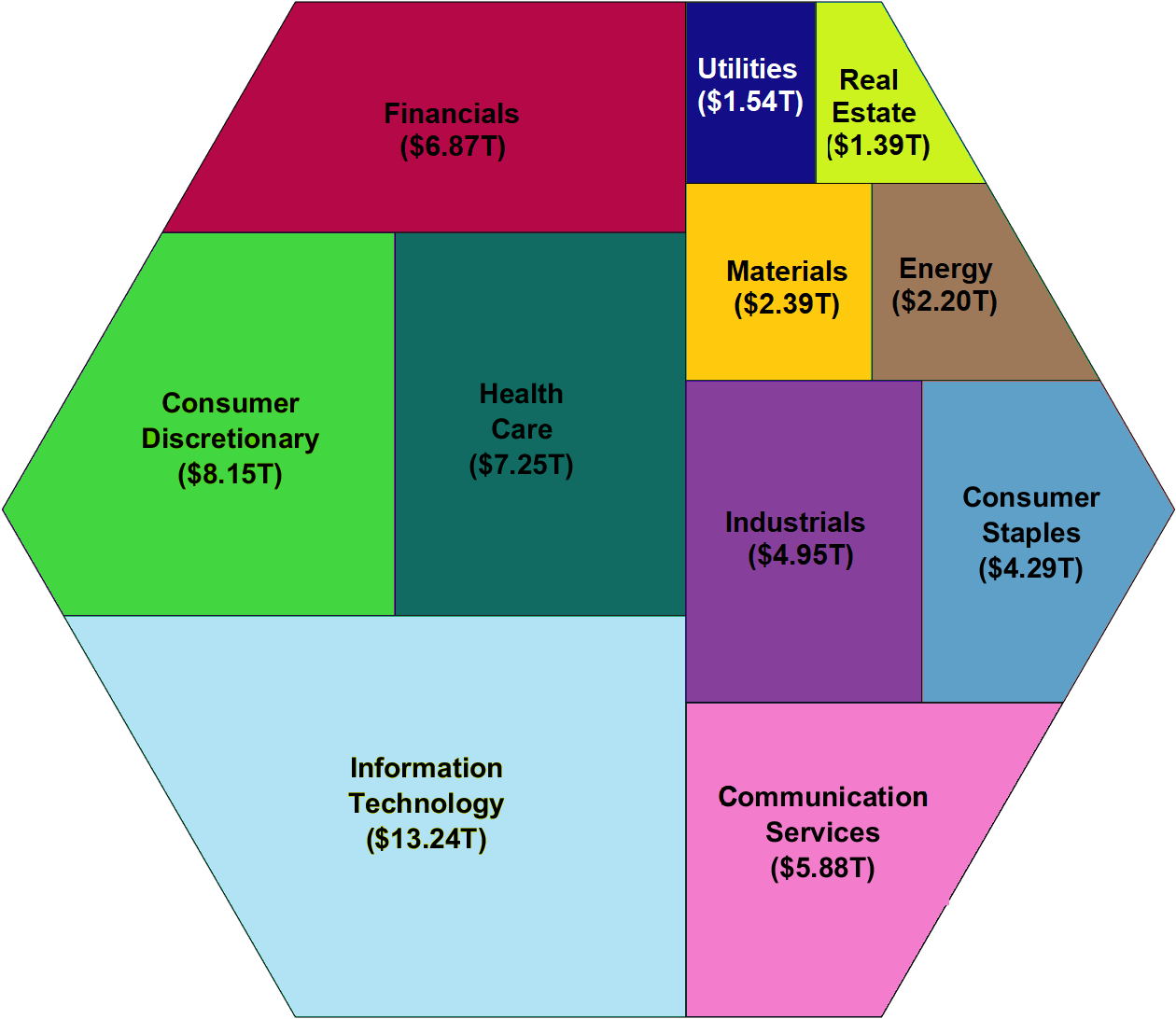}%
        \caption{}%
        \label{fig:StockMarketHex}%
    \end{subfigure}%
    \\%
    \begin{subfigure}[b]{.35\textwidth}%
        \includegraphics[width=\textwidth]{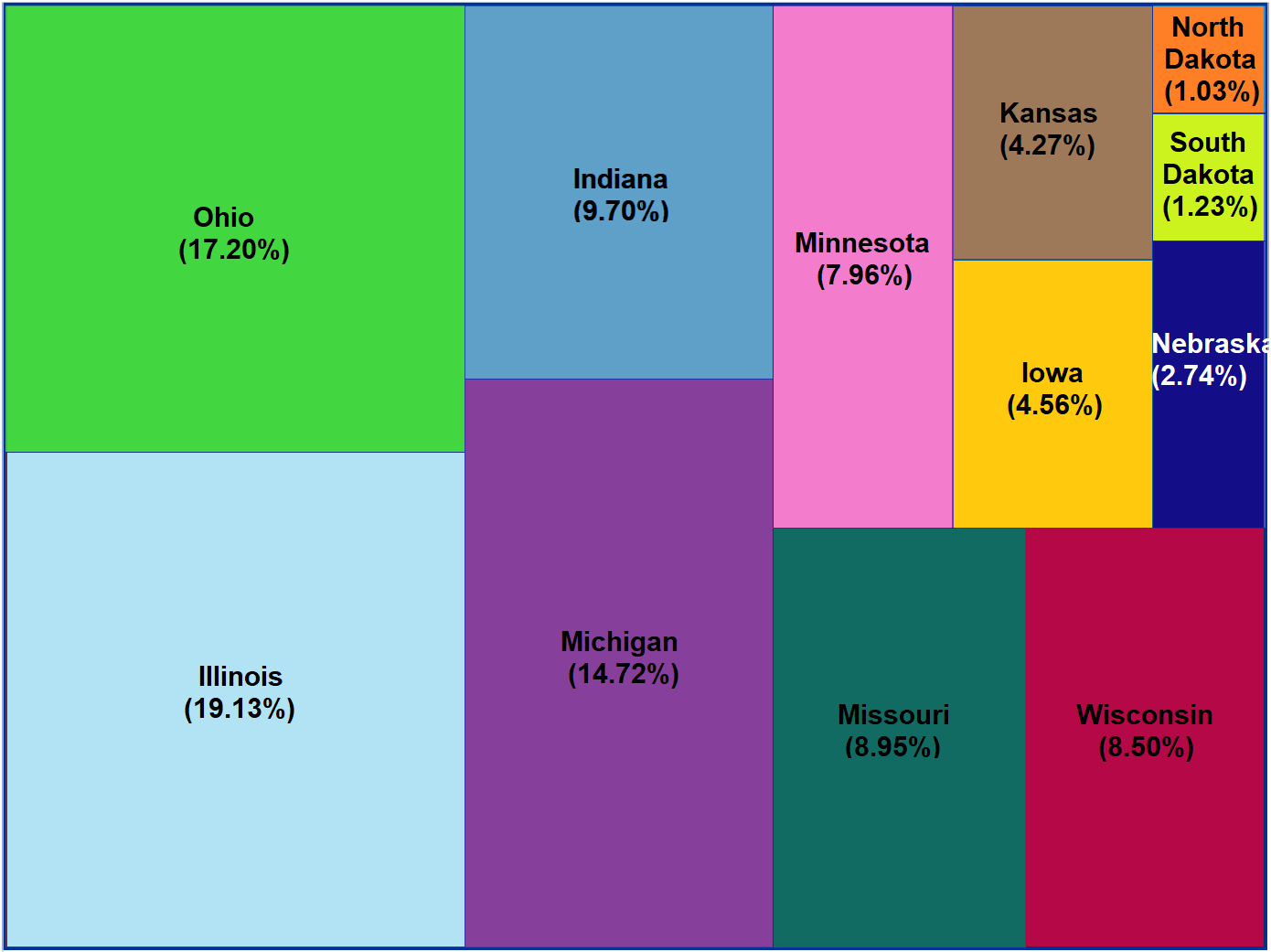}%
        \caption{}%
        \label{fig:MidwestPopulationRect}%
    \end{subfigure}%
    %
   \hspace{0.05\textwidth}
    \begin{subfigure}[b]{.3\textwidth}%
        \includegraphics[width=\textwidth]{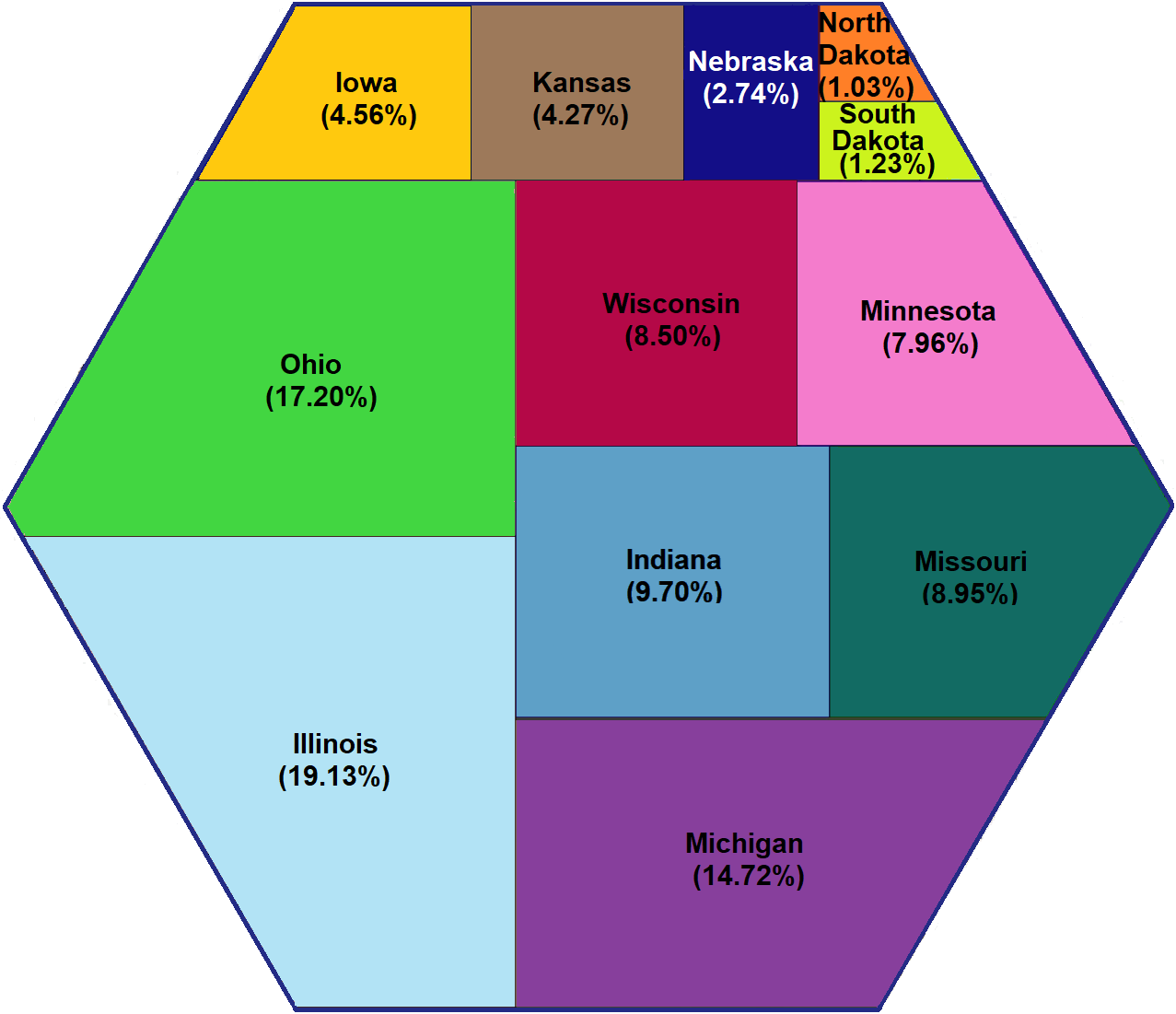}%
        \caption{}%
        \label{fig:MidwestPopulationHex}%
    \end{subfigure}%
    \subfigsCaption{\protect
        Treemaps generated using our \modelDynamicProg approach laid out in a rectangle or hexagon.
        \subref{fig:StockMarketRect} and \subref{fig:StockMarketHex} show the market capitalization of each sector of the U.S.\ Stock Market after closing on 2020-12-18.
        \subref{fig:MidwestPopulationRect} and \subref{fig:MidwestPopulationHex} display ratios of population of 12 states in the Midwest based on 2010 U.S.\ Census data.%
    }%
    \label{fig:DynamicDivideConquer}%
\end{figure*}

\Cref{alg:DynamicProg} shows the detailed steps of this approach.
The direction of the cuts, similar to \modelDC \cite{liang2015divide}, could be rectangular, angular, or polygonal.
Here we implement the rectangular approach with vertical and horizontal cuts, but replacing the type of cut is an easy generalization.
Our algorithm can also be easily modified to handle non-rectangular layout containers to partition them into other polygonal sub-regions.
Example treemaps made with \modelDynamicProg are shown in \cref{fig:StockMarketRect,fig:StockMarketHex} for stock market sectors and \cref{fig:MidwestPopulationRect,fig:MidwestPopulationHex} for state populations.
There, we use both rectangular and hexagonal layout containers.

Besides being able to handle cases with extreme areas, our computational results in \cref{sec:ComputationalResults} show that our \modelDynamicProg algorithm performs significantly better than \modelDC \cite{liang2015divide} with respect to aspect ratio and stability metrics.

\subsection{Comparison of Recursive Subdivision Approaches}

We compare the results of the divide and conquer algorithms and dynamic programming algorithm with \modelOptimization using several examples.

Consider a tree with 7 leaves with weights $(w_1,...,w_7) =$ (0.1277, 0.0837, 0.0922, 0.2235, 0.2845, 0.0994, 0.0890), that add up to 1.
Assume, without loss of generality, that we want to visualize it in a treemap with a unit square as the layout container in a way to minimize the total perimeter of the rectangles.
\Cref{fig:Random_Optimal}--\cref{fig:Random_DC} show the results using \modelOptimization, our dynamic programming and divide and conquer algorithms, and \modelDC \cite{liang2015divide}.
\modelOptimization results in the lowest total perimeter, followed by \modelDynamicProg.

\Cref{fig:Random_Optimal_Extreme}--\cref{fig:Random_ModifiedDC_Extreme} compare the same approaches for an extreme random example $(w_1,...,w_7) =$ (0.0795, 0.0709, 0.1074, 0.1121, 0.3980, 0.1023, 0.1298), where one weight (area) is much larger.
In this extreme case, \modelDynamicProg performs very well and in fact found the optimal solution.

We repeat the comparison for the treemap of market capitalization of different sectors in the U.S.\ Stock Market (\cref{fig:StockMarket_Unitbox_Optimal}--\cref{fig:StockMarket_Unitbox_ModifiedDC}) and ratios of population of the states in the U.S.
Midwest (\cref{fig:Midwest_Unitbox_Optimal}--\cref{fig:Midwest_Unitbox_ModifiedDC}).
In both cases, \modelOptimization performs best followed by \modelDynamicProg.

We have only been comparing the results by the total perimeter metric---the objective function of \modelOptimization. It should be  mentioned that when the optimal solution was sliceable \modelDynamicProg was able to find it (\cref{fig:Random_DynamicDC_Extreme}). We can generally say that when optimal solution is sliceable \modelDynamicProg can either find it or get very close to it. However, just like most other treemap algorithms it does not explore non-sliceable layouts. \modelDC and \modelModifiedDC performed more or less similar to each other on these examples.
\cref{sec:ComputationalResults} details more comprehensive comparisons that show both \modelDynamicProg and \modelModifiedDC beat \modelDC of \cite{liang2015divide}.

\renewcommand\tabCellImage[3]{
    \begin{subfigure}[c]{.2\textwidth}%
        \includegraphics[sqrtofarea=.95\textwidth]{#2}%
        \caption{Perimeter = #1}%
        \label{#3}%
    \end{subfigure}%
}

\begin{figure*}[tbp]
    \centering
    \begin{tabular}{ccccc}
            & \modelOptimization & \modelDynamicProg  & \modelModifiedDC & \modelDC \cite{liang2015divide} \\
        
        {\scriptsize \begin{tabular}{@{}l@{}}
        {\normalsize Random}\\
        \quad $L=\{$\\
        \qquad $0.1277,$\\
        \qquad $0.0837,$\\
        \qquad $0.0922,$\\
        \qquad $0.2235,$\\
        \qquad $0.2845,$\\
        \qquad $0.0994,$\\
        \qquad $0.0890$\\
        \quad $\}$\end{tabular}}
            & \tabCellImage{\best {10.3906}}    
            {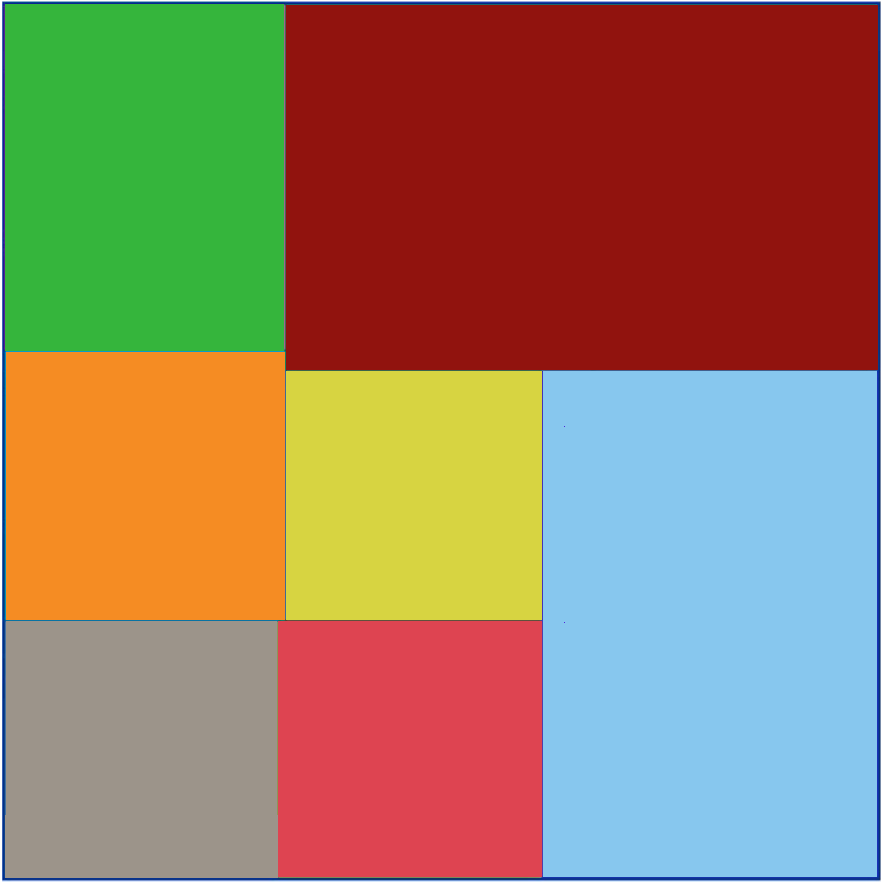}
            {fig:Random_Optimal}
            & \tabCellImage{\rnup {10.4649}}   
            {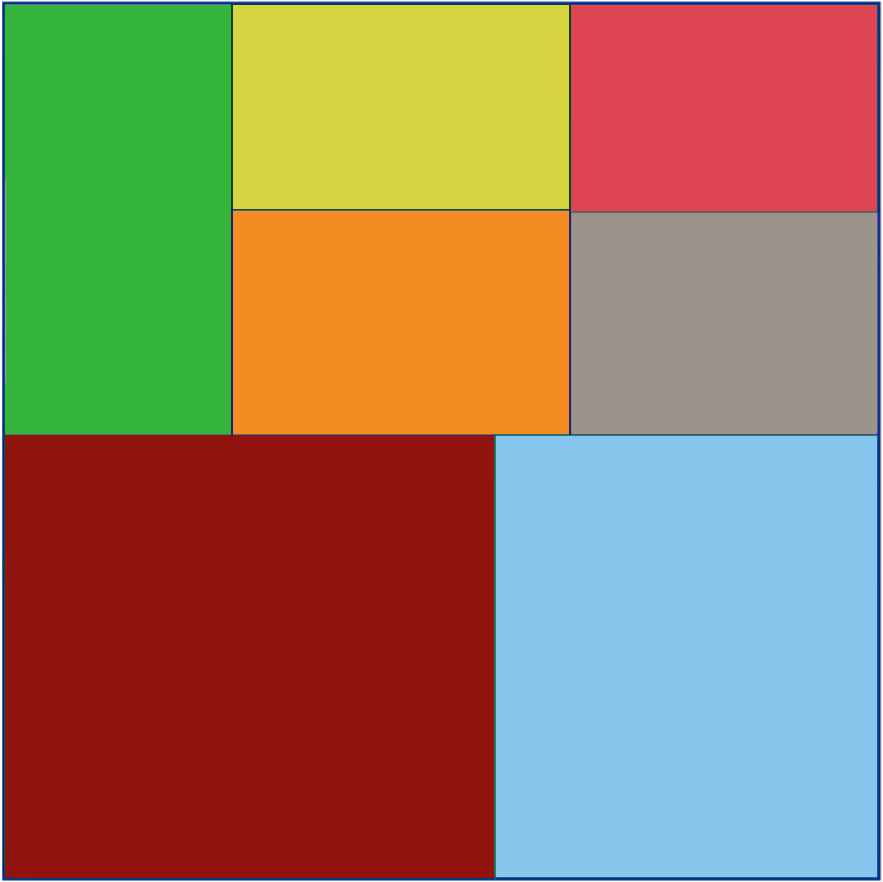} 
            {fig:Random_DynamicDC}
            & \tabCellImage{\rnup{10.4649}}  
            {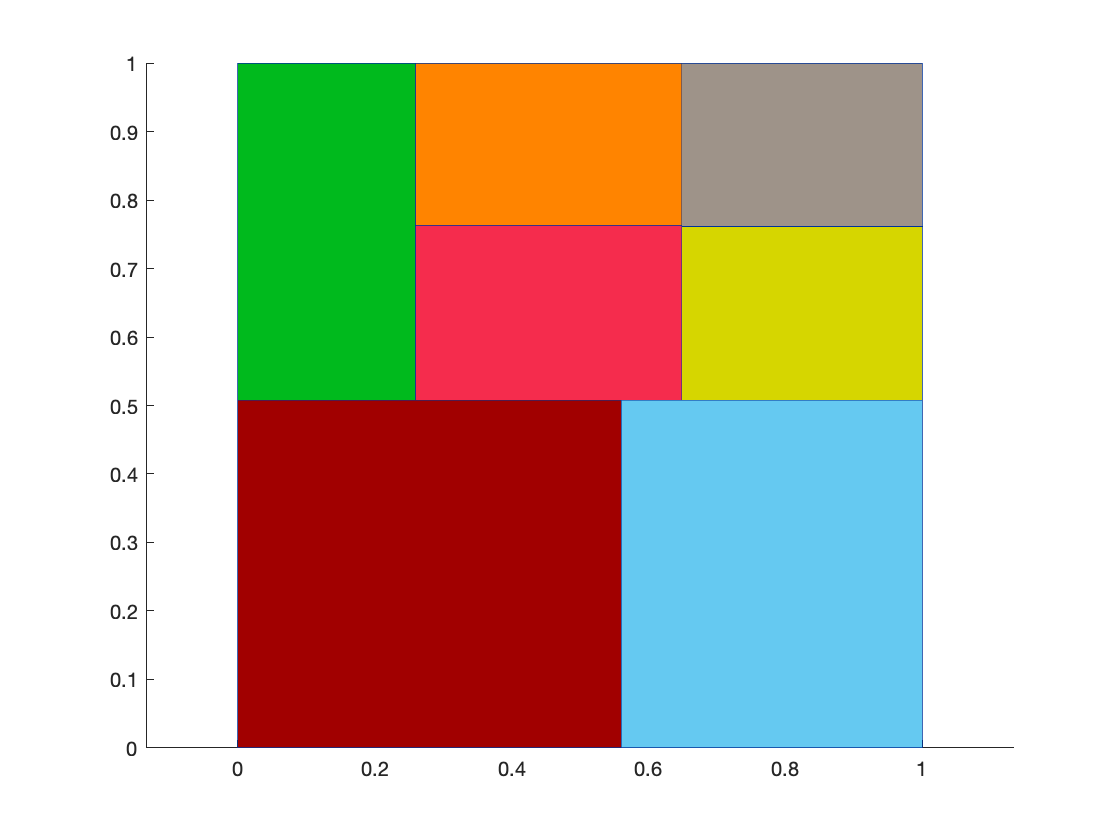}
            {fig:Random_ModifidDC}
            & \tabCellImage       {10.6092}    
            {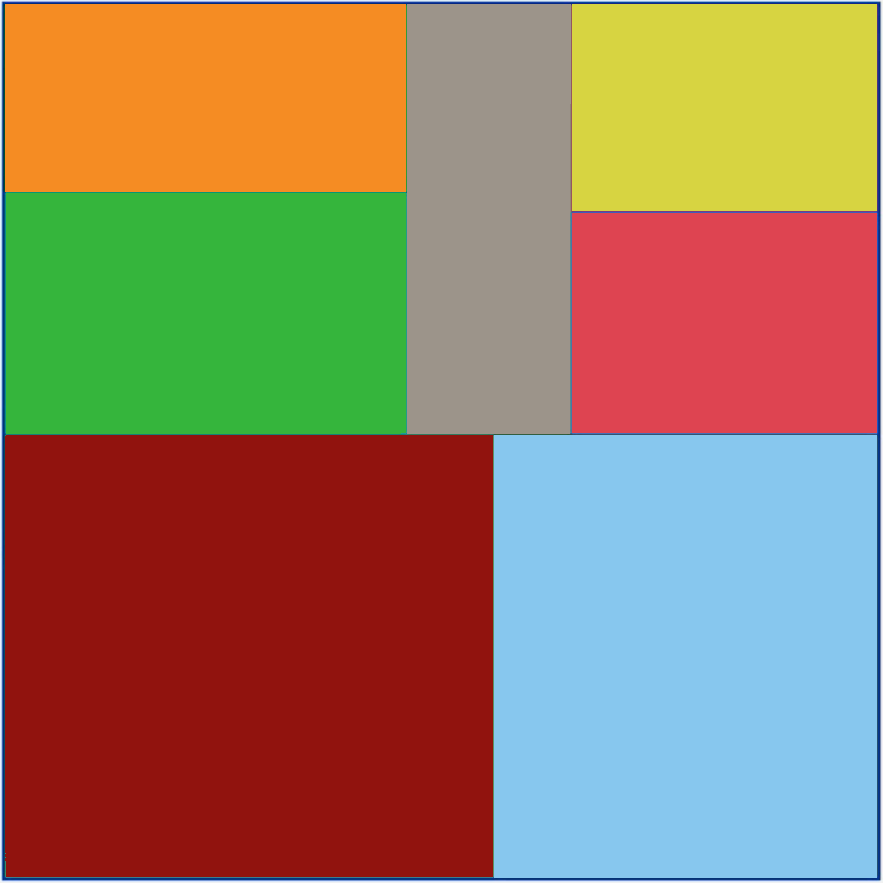}
            {fig:Random_DC}
        \\
        {\scriptsize \begin{tabular}{@{}l@{}}
        {\normalsize Random}\\
        {\normalsize Extreme}\\
        \quad $L=\{$\\
        \qquad $0.0795,$\\
        \qquad $0.0709,$\\
        \qquad $0.1074,$\\
        \qquad $0.1121,$\\
        \qquad $0.3980,$\\
        \qquad $0.1023,$\\
        \qquad $0.1298$\\
        \quad $\}$\end{tabular}}
            & \tabCellImage{\best {10.1965}}   {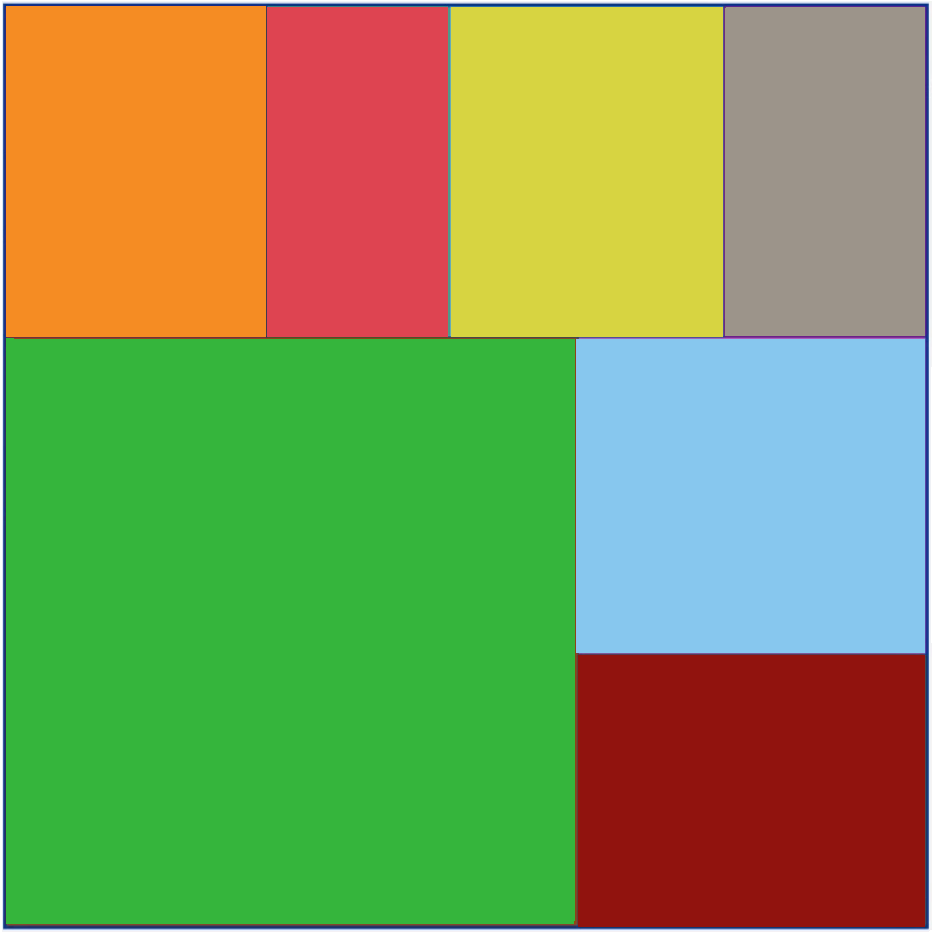}            {fig:Random_Optimal_Extreme} 
            & \tabCellImage{\best {10.1965}}   {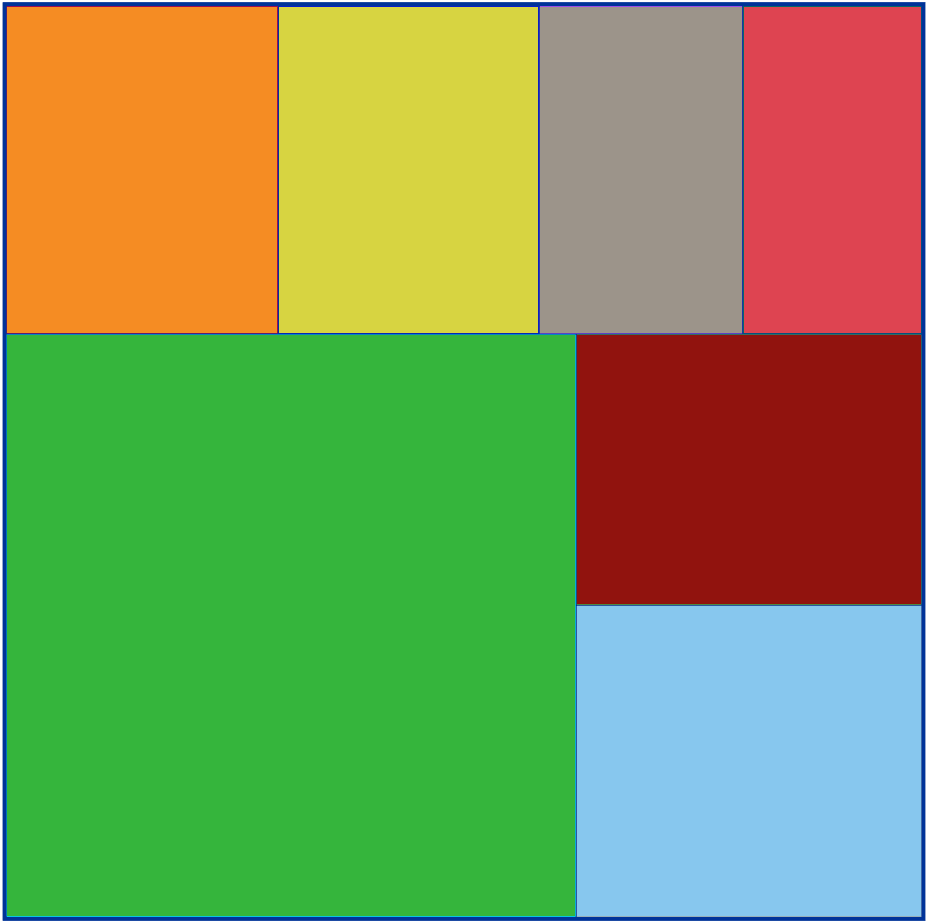}          {fig:Random_DynamicDC_Extreme} 
            & \tabCellImage{\best{10.1965}} 
             {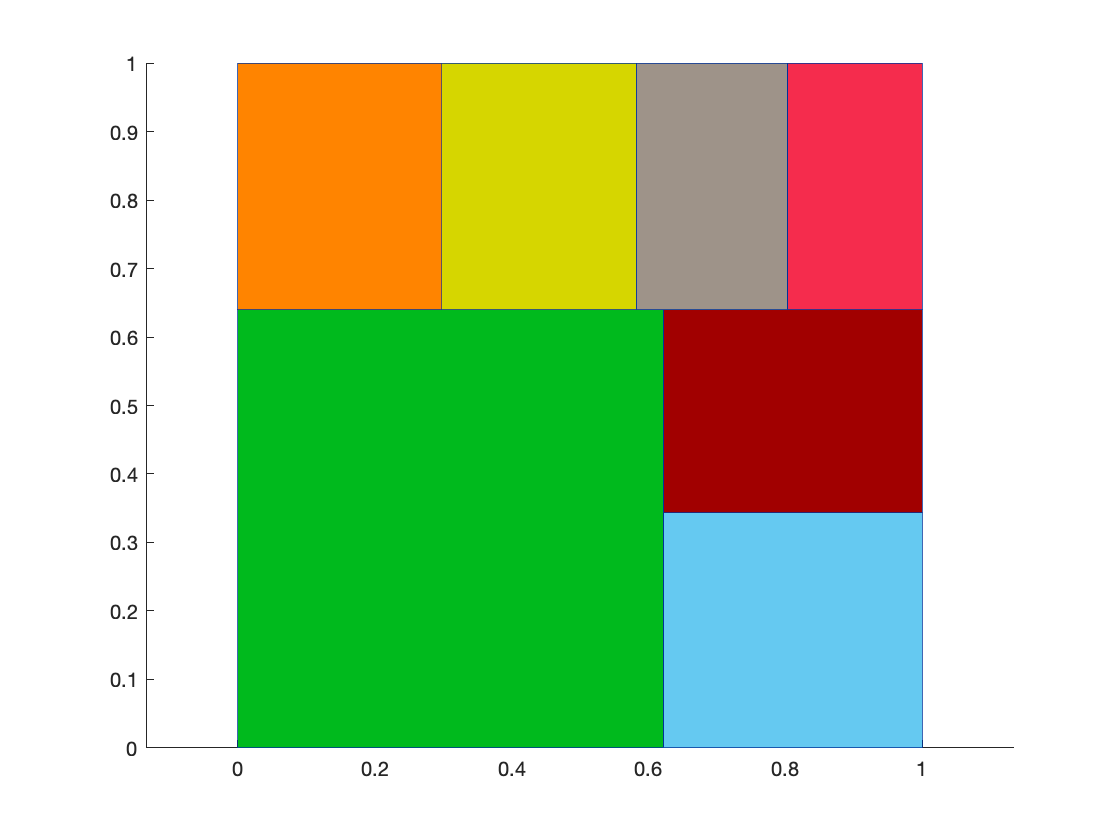}         {fig:Random_ModifiedDC_Extreme}
            & \tabCellImage       {10.5111}    {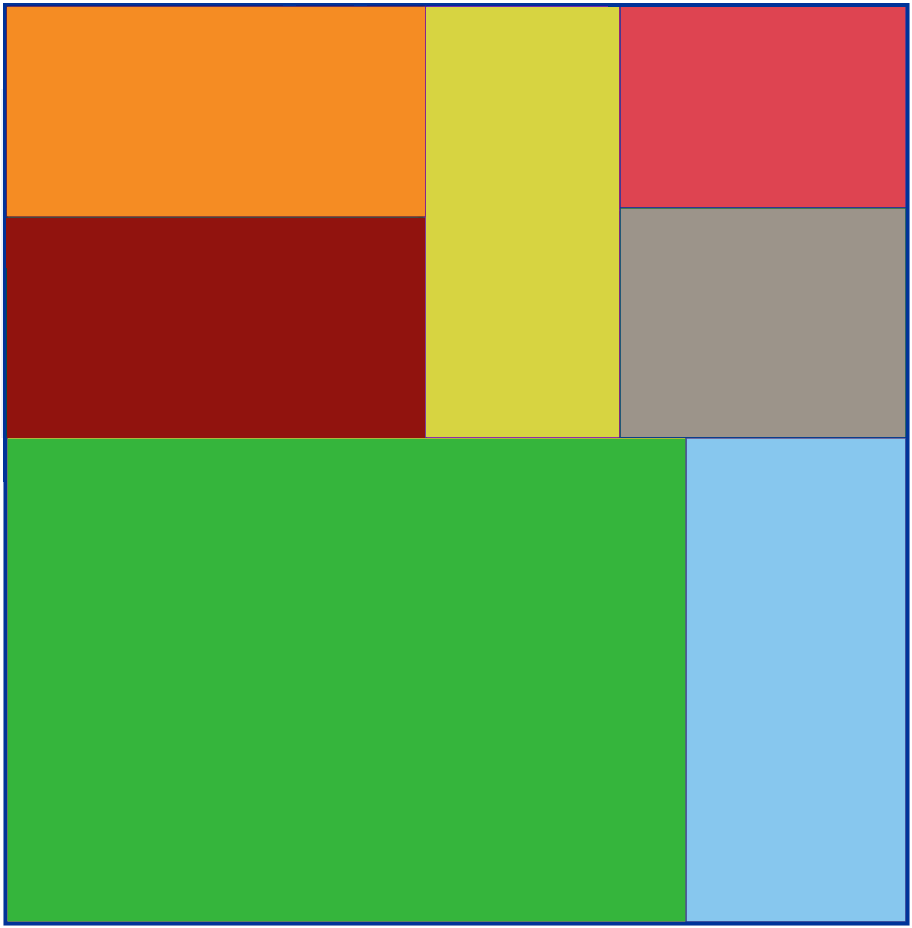}                 {fig:Random_DC_Extreme} 
        \\
        \begin{tabular}{@{}l@{}}
        Stock\\
        Market
        \end{tabular}
 
            & \tabCellImage{\best {12.6130}}  {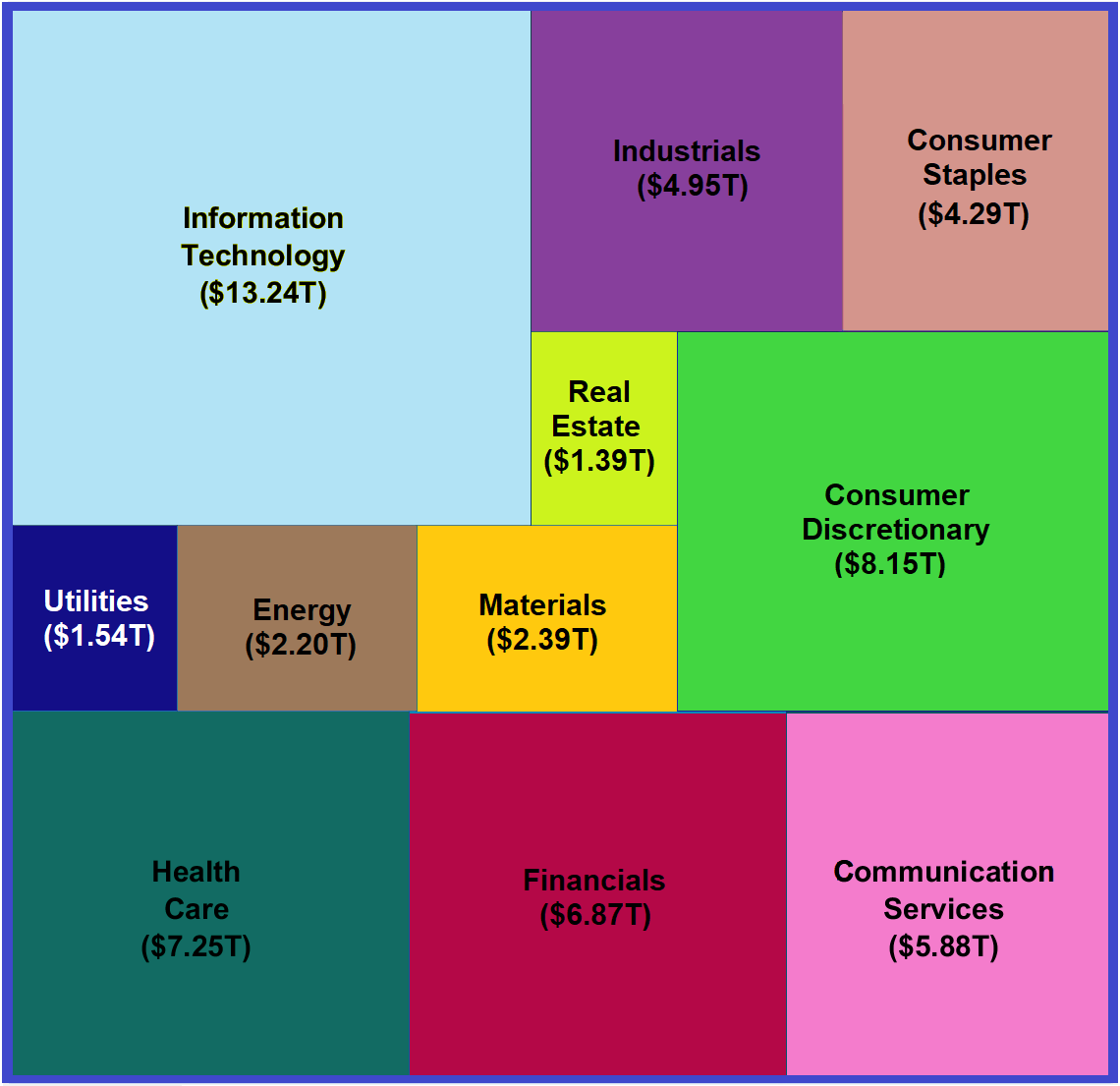}     {fig:StockMarket_Unitbox_Optimal} 

            & \tabCellImage{\rnup {12.7082}} {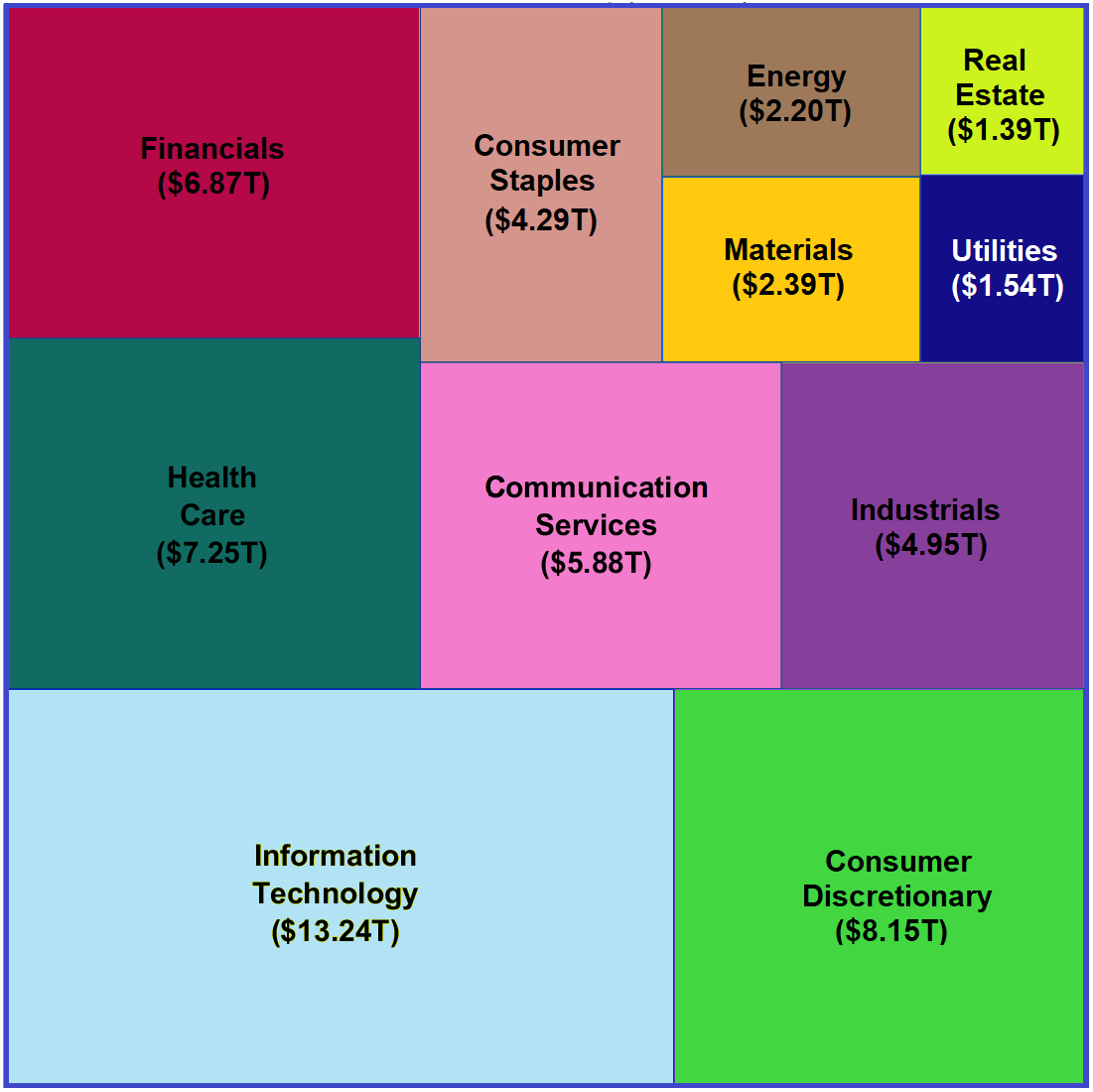}   {fig:StockMarket_Unitbox_DynamicDC} 
           & \tabCellImage       {12.9822} 
           {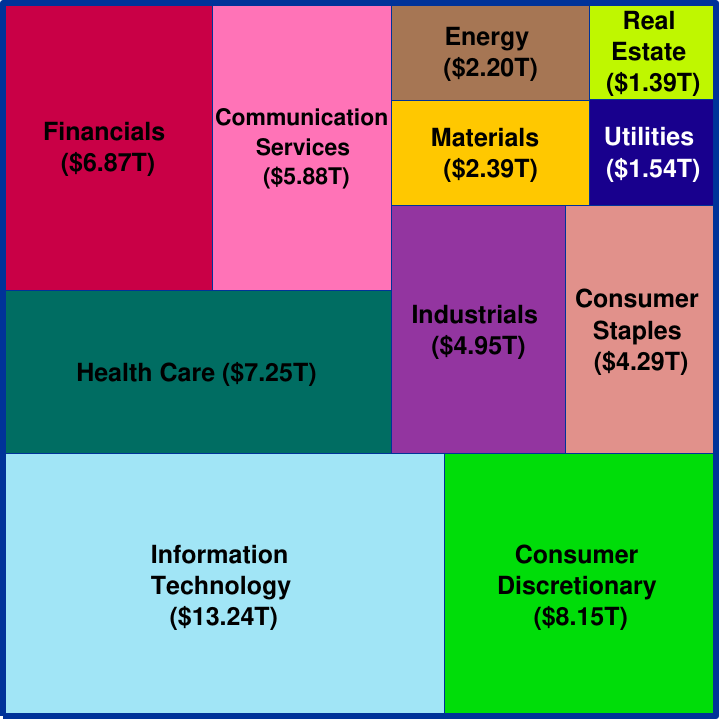}  {fig:StockMarket_Unitbox_ModifiedDC}
            & \tabCellImage       {12.9051}    {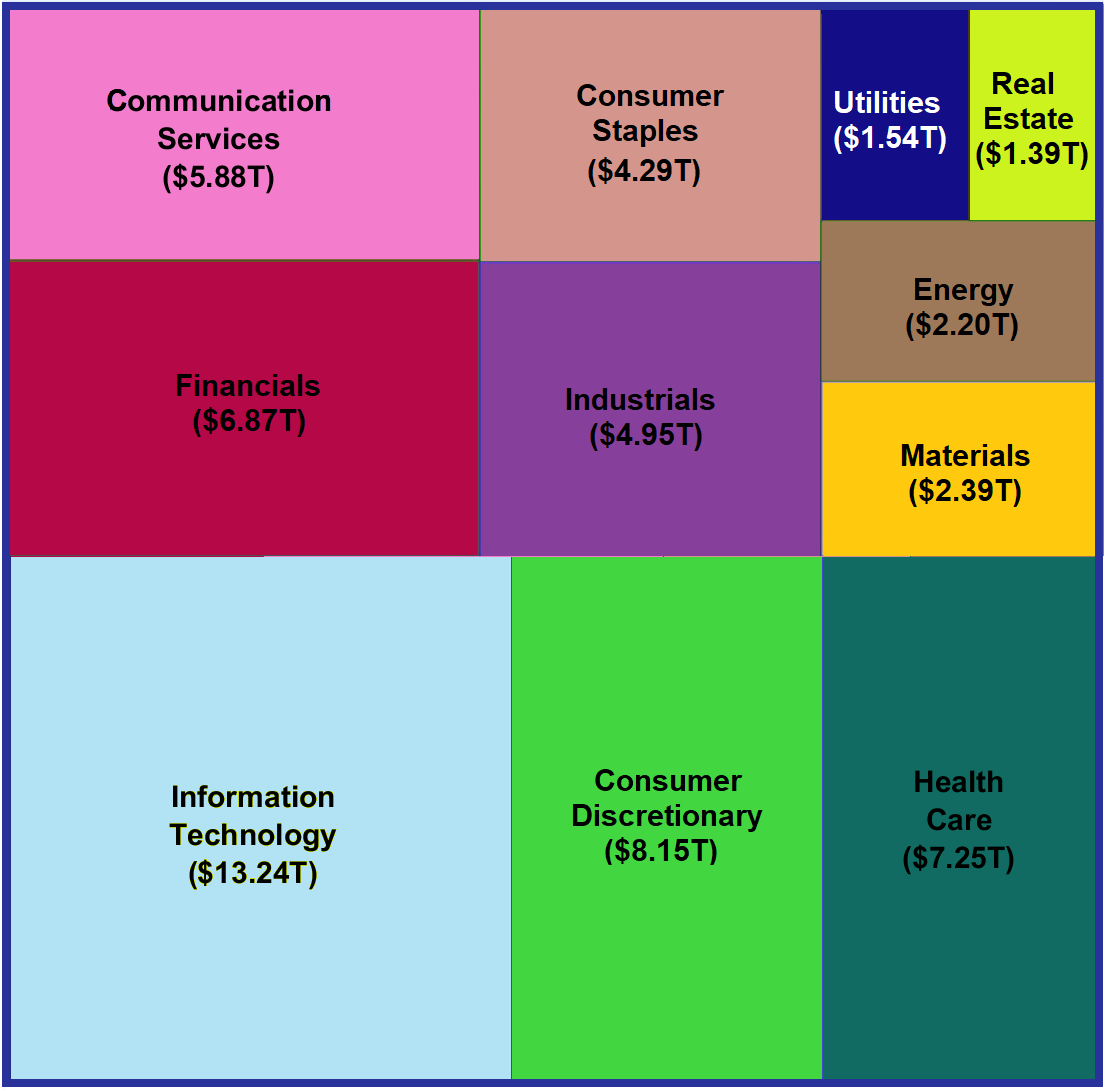}          {fig:StockMarket_Unitbox_DC} 
        \\
        \begin{tabular}{@{}l@{}}
        State\\
        Pop.
        \end{tabular}
            & \tabCellImage{\best {12.8988}}   {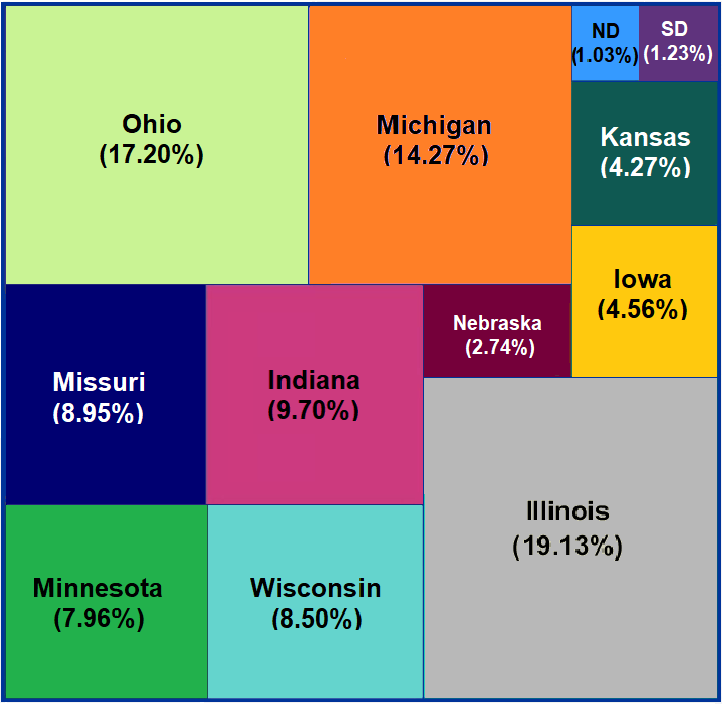}         {fig:Midwest_Unitbox_Optimal} 
            & \tabCellImage{\rnup {13.0456}}   {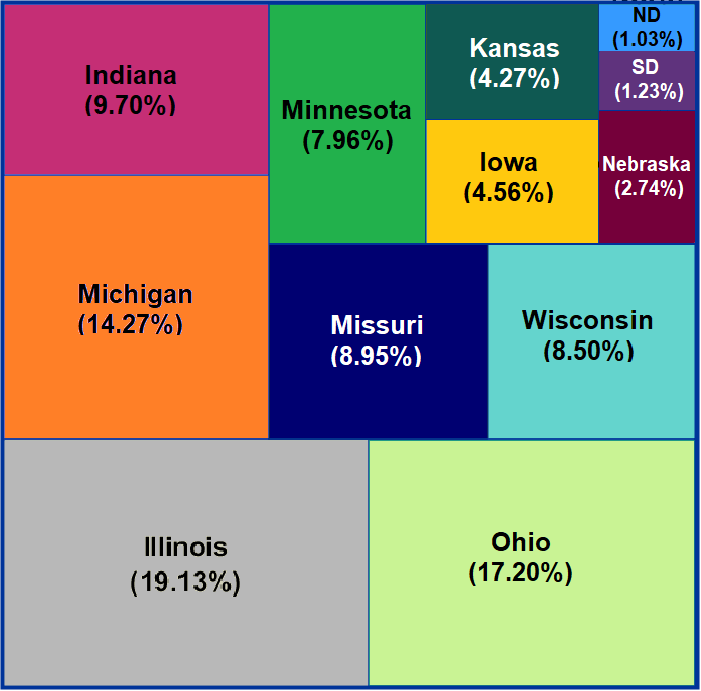}       {fig:Midwest_Unitbox_DynamicDC} 
            & \tabCellImage       {13.2752} 
            {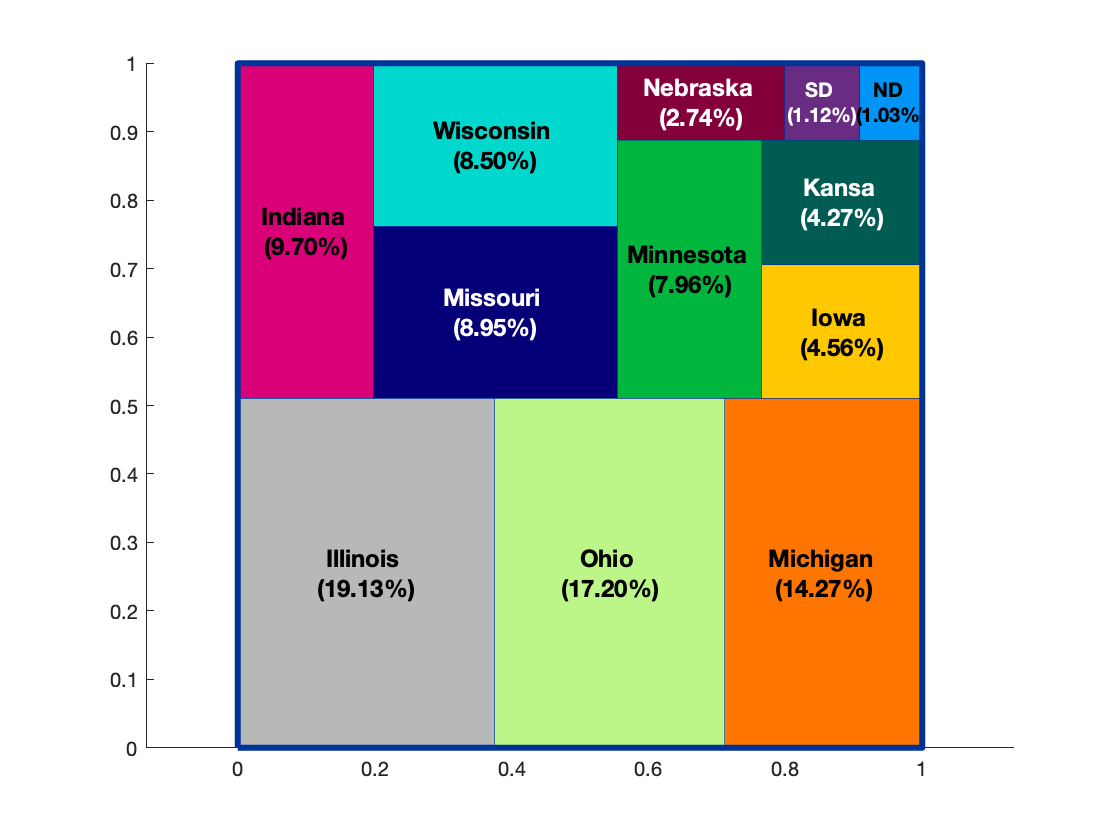}      {fig:Midwest_Unitbox_ModifiedDC}
            & \tabCellImage       {13.2155}    {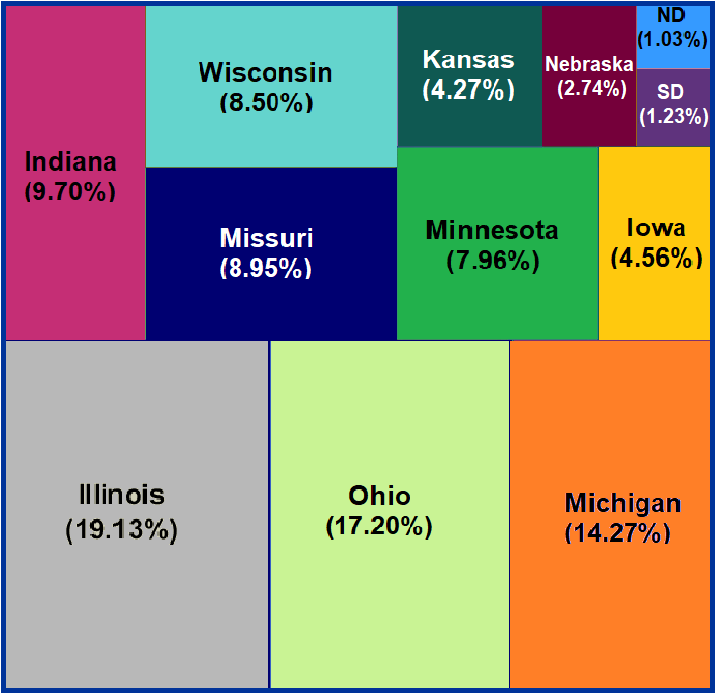}              {fig:Midwest_Unitbox_DC} 
        \\
    \end{tabular}
    \caption{
        These visualizations show the effect of applying four treemap layout algorithms to four weighted, single level trees.
        The treemaps are generated and laid out in a unit box using either
        our \modelOptimization,
        our dynamic divide and conquer algorithm (\modelDynamicProg),
        our modified divide and conquer algorithm (\modelModifiedDC), or
        the divide and conquer approach of Liang \etal \cite{liang2015divide} (\modelDC).
        Two of the trees have seven leaves with random weights (areas, given by $L=\{\ldots\}$), with the second (Extreme) example containing a weight much larger than the others.
        The third tree shows the market capitalization of different sectors in the U.S.\ Stock Market after closing on 2020-12-18.
        The final tree shows ratios of state population in the U.S.\ Midwest from the 2010 U.S.\ Census.
        For each treemap, the total perimeter of the component subrectangles is shown with the \bestName (shortest) and \rnupName values shown using color.
        Note that \modelOptimization results in the shortest perimeter. Our \modelDynamicProg performs the best among the algorithms. Our \modelModifiedDC gives better results than \modelDC on two instances including the extreme example (for which it is designed), and performs weaker on the other two instances. The extreme example shows the largest gap between the two algorithms as expected. It is clear that the gap on maximum aspect ratio metric is even larger.
        Moreover, it produces \emph{non-sliceable} layouts in three cases: \subref{fig:Random_Optimal}, \subref{fig:StockMarket_Unitbox_Optimal}, and \subref{fig:Midwest_Unitbox_Optimal}.
    }
    \label{fig:comparisonGrid}
\end{figure*}

\section{Constructive Algorithms for Spiral Treemaps}
\label{sec:SpiralAlgorithms}

Most of the existing algorithms, including the algorithms we introduced in the previous section, generate treemaps that lack an overall pattern. Such patterns could make treemaps more appealing visually, make it easier to compare the size of different sub-regions, and facilitate spotting specific sub-regions (increasing the readability as described in Bederson \etal \cite{bederson2002ordered}).  

In contrast, spiral treemaps place the sub-regions in a way that the overall layout resembles a spiral centered inside the container.
As opposed to treemaps created by guillotine cuts or by sequential placement of rectangles in horizontal or vertical directions, spiral treemaps have not been adequately studied \cite{tu2007visualizing}.
Here we present three new algorithms for generating spiral treemaps with rectangular sub-regions.

An experiment conducted by Kong \etal \cite{kong2010perceptual} suggests that rectangles with large aspect ratios $\ge$ 9/2 reduce the accuracy of area comparison tasks, particularly when the rectangles have different orientations.
They also show that the accuracy of the perception is equally poor when comparing squares.
What actually seems to help is having a distribution of reasonable non-square aspect ratios.
Their results particularly show that optimizing towards a 3/2 aspect ratio performs better than \modelSquarified.
Lu \etal \cite{lu2017golden} present an algorithm that, instead of targeting squares like \modelSquarified, tries to reach a layout with rectangles having aspect ratios as close as possible to the golden ratio $\phi=\frac{1+\sqrt{5}}{2}\simeq 1.618$---so-called golden rectangles.
This is also close to the 3/2 ratio conjectured by Kong \etal.

Golden ratios, golden rectangles, and golden spirals have been discovered in nature, human body, and galaxies, and have been used frequently in architecture (e.g., \emph{Great Mosque of Kairouan} in Tunisia \cite{boussora2004use}, and \emph{Naqsh-e Jahan Square} in Iran \cite{dahar2013geometrical,elliot2007mirrors})\footnote{The popular golden ratio example of the Parthenon temple in ancient Greece is disputed and seems to be without foundation \cite{foutakis2014did}.}, in art (e.g., the painting ``\emph{The Sacrament of the Last Supper}'' by Salvador Dali \cite{GoldenRatioDali}), in finance (e.g., the use of \emph{Fibonacci retracement levels} in prediction of stock price changes \cite{frost1995elliott}), and in optimization (e.g., \emph{golden section search} \cite{kiefer1953sequential}).
Here, we follow the same intuition and generate treemaps that mimic the \emph{golden spiral} or its approximation the \emph{Fibonacci spiral}.

\subsection{Symmetric Spiral Treemap Algorithm}

Our new \modelSymmetricSpiral layout first sorts the input areas in non-descending order.
Then, starting with the first two sub-rectangles ($R_1$ and $R_2$) in the sorted list, we juxtapose the sub-rectangles ($R_2$ to the left of $R_1$) with the same height so that their union forms a larger rectangle with aspect ratio of 2.
The choice of aspect ratio of these two initial placements is driven by the way the \emph{Fibonacci spiral} is constructed. We denote this aspect ratio with $\rho_s$ and treat it as a configurable parameter.
The direction then changes and the next rectangle is placed on top of the union of the current rectangles, forming a larger rectangle.
Again, direction changes and the next rectangle is placed to the right of the union of already-placed rectangles.
This procedure is repeated until all rectangles are placed.
A new rectangle is added in each step and its union with previously-placed rectangles forms a larger rectangle.
Direction turns clockwise between left, top, right, and bottom.

Pseudocode for \modelSymmetricSpiral is shown in \cref{alg:FibonacciSpiralTreemap}, and its computational complexity is  $\mathcal{O}(n\log n)$ due to the sorting step; the rest of the algorithm runs in $\mathcal{O}(n)$. In \cref{alg:FibonacciSpiralTreemap} we follow a clockwise-outward-growing rotation for forming the spiral structure.
Unfortunately, \modelSymmetricSpiral is wont to create skinny rectangles as the spiral gets closer to the outermost layer.
This is due to placing each new sub-rectangle on the longer side of the union of currently-placed sub-rectangles. Therefore, the aspect ratios of the sub-rectangles could get worse as the spiral is growing. However, our goal in this algorithm is really not to produce high quality treemaps with respect to aspect ratio metrics. \modelSymmetricSpiral serves a purpose like \modelSliceDice as our basis for the next two and future spiral treemaps. It also generally produces more \emph{readable} and aesthetically appealing visualization by maximizing \emph{symmetry} and \emph{fractal-like patterns} in the treemaps. One could define evaluation metrics for these properties, e.g., readability measure introduced in \cite{bederson2002ordered} and the Fractal Value as defined in Eq. (1) in \cite{dAmbros2005fractal}, and evaluate the quality of this algorithm that way, which is out of the scope of this paper.

\begin{algorithm}[tbp]
    \SetAlgoLined
    \BlankLine
    \KwIn{A list of $n$ areas $L=\{A_1,A_2,...,A_n\}$.}
    \KwOut{A rectangle $R$ partitioned into $n$ sub-rectangles $R_1,R_2,...,R_n$ with areas $A_1,A_2,...,A_n$ such that $\Area (R) = \sum_{i=1}^n A_i$.}
    \BlankLine
    \tcc{***********************************************}
    \eIf{$n=1$}{ \Return{$R_1$}\;}{ 
    Sort the list of areas in ascending order and reindex the sorted areas as $A_1,A_2,...,A_n$\;
    \tcc{ Ensuring $R_1$ and $R_2$ together form a bigger rectangle with aspect ratio $\rho_s = 2$.}
    Set $h_1=h_2=\sqrt{(A_1+A_2)/2}$, \,\, $w_1=A_1/h_1$, and $w_2=A_2/h_2$\;
    Place $R_2$ on the left of $R_1$\;
    Let $Q_2=\Box(R_1 \cup R_2)$\;
    \eIf{$n=2$}{\Return{$Q_2,R_1,R_2$\;}}{
    Set $\mbox{direction} =$ ``top''\;
    \For{$i \in \{3,...,n\}$}{
    	\uIf{$\mbox{direction} =$``top''}{Set width $w_i=\width(Q_{i-1})$, height $h_i=A(i)/w_i$, and place rectangle $R_i$ exactly on top of $Q_{i-1}$\; 
    	Set $\mbox{direction} =$ ``right''\;}
    	\uElseIf{$direction=$ ``right''}{Set height $h_i=\height(Q_{i-1})$, width $w_i=A(i)/h_i$, and place rectangle $R_i$ exactly on the right side of $Q_{i-1}$\; 
    	Set $\mbox{direction} =$ ``bottom''\;}
    	\uElseIf{$direction =$ ``bottom''}{Set width $w_i=\width(Q_{i-1})$, height $h_i=A(i)/w_i$, and place rectangle $R_i$ exactly on top of $Q_{i-1}$\; 
    	Set $\mbox{direction} =$ ``left''\;}
    	\Else{Set height $h_i=\height(Q_{i-1})$, width $w_i=A(i)/h_i$, and place rectangle $R_i$ exactly on the right side of $Q_{i-1}$\; 
    	Set $\mbox{direction} =$ ``top''\;}
    	Set $Q_i = Q_{i-1}\cup R_i$\;
    }
    Set $R = Q_n$\;
    Rotate $R$ and the partition $R_1,...,R_n$ for $90^{\circ}$ counterclockwise if $\width(R)<\height(R)$\;
    \Return{ $R,R_1,...,R_n$}\; }
    }
    \protect\caption{
        \label{alg:FibonacciSpiralTreemap}
        \modelSymmetricSpiral$(L)$
        Generates a treemap of rectangles with the given areas in a clockwise-outward-growing spiral pattern.}
\end{algorithm}

\subsection{Square-Bundled Spiral Treemap Algorithm}
\label{subsec:SquareBundledSpiralAlg}

Our next algorithm, \modelSquareBundle, remedies the skinny rectangle issue by bundling a subset of rectangles throughout the algorithm and creating square-like bundles. As in \modelSymmetricSpiral, first sort the input areas in an ascending order and place the first two rectangles with $R_2$ to the left of $R_1$ in a way that their union has aspect ratio of 2 (i.e., $\rho_s=2$).
The changes in direction also follow the same procedure.
However, instead of placing each rectangle in a spiral pattern we place bundled square-like rectangles in a spiral pattern.
Bundles are formed in each step by finding a subset of remaining rectangles that makes the bundle as close as possible to a square---given that one side of the bundled rectangle is already fixed to one side of the union of already-placed rectangles. 
This makes our treemap as close as possible to the Fibonacci spiral, where each new box is a square with an edge length equal to the width (height) of the union of the previously-placed rectangles.
The layout inside a bundle can be created using any rectangular treemapping algorithm as a subroutine.
In our implementation we use \modelSquarified. Our choice comes from our observations that for square or square-like input regions \modelSquarified has an advantage.

Pseudocode for \modelSquareBundle is shown in \cref{alg:SquareBundledSpiralTreemap} and its computational complexity is  $\mathcal{O}(n\log n)$ due to the sorting step; the rest of the algorithm runs in $\mathcal{O}(n)$. In \cref{alg:SquareBundledSpiralTreemap} we also follow a clockwise-outward-growing rotation for forming the spiral structure.

\begin{algorithm}[tbph]
    \SetAlgoLined
    \BlankLine
    \KwIn{A list of $n$ areas $L=\{A_1,A_2,...,A_n\}$.}
    \KwOut{A rectangle $R$ partitioned into $n$ sub-rectangles $R_1,R_2,...,R_n$ with areas $A_1,A_2,...,A_n$ such that $\Area (R) = \sum_{i=1}^n A_i$.}
    \BlankLine
    \tcc{***********************************************}
    \If{$n=1$}{
        \Return{$R_1$}\;
    }
    Sort the list of areas in ascending order and reindex the sorted areas as $A_1,A_2,...,A_n$\;
    \tcc{Ensuring $R_1$ and $R_2$  together form a bigger rectangle with aspect ratio $\rho_s = 2$.}
    Set $h_1=h_2=\sqrt{(A_1+A_2)/2}$, \,\, $w_1=A_1/h_1$, and $w_2=A_2/h_2$\;
    Place $R_2$ on the left of $R_1$\;
    Let $Q_2=\Box(R_1 \cup R_2)$\;
    \If{$n=2$}{
        \Return{$Q_2,R_1,R_2$\;}
    }
    Set $\mbox{direction} =$ ``top''\;
    \While{$i \mathrel{\mathtt{!=}} n$}{
    	\eIf{$\mbox{direction} =$``top'' {\bf or} ``bottom''}{
    	    Set $w_b = \width(Q_{i-1})$\; 
        	Set $k = \argmin_{k \in \{i,...,n\}} \left\vert w_b - \frac{\sum_{j=i}^{k} A_j}{w_b}\right\vert$\;
        	Set $h_b = \sum_{j=i}^{k} A_j/w_b$\;
        	\eIf{$\mbox{direction} =$``top''}{
            	Place rectangle $B = w_b \times h_b$ exactly on top of $Q_{i-1}$\; 
            	Set $\mbox{direction} =$ ``right''\;
        	}{
            	Place rectangle $B = w_b \times h_b$ exactly below $Q_{i-1}$\; 
            	Set $\mbox{direction} =$ ``left''\;
        	}
    	}
    	{ 
        	\tcc{$\mbox{direction} =$ ``right'' {\bf or} ``left''}
        	Set $h_b = \height(Q_{i-1})$\; 
        	Set $k = \argmin_{k \in \{i,...,n\}} \left\vert h_b - \frac{\sum_{j=i}^{k} A_j}{h_b}\right\vert$\;
        	Set $w_b = \sum_{j=i}^{k} A_j/h_b$\;
        	\eIf{$\mbox{direction}=$ ``right''}{
            	Place rectangle $B = w_b \times h_b$ exactly on the right side of $Q_{i-1}$\; 
            	Set $\mbox{direction} =$ ``bottom''\;
        	}{
            	Place rectangle $B = w_b \times h_b$ exactly on the left side of $Q_{i-1}$\; 
            	Set $\mbox{direction} =$ ``top''\;
        	}
    	}
    	Let $(R_i,...,R_k) = {\sf Squarified}(B,\{A_i,...,A_k\})$\;
    	Set $Q_i = Q_{i-1}\cup \left(\cup_{j=i}^k R_j\right)$\;
    	Set $i=k+1$\;
    }
    Set $R = Q_n$\;
    Rotate $R$ and the partition $R_1,...,R_n$ for $90^{\circ}$ counterclockwise if $\width(R)<\height(R)$\;
    \Return{ $R,R_1,...,R_n$}\;
    \caption{\protect
    \label{alg:SquareBundledSpiralTreemap}%
    \modelSquareBundle$(L)$ \hfill \break
        Generates a treemap of rectangles with the given areas.
        In each iteration the algorithm bundles a group of areas and places them all together in a way that (1) the blocks of bundled rectangles form a clockwise outward growing spiral pattern and (2) the layout inside each block is formed using a subroutine such as \modelSquarified.%
        }%
\end{algorithm}

\subsection{Strip-Bundled Spiral Treemap Algorithm}
\label{subsec:StripBundledSpiralAlg}

As an alternative approach for improving \modelSymmetricSpiral, we also develop \modelStripBundle.
The key difference is how many rectangles we place at each change in direction.
By considering the side of the current union where the next rectangle has to be placed as a strip with its width (height) fixed with this side of the current union and its height (width) being flexible, we can add additional rectangles one-by-one, using the flexibility of the strip, to optimize the aspect ratio before changing the direction and moving to the next side.
As in \modelSymmetricSpiral and \modelSquareBundle, we first sort the areas in ascending order and add the first two rectangles with $\rho_s=2$.
In each step, we create a strip on the top, right, bottom, or left side of the union of already-placed rectangles.
We then add the remaining rectangles to the strip in order until the 
maximum aspect ratio of the rectangles does not further improve.
Then we change direction, create the next strip, and repeat this procedure until all rectangles are placed. 

Pseudocode for \modelStripBundle is shown in \cref{alg:StripBundledSpiralTreemap} and its computational complexity is also $\mathcal{O}(n\log n)$ due to the sorting step; the rest of the algorithm runs in $\mathcal{O}(n)$. In \cref{alg:StripBundledSpiralTreemap} we again follow a clockwise-outward-growing rotation for forming the spiral structure.

\begin{algorithm}[tbph]
    \SetAlgoLined
    \BlankLine
    \KwIn{A list of $n$ areas $L=\{A_1,A_2,...,A_n\}$.}
    \KwOut{A rectangle $R$ partitioned into $n$ sub-rectangles $R_1,R_2,...,R_n$ with areas $A_1,A_2,...,A_n$ such that $\Area (R) = \sum_{i=1}^n A_i$.}
    \BlankLine
    \tcc{***********************************************}
    \If{$n=1$}{
        \Return{$R_1$}\;
    }
    Sort the list of areas in ascending order and reindex the sorted areas as $A_1,A_2,...,A_n$\;
    \tcc{ Ensuring $R_1$ and $R_2$  together form a bigger rectangle with aspect ratio $\rho_s = 2$.}
    Set $h_1=h_2=\sqrt{(A_1+A_2)/2}$, \,\, $w_1=A_1/h_1$, and $w_2=A_2/h_2$\;
    Place $R_2$ on the left of $R_1$\;
    Let $Q_2=\Box(R_1 \cup R_2)$\;
    \If{$n=2$}{
        \Return{$Q_2,R_1,R_2$\;}
    }
    Set $\mbox{direction} =$ ``top''\;
    \While{$i \mathrel{\mathtt{!=}} n$}{
    	\eIf{$\mbox{direction} =$``top'' {\bf or} ``bottom''}{
        	Set $w_s = \width(Q_{i-1})$\; 
        	Set $k = \argmin_{k \in \{i,...,n\}} \max_{j \in \{i,...,k\}}$\\$
        	    \quad \left\{\frac{(\sum_{j=i}^{k} A_j/w_s)^2}{A_j}\, , \, \frac{A_j}{(\sum_{j=i}^{k} A_j/w_s)^2} \right\}$\; 
        	Set $h_s = \sum_{j=i}^{k} A_j/w_s$\;
        	\eIf{$\mbox{direction} =$``top''}{
            	Place rectangle $S = w_s \times h_s$ exactly on top of $Q_{i-1}$ and form $Q_{i}$\; 
            	Slice $S$ into rectangles $R_i,...,R_k$ with vertical cuts from left to right with areas $A_i,...,A_k$\;
            	Set $\mbox{direction} =$ ``right''\;
        	}{
            	Place rectangle $S = w_s \times h_s$ exactly below $Q_{i-1}$ and form $Q_{i}$\; 
            	Slice $S$ into rectangles $R_i,...,R_k$ with vertical cuts from right to left with areas $A_i,...,A_k$\;
            	Set $\mbox{direction} =$ ``left''\;
        	}
    	}{ 
        	\tcc{$\mbox{direction} =$ ``right'' {\bf or}  ``left''}
        	Set $h_s = \height(Q_{i-1})$\; 
        	Set $k = \argmin_{k \in \{i,...,n\}} \max_{j \in \{i,...,k\}}$\\$
        	    \quad \left\{\frac{(\sum_{j=i}^{k} A_j/h_s)^2}{A_j}\, , \, \frac{A_j}{(\sum_{j=i}^{k} A_j/h_s)^2} \right\}$\; 
        	Set $w_s = \sum_{j=i}^{k} A_j/h_s$\;
        	\eIf{$\mbox{direction} =$ ``right''}{
        	Place rectangle $S = w_s \times h_s$ exactly on the right side of $Q_{i-1}$ and form $Q_{i}$\; 
        	Slice $S$ into rectangles $R_i,...,R_k$ with horizontal cuts from top to bottom with areas $A_i,...,A_k$\;
        	Set $\mbox{direction} =$ ``bottom''\;}{
        	Place rectangle $S = w_s \times h_s$ exactly on the left side of $Q_{i-1}$ and form $Q_{i}$\;
        	Slice $S$ into rectangles $R_i,...,R_k$ with horizontal cuts from bottom to top with areas $A_i,...,A_k$\; 
        	Set $\mbox{direction} =$ ``top''\;}
    	}
    }
    Set $R = Q_n$\;
    Rotate $R$ and the partition $R_1,...,R_n$ for $90^{\circ}$ counterclockwise if $\width(R)<\height(R)$\;
    \Return{ $R,R_1,...,R_n$}\; 
    
    \protect\caption{
        \label{alg:StripBundledSpiralTreemap}
        \modelStripBundle$(L)$ \hfill \break
        Generates a treemap of rectangles with the given areas drawn in a clockwise outward growing spiral pattern.
        In each iteration, rectangles are added as needed to minimize the total perimeter.
    }
\end{algorithm}

\subsection{Comparison of Spiral Algorithms}

Here we compare the results of the above three spiral algorithms with \modelSquarified, i.e., one of the best existing algorithms, with respect to aspect ratio, as shown in e.g., \cite{sondag2018stable,vernier2020quantitativeJ}. As previously see in \cite{sondag2018stable} and we observe it here too it can also perform reasonably well on stability depending on the data set and the measure of stability.
Since the main objective of all four algorithms is to minimize aspect ratio, we examine the maximum aspect ratio (maxAR).
We also compare their stability using the maximum Hausdorff distance (maxHD).
(\cref{sec:ComputationalResults} provides a more comprehensive comparison.)

\Cref{fig:Spiral_illustration1} illustrates the output treemaps of our spiral algorithms compared with \modelSquarified on a single level tree of 60 nodes with random weights.
It is clear that \modelSquareBundle and \modelStripBundle are more visually appealing than \modelSymmetricSpiral.
Here, both improved spiral algorithms score better on aspect ratio and stability than \modelSquarified.
\modelSymmetricSpiral is particularly bad on aspect ratio, as expected, but better than \modelSquarified and the other two spiral algorithms on stability.
 
 \begin{figure}[tbp]%
    \centering%
    \begin{subfigure}[b]{.47\textwidth}%
        \centering
        \includegraphics[sqrtofarea=.95\textwidth]{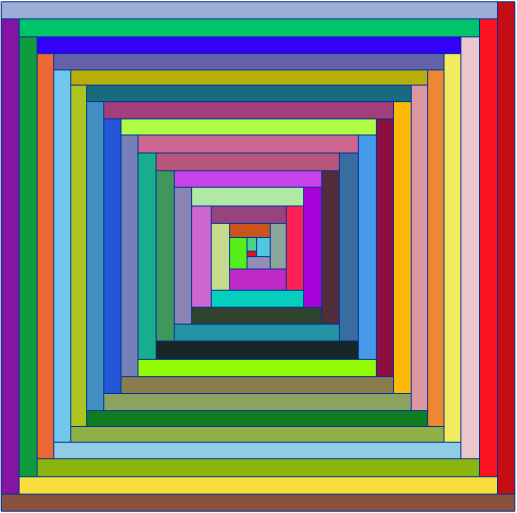}%
        \caption*{
            \modelSymmetricSpiral\\
            maxAR = 30.01\\
        }%
        \label{fig:Spiral_Fibonacci}%
    \end{subfigure}%
    \hfill
    \begin{subfigure}[b]{.53\textwidth}%
        \centering
        \includegraphics[sqrtofarea=.84\textwidth]{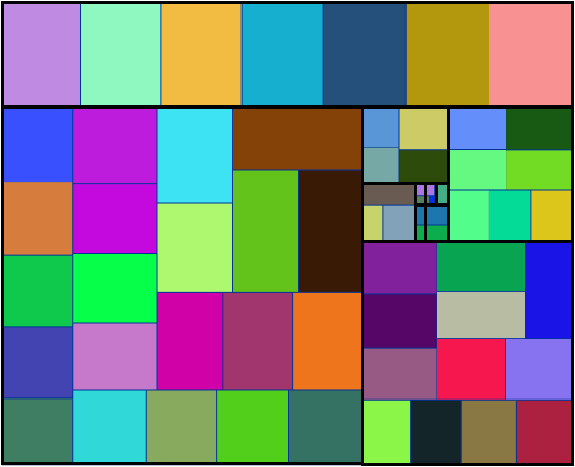}%
        \caption*{
            \modelSquareBundle\\
            \rnup{maxAR = 2.35}\\
        }%
        \label{fig:Spiral_SquareBundled}%
    \end{subfigure}%
    \\
    \begin{subfigure}[b]{.47\textwidth}%
        \centering
        \includegraphics[sqrtofarea=.95\textwidth]{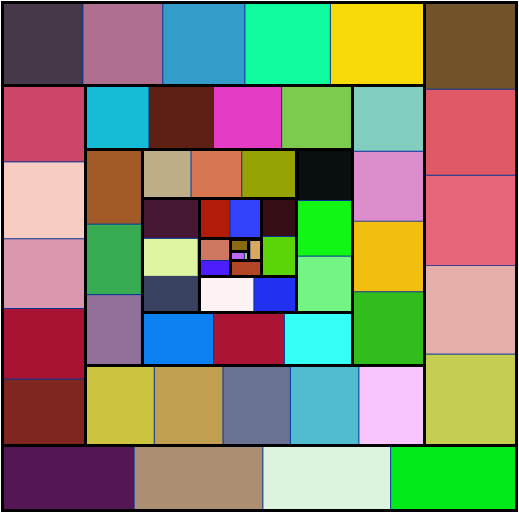}%
        \caption*{
            \modelStripBundle\\
            \best{maxAR = 1.98}\\
        }%
        \label{fig:Spiral_StripBundled}%
    \end{subfigure}%
    \hfill
    \begin{subfigure}[b]{.53\textwidth}%
        \centering
        \includegraphics[sqrtofarea=.84\textwidth]{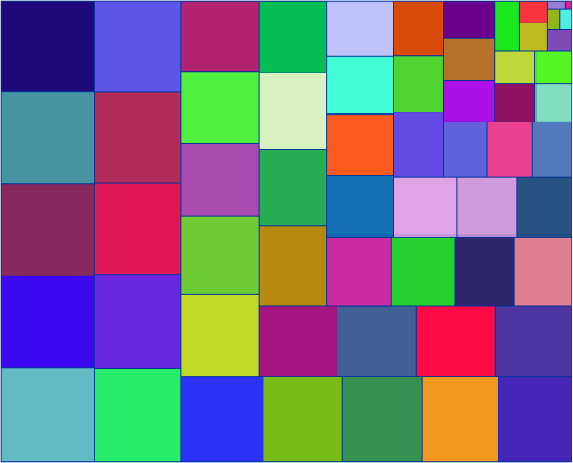}%
        \caption*{
            \modelSquarified\\
            maxAR = 2.69\\
        }%
        \label{fig:Spiral_vs_Squarified}%
    \end{subfigure}%
    \subfigsCaption{\protect
        Our three spiral treemap algorithms compared with \modelSquarified for making a rectangular treemap.
        The data shown is a single level tree of 60 randomly-weighted nodes.
        Bundled rectangles are shown using thicker lines.
        \modelSymmetricSpiral has the worst maximum aspect ratio (maxAR), but our other spiral algorithms beat \modelSquarified.
        They also have better maximum Hausdorff distances (maxHD) than \modelSquarified, with \modelSymmetricSpiral doing best.
        The \bestName and \rnupName values are shown using color.%
    }%
    \label{fig:Spiral_illustration1}%
\end{figure}
 
\begin{figure}[tbp]%
    \centering%
    \begin{subfigure}[b]{.55\textwidth}%
        \centering
        \includegraphics[sqrtofarea=.7\textwidth]{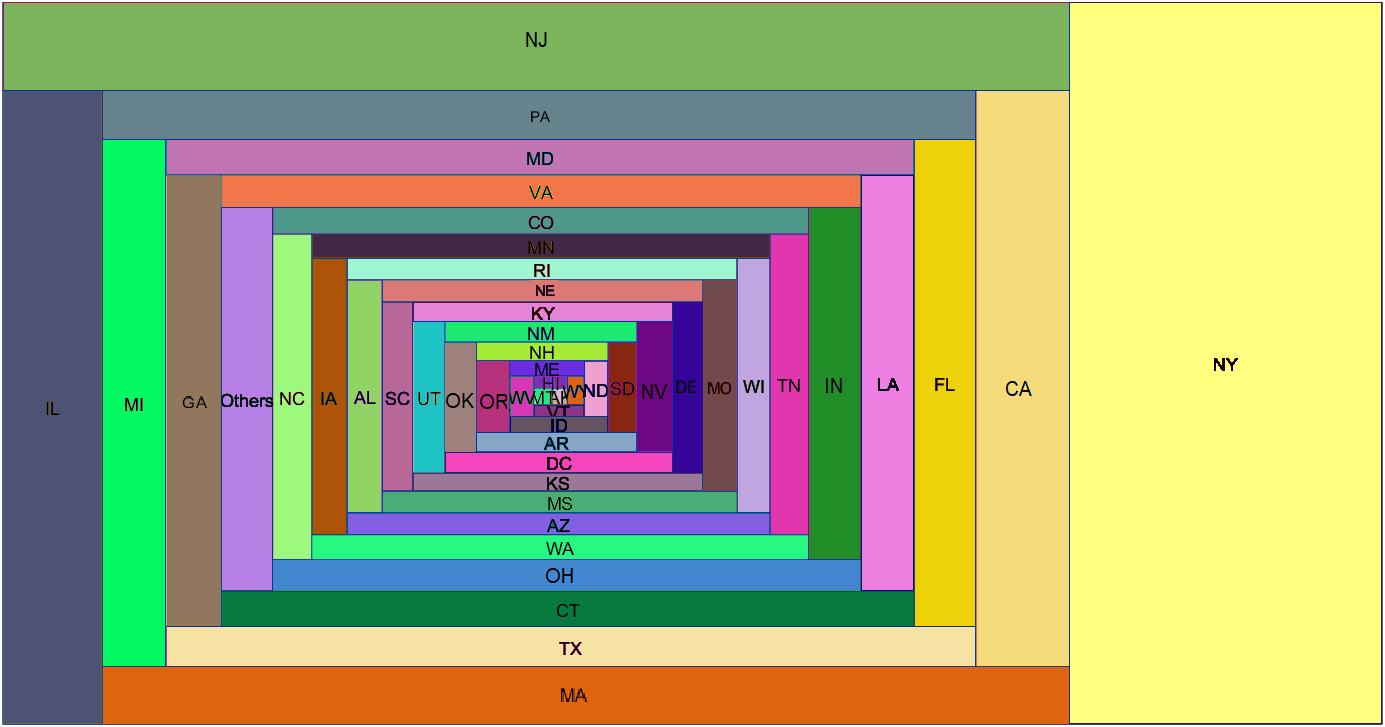}%
        \caption*{
            \modelSymmetricSpiral\\
            maxAR = 21.07\\
        }%
        \label{fig:COVID_FibonacciSpiral}%
    \end{subfigure}%
    \hfill
    \begin{subfigure}[b]{.45\textwidth}%
        \centering
        \includegraphics[sqrtofarea=.85\textwidth]{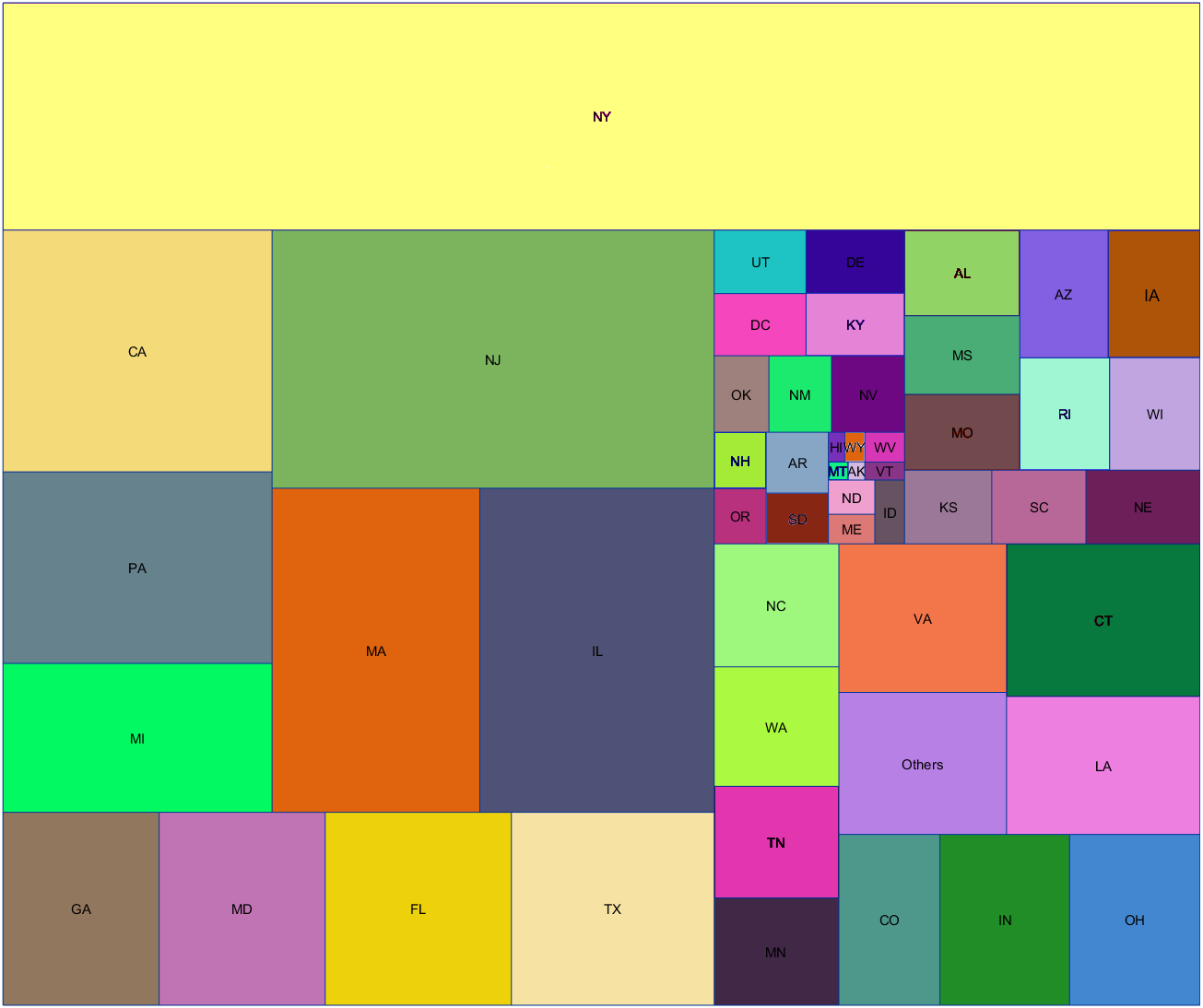}%
        \caption*{
            \modelSquareBundle\\
            maxAR = 5.26\\
        }%
        \label{fig:COVID_SquareBundledSpiral}%
    \end{subfigure}%
    \\
    \begin{subfigure}[b]{.55\textwidth}%
        \centering
        \includegraphics[sqrtofarea=.7\textwidth]{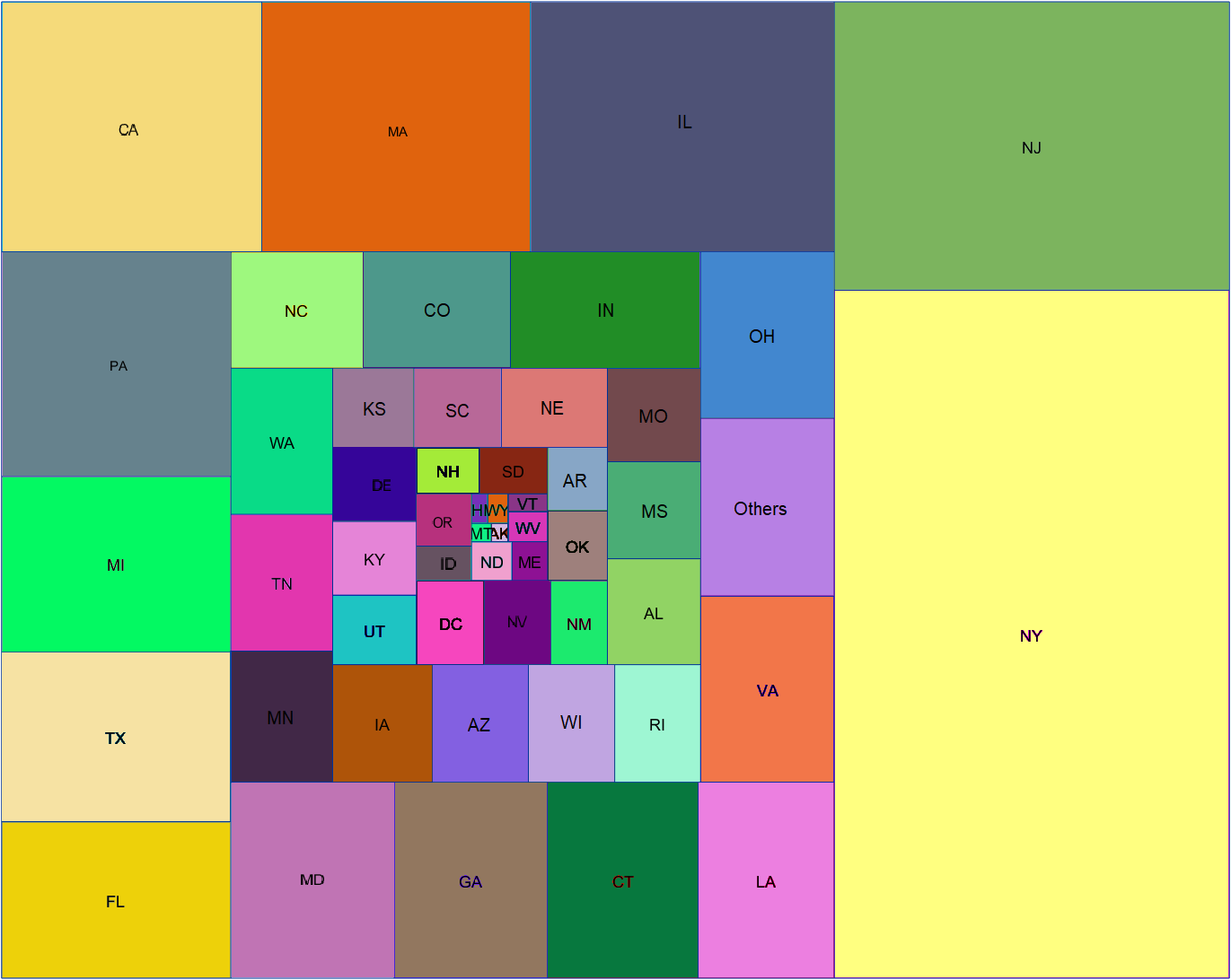}%
        \caption*{
            \modelStripBundle\\
            \rnup{maxAR = 2.20}\\
        }%
        \label{fig:COVID_StripBundledSpiral}%
    \end{subfigure}%
    \hfill
    \begin{subfigure}[b]{.45\textwidth}%
        \centering
        \includegraphics[sqrtofarea=.85\textwidth]{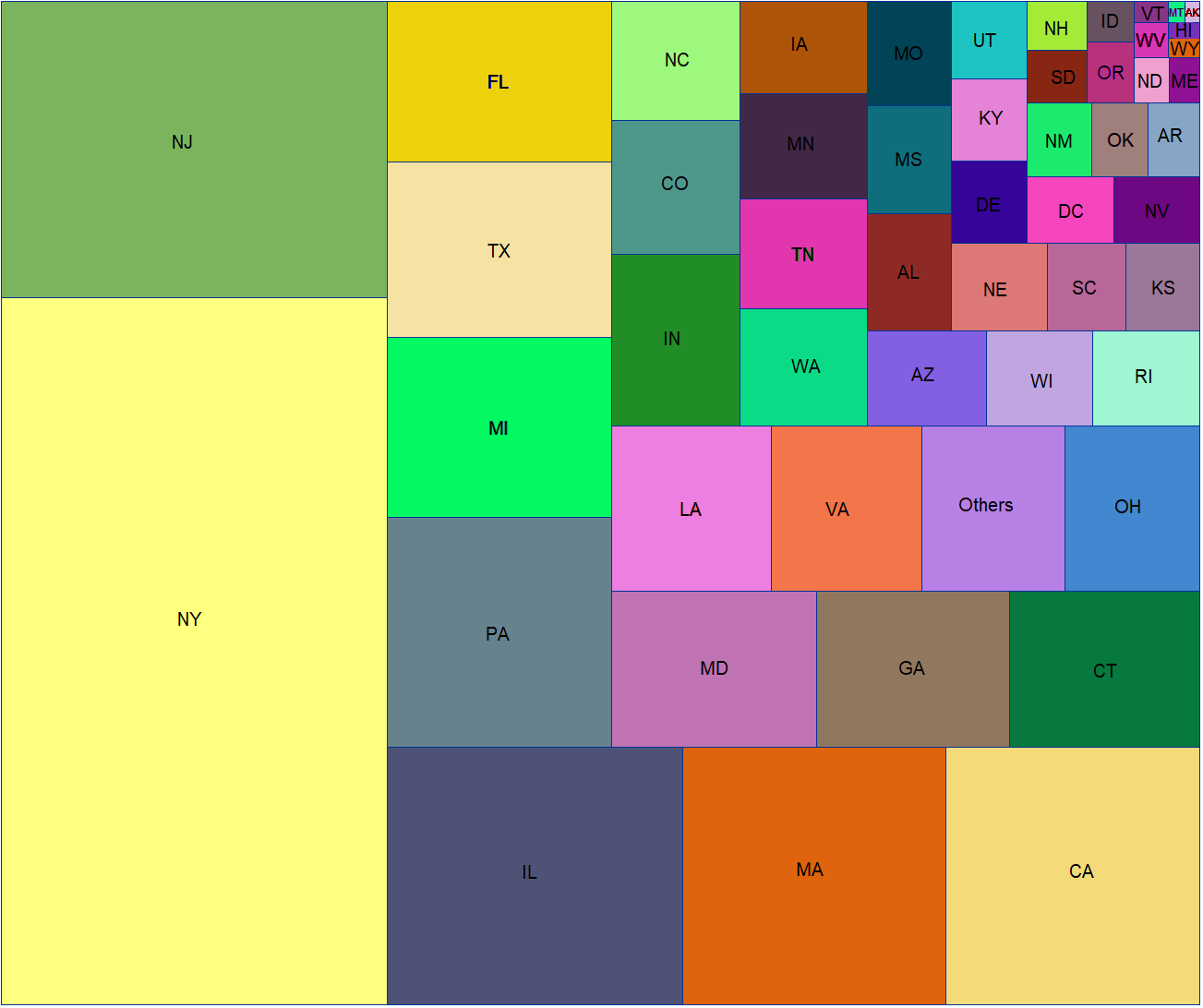}%
        \caption*{
            \modelSquarified\\
            \best{maxAR = 2.00}\\
        }%
        \label{fig:COVID_Squarified}%
    \end{subfigure}%
    \subfigsCaption{\protect
        Our three spiral treemap algorithms compared with \modelSquarified for making a rectangular treemap.
        The data shown is the count of diagnosed cases of COVID-19 in 52 different U.S. states and territories from the CDC as of 2020-05-20.
        \modelSquarified has the best maximum aspect ratio (maxAR), followed by \modelStripBundle.
        However, our spiral algorithms have better maximum Hausdorff distances (maxHD).
        The \bestName and \rnupName values are shown using color.
        Note: we are not advocating this visualization for pandemic data.%
    }%
    \label{fig:Spiral_illustration2}%
\end{figure}

Another example in \Cref{fig:Spiral_illustration2} shows treemaps of the number of COVID-19 diagnosed cases in 52 different U.S.\ states and territories as of 2020-05-20, generated using the same four algorithms.
The data is collected from the Center for Disease Control and Prevention (CDC).
In this example, \modelSquarified has the best aspect ratio with \modelStripBundle a close second.
All of our spiral algorithms score better than \modelSquarified for stability.
Although \modelSymmetricSpiral created poor aspect ratios, in this application it may be preferred to the other approaches as it may be easier to follow the sorted order that actually matters more in this application.

Finally, these examples also demonstrate our expectation that \modelSquareBundle is the closest to the Fibonacci spiral as the the distribution of the edge length of the blocks are closer to the Fibonacci sequence. One could try to study the impact of choices other than \modelSquarified for the layout inside each bundled block.

\section{Experiments and Results}
\label{sec:ComputationalResults}

Moving beyond usage examples, we here computationally compare treemap algorithms based on aspect ratio and stability.
We consider the following evaluation criteria.
For aspect ratio: total perimeter, maximum and average aspect ratio (maxAR and avgAR), and area-weighted average aspect ratio (AWAR).
For stability: the maximum and average Hausdorff distance (maxHD and avgHD).

As discussed in \cref{subsec:literature}, we compare our algorithms vs. \modelOptimization and the best-of-breed alternative: \modelSquarified \cite{bruls2000squarified}, that its superior quality, considering aspect ratio measure, is demonstrated by extensive computational results summarized in Fig. 8 -- Fig. 11 in \cite{vernier2020quantitativeJ} and Fig. 21 and Fig. 26 in the paper \cite{sondag2018stable}. It is also clear from \cite{sondag2018stable}, as we observe it here too, it can also perform reasonably well on stability depending on the data set and the measure of stability. We also compare our results with the more recent algorithm \modelDC \cite{liang2015divide} that follows a similar approach and performs well when compared to \modelSquarified. Moreover, again as explained \cref{subsec:literature}, we skip space-filling curve-based algorithms for this comparison as they aim to to preserve order and maximize readability and stability and generally perform poorly with respect to aspect ratio metrics, as shown in \cite{tu2007visualizing,tak2012enhanced,sondag2018stable,vernier2020quantitativeJ}. To make the optimization results comparable with the algorithmic results, we set all parameters in our \modelOptimization to be zero.

We generated 25 random test problems for the comparison.
These are listed in {\tt synthetic\_data.csv} in the supplemental material at \anonymizeOSF{\OSFSupplementText}\footnote{\url{osf.io/y8sgm/view_only=4e5edb49f893410a866f99fcf4e71cb5}} and
\iflabelexists
  {tab:testProblems}
  {\cref{tab:testProblems}}
  {a table in the appendices at \anonymizeOSF{\OSFSupplementText}}.
Since treemapping is NP-hard and finding an optimal solution for large problems is challenging, we avoided cases with more than 12 sub-regions.
This is to ensure we can include \modelOptimization in the comparison.
To further ease visual comparison, each weighted tree in our problems has a height of 1.
Deeper trees can be easily laid out by running each of the algorithms recursively, as shown in \cref{sec:flare}.

In order to analyze the stability of treemaps constructed by each algorithm, we build the treemaps for each sample, then perturb the input areas for a few rounds in three level of perturbation, small, medium, and high. Furthermore, we built new treemaps with new areas and calculated the Hausdorff Distance between the pairs of rectangles in two treemaps. In small level of perturbation we added a random variable between 0 and 1 multiplied by 0.01 to the previous area then we normalized them to have area equal to 1. For medium and high level of perturbation we did the same but instead of 0.01 we used 0.05 and 0.1.

\begin{table*}[tbp]
    \caption{The \bestName and \rnupName value in each column is shown using color. Detailed results can be found in 
        \iflabelexists
            {tab:DetailedComputationalResults}
            {\cref{tab:DetailedComputationalResults}.}
            {a table in the appendices at \anonymizeOSF{\OSFSupplementText}.}
        	All optimization parameters are set to be zero. We set $c=2$ in our modified divided and conquer approach. The three spiral algorithms are considered for two cases, when $\rho_s=2$ and $\rho_s=\phi=1.618$. The last three rows show the difference between the spiral averages (average for $\rho_S= \phi=1.618$ $-$ average for $\rho_s=2$) in percentage. The green color shows improvement and the red color shows deterioration.
    }
    \label{tab:AggregateComputationalResults}
    \centering
    \begin{tabular}{lrrrrrr}
        \textbf{Approach}   & \textbf{Perimeter}& \textbf{maxAR}& \textbf{avgAR}& \textbf{AWAR} & \textbf{maxHD}& \textbf{avgHD}\\
        \hline
    \modelOptimization  &  \best{10.7257} & \best{1.7584} & \best{1.3211} & \best{1.3014} &   --    & --  \\
    \modelSquarified \cite{bruls2000squarified} & 10.8738 & 2.9940 & 1.6452 & 1.4706 & \best{0.4512} & \best{0.2135} \\
    \modelDC \cite{liang2015divide}  & 10.9392 & 3.0372 & 1.6906 & 1.5735 & 1.2278 & 0.7535 \\
    \modelModifiedDC ($c=2$)    & 10.9021 & \rnup{2.5931} & 1.6231 & 1.5374 & 1.2268 & 0.7557 \\
    \modelDynamicProg     & \rnup{10.8136} & 2.6512 & \rnup{1.5221} & \rnup{1.3935} & 0.5222 & 0.2346 \\
      &       &      &	    &       &       &  \\  
     \modelSymmetricSpiral $(\rho_s=2)$    & 12.3607 & 5.4034 & 2.7052 & 3.0156 & 0.5325 & 0.2410 \\
     \modelSquareBundle $(\rho_s=2)$  &  11.2379 & 3.2675 & 1.7874 & 1.8639 & 0.6808 & 0.2984 \\
    \modelStripBundle $(\rho_s=2)$  & 11.2122 & 3.5765 & 1.7559 & 1.8089 & 0.5670 & 0.2682 \\
    	 &       &      &	    &       &       &  \\ 
      \modelSymmetricSpiral  $(\rho_s=1.618)$  &  12.2727 & 5.0748 & 2.6751 & 2.9747 & \rnup{0.4883} & \rnup{0.2258} \\
     \modelSquareBundle $(\rho_s=1.618)$ &  11.2149 & 3.6698 & 1.8116 & 1.8135 & 0.8287 & 0.3607 \\
    \modelStripBundle $(\rho_s=1.618)$  & 11.2908 & 3.9746 & 1.8568 & 1.7897 & 0.5481 & 0.2664 \\
          &       &      &	    &       &       &  \\       
     \modelSymmetricSpiral  $(\rho_s=1.618 - \rho_s=2)$   &  \textcolor[rgb]{ .329,  .51,  .208}{-0.71\%} & \textcolor[rgb]{ .329,  .51,  .208}{-6.08\%} & \textcolor[rgb]{ .329,  .51,  .208}{-1.11\%} & \textcolor[rgb]{ .329,  .51,  .208}{-1.36\%} & \textcolor[rgb]{ .329,  .51,  .208}{-8.30\%} & \textcolor[rgb]{ .329,  .51,  .208}{-6.31\%} 
\\
    \modelSquareBundle $(\rho_s=1.618 - \rho_s=2)$ & \textcolor[rgb]{ .329,  .51,  .208}{-0.20\%} & \textcolor[rgb]{ .753,  0,  0}{12.31\%} & \textcolor[rgb]{ .753,  0,  0}{1.35\%} & \textcolor[rgb]{ .329,  .51,  .208}{-2.70\%} & \textcolor[rgb]{ .753,  0,  0}{21.72\%} & \textcolor[rgb]{ .753,  0,  0}{20.87\%}
 \\
    \modelStripBundle $(\rho_s=1.618 - \rho_s=2)$  & \textcolor[rgb]{ .753,  0,  0}{0.70\%} & \textcolor[rgb]{ .753,  0,  0}{11.13\%} & \textcolor[rgb]{ .753,  0,  0}{5.75\%} & \textcolor[rgb]{ .329,  .51,  .208}{-1.06\%} & \textcolor[rgb]{ .329,  .51,  .208}{-3.34\%} & \textcolor[rgb]{ .329,  .51,  .208}{-0.64\%} \\
    \end{tabular}
\end{table*}

\begin{figure*}[tbp]
    \centering
    \includegraphics[width=0.65\textwidth]{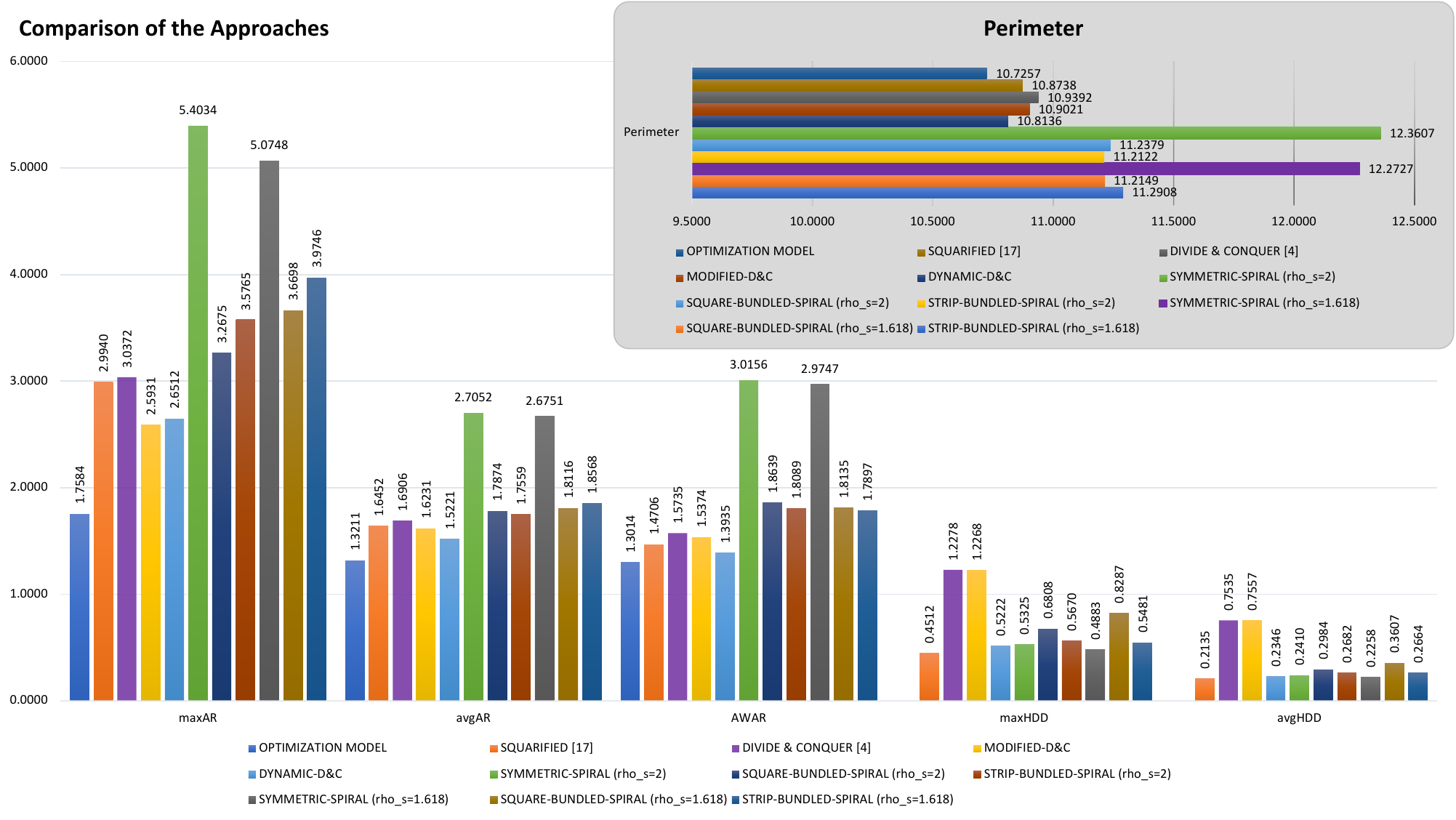}
    \caption{%
        Visual representation of the values in \cref{tab:AggregateComputationalResults} to compare the size of differences between treemap approaches with respect to metrics for aspect ratio (Perimeter, maxAR, avgAr, AWAR) and stability (maxHD, avgHD).%
    }%
    \label{fig:comparisonBarchart}%
\end{figure*}

 \begin{figure}[tbp]%
    \centering%
    \begin{subfigure}[b]{.49\textwidth}%
        \centering
        \includegraphics[sqrtofarea=.99\textwidth]{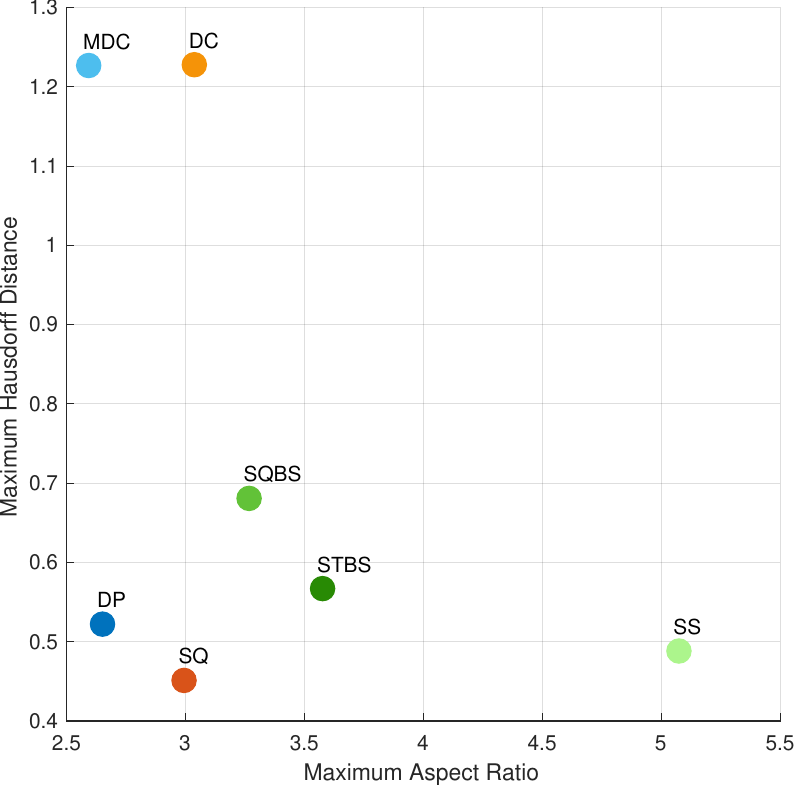}%
        \caption*{ }%
        \label{fig:maxARHDD}%
    \end{subfigure}%
    \hfill
    \begin{subfigure}[b]{.49\linewidth}%
        \centering
        \includegraphics[sqrtofarea=.99\textwidth]{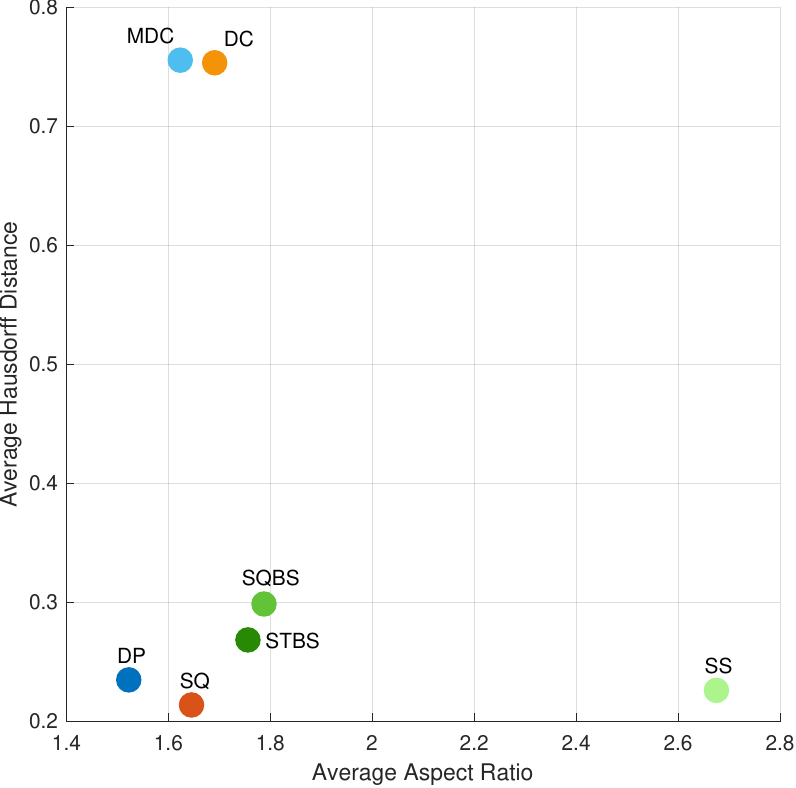}%
        \caption*{ }%
        \label{fig:aveARHDD}%
    \end{subfigure}%
    \\
    \vspace{-16pt}
     \begin{subfigure}[b]{.49\textwidth}%
        \centering
        \includegraphics[sqrtofarea=.99\textwidth]{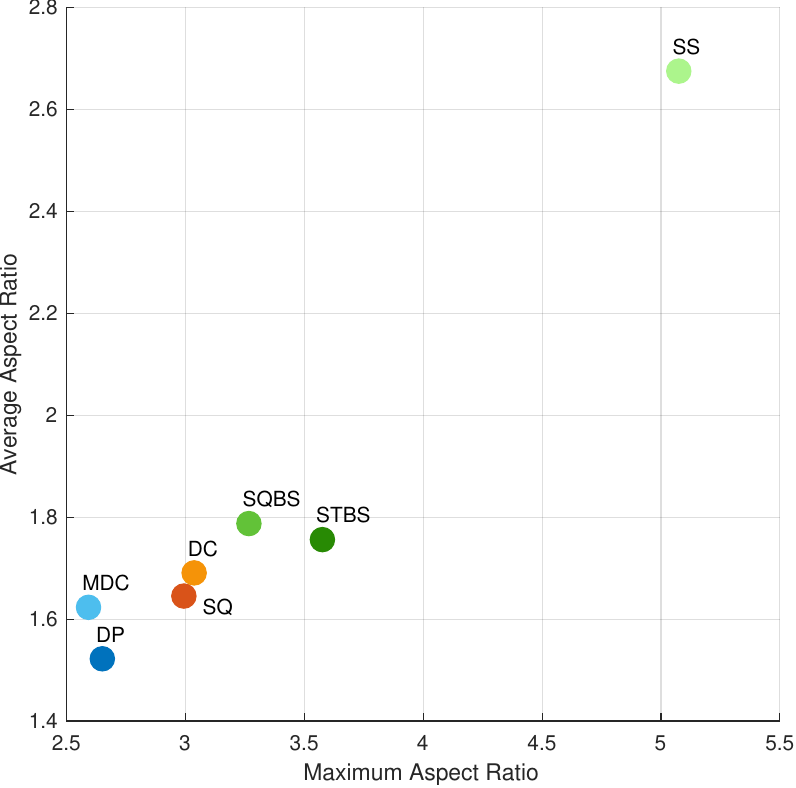}%
        \caption*{ }%
        \label{fig:maxaveAR}%
    \end{subfigure}%
    \hfill
    \begin{subfigure}[b]{.49\linewidth}%
        \centering
        \includegraphics[sqrtofarea=.99\textwidth]{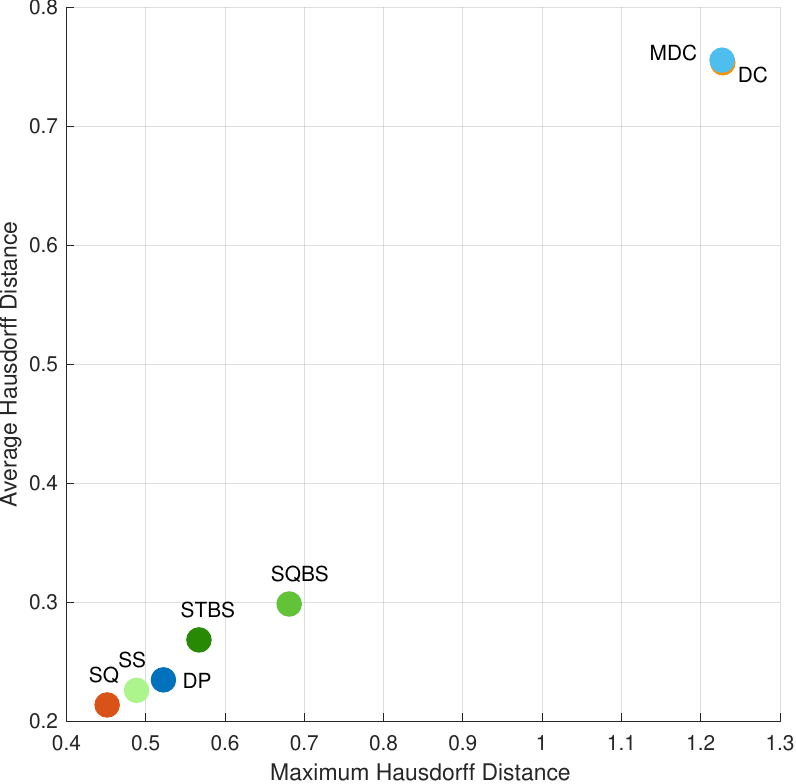}%
        \caption*{ }%
        \label{fig:maxaveHDD}%
    \end{subfigure}%
    \\
    \vspace{-10pt}
    \subfigsCaption{\protect
        Comparison of different algorithms regarding maximum and average aspect ratios and maximum and average Hausdorff distances. 
        The algorithms are color coded and labeled with SQ (\modelSquarified \cite{bruls2000squarified}), DC (\modelDC \cite{liang2015divide}), MDC (\modelModifiedDC), DP (\modelDynamicProg), SS (\modelSymmetricSpiral), SQBS (\modelSquareBundle), STBS (\modelStripBundle). We see that taking both metrics into consideration, our DP performs the best. Our MDC beats both DC and SQ on aspect ratio measure but performs weaker that SQ on stability. Our two bundled spiral algorithms perform relatively good with the respect to both maximum and average criteria. The symmetric spiral algorithm performs really well on stability but really weak on aspect ratio. For SS we used $\rho_s=\rho_s=\phi=\frac{1+\sqrt{5}}{2}\simeq 1.618$ and for the bundled spiral we used the results from $\rho_s=2$.%
    }%
    \label{fig:Spiral_illustration1}%
\end{figure}

The experimental results are summarized in \cref{tab:AggregateComputationalResults} and \cref{fig:comparisonBarchart}, with individual problem results in
\iflabelexists
  {tab:DetailedComputationalResults}
  {\cref{tab:DetailedComputationalResults}.}
  {a table in the appendices at \anonymizeOSF{\OSFSupplementText}.}
Our results show that \modelOptimization is the best performer on all aspect ratio measures---in some cases by wide margins. 
This is despite only optimizing for total perimeter.
Thus, we recommend \modelOptimization be used as a benchmark against which more computationally efficient approaches to optimizing for aspect ratio can be compared, at least for reasonably sized problems. Since our objective function is defined for perimeter (and consequently for aspect ratios) and does not aim to produce stable treemaps, we skip the stability performance of  \modelOptimization. However, as we discussed the flexibility of our optimization model in \cref{sec:optModel}, it could be possible to make a trade-off between perimeter (aspect ratio) and stability but it is beyond the scope of this paper. 

For all measures, our \modelDynamicProg outperforms \modelDC \cite{liang2015divide} by wide margins. This makes sense because it explores a larger set of possible layouts.
Our \modelDynamicProg  performs considerably better than \modelSquarified \cite{bruls2000squarified} on aspect ratio measures and only slightly weaker on stability measures. This is besides the fact that the input region in 52\% of our test problems was a unit box which favors \modelSquarified \cite{bruls2000squarified}. \modelDynamicProg found the optimal solution in 3 out 25 test problems, which is also a certificate to show that these 3 problems had sliceable optimal solutions.
Our \modelModifiedDC outperforms \modelDC \cite{liang2015divide} on all but one measure and beats \modelSquarified \cite{bruls2000squarified} by a wide margin on maxAR measure. As expected our modification to the divide and conquer approach of \cite{liang2015divide} shows its most significance on the maximum aspect ratio which amounts to more than 17\%. Our \modelDynamicProg and \modelModifiedDC significantly improve on \modelDC \cite{liang2015divide}, despite the fact that in our test problems we did not consider any extreme instance, as described in \cref{sec:DP}, for which \modelDC of \cite{liang2015divide} would perform particularly weak.

As we expected, our spiral algorithms perform much better on stability measures than aspect ratio measures. \modelSymmetricSpiral performs particularly weak on total perimeter and maxAR, due to stretching some rectangles. The modified spiral algorithms \modelSquareBundle and \modelStripBundle perform better on the aspect ratio measures. However, they perform weaker than the non-spiral approaches. This is because (1) the sub-regions placed last are more prone to poor aspect ratios and (2) the sub-regions are sorted in ascending order of their areas giving more weight to those sub-regions placed last.
Stability analysis shows that all of our spiral algorithms generally perform very well, which was again expected due to the packing framework and spiral structure. By following a spiral structure, we believe they also provide additional aesthetic appeal vs. other approaches. They maximize \emph{symmetry} and \emph{fractal-like patterns} in the treemaps. It should be also mentioned that removing the sorting step, we can easily modify these algorithms to make them order-preserving. However, this may come at some cost on the side of aspect ratios. One could consider evaluation metrics such as readability measure introduced in \cite{bederson2002ordered} and the Fractal Value as defined in Eq. (1) in \cite{dAmbros2005fractal}, and evaluate the quality of our spiral algorithms that way.

We should also mention that in some of our test problems, our spiral algorithms performed better than all other approaches including \modelOptimization even on the optimization objective (total perimeter). This is because spiral algorithms are not restricted to the input container. Since this lack of restriction to input region can lead to both significant improvements or deterioration, we can claim that the performance of our spiral algorithms highly depend on the problem instances. They performed weaker on average on our test problems, although they were able to find solutions even better than our \modelOptimization on some instances. However, they should be seen as alternatives that have the potential to generate superior results depending on the problem instance.  

In our spiral algorithms, an adjustable parameter $\rho_s$ sets the aspect ratio of the first two placed rectangles.
To mimic the Fibonacci/Golden spiral we set $\rho_s = 2$.
To check the sensitivity of the algorithm to $\rho_s$---while still approximating a golden spiral---we tried the golden ratio, i.e., $\rho_s=\phi=\frac{1+\sqrt{5}}{2}\simeq 1.618$, and tested the aforementioned data \& measures.
The last three rows of \Cref{tab:AggregateComputationalResults} show that this change led to mixed results; \modelSymmetricSpiral is improved on all metrics, while \modelSquareBundle and \modelStripBundle are improved on some metrics and deteriorated on some others. The changes in stability metrics on both sides were significant.
More sensitivity analysis could be done by setting $\phi \leq \rho_s \leq 2$, while maintaining the overall golden spiral structure. 

Besides the quality metrics, run time is also important to consider.
Our \modelModifiedDC, \modelSymmetricSpiral, \modelSquareBundle, and \modelStripBundle have a time complexity of $\mathcal{O}(n\log n)$, as does the best-of-breed \modelSquarified \cite{bruls2000squarified} and \modelDC \cite{liang2015divide}.
The running time of our \modelDynamicProg, which as discussed earlier is designed for quality not computational efficiency, is  $\mathcal{O}(n^3)$. 
All of our algorithms scale very well, although to lesser degree for \modelDynamicProg, as apparent by these worst-case running times and the fact that the implementation time of the quasilinear algorithms on all of our test cases including the 25 random test problems, the examples of size 60 and 52 in \cref{fig:Spiral_illustration1} and \cref{fig:Spiral_illustration2} and the a famous 220 leaf node example in Section \ref{sec:flare}, was between fractions of a second to a few seconds, similar to that of \modelDC \cite{liang2015divide} and \modelSquarified \cite{bruls2000squarified}. This time for \modelDynamicProg increases to a few minutes.

\section{Usage Examples}
\label{sec:UsageExamples}

Here we present several usage examples to demonstrate the utility of our treemap algorithms on realistic data, in contrast to the random data in our computational experiments.
All of these datasets are available at \anonymizeOSF{\OSFSupplementText}.

\subsection{Single Level Weighted Trees}
We previously showed treemaps of single level weighted trees.
This is to ease comparisons, but deeper trees can be laid out via recursive application as we will see in \cref{sec:flare}.

\textbf{Stock Market}---Stock market capitalization by sector has long been used as an example for treemap algorithms, starting with Wattenberg's Map of the Market \cite{Wattenberg1998MarketMap}.
Our data is from the U.S.\ Stock Market after closing on 2020-12-18.
\cref{fig:StockMarketRect,fig:StockMarketHex} and \cref{fig:StockMarket_Unitbox_Optimal,fig:StockMarket_Unitbox_DynamicDC,fig:StockMarket_Unitbox_DC,fig:StockMarket_Unitbox_ModifiedDC} show the associated treemaps using our \modelOptimization and the three divide and conquer approaches.

\textbf{U.S.\ Census}---From the 2010 U.S.\ Census data we extracted 12 states in the Midwest.
The treemaps in \cref{fig:MidwestPopulationRect,fig:MidwestPopulationHex} and
\cref{fig:Midwest_Unitbox_Optimal,fig:Midwest_Unitbox_DynamicDC,fig:Midwest_Unitbox_DC,fig:Midwest_Unitbox_ModifiedDC} show the proportion of the total population that each of these states contributes.
The treemaps again use \modelOptimization and the three divide and conquer approaches.

\textbf{COVID-19}---Similar to the state population data, this example shows the proportion of the total diagnosed cases of COVID-19 there were in each of the 52 U.S.\ States and territories as of 2020-05-20, according to the CDC.
We show this data in \cref{fig:Spiral_illustration2} using our three spiral approaches.

\subsection{Multiple Level Weighted Trees}
\label{sec:flare}
Similar to most other treemap algorithms, our proposed algorithms are proposed for the single level trees but can be easily applied to multi-level trees as well. This is straightforward for our divide and conquer and dynamic programming algorithms as they are based on a subdivision approach and the space for the parent is formed prior to forming the space for children. However, our spiral algorithms that follow a packing approach require a bit of explanation. We start from the bottom level and construct the rectangles for immediate parents of the the leaf nodes following our spiral structure. Once we have all the rectangles in the next level constructed, we make the rectangles for their parents and recursively iterate this until the rectangle for the root is formed. The only thing to consider here is that we will only set $\rho_s=2$ or $\rho_s=\frac{1+\sqrt{5}}{2}$ at the bottom level and relax this step for the higher levels. 
Note that we can also do this in reverse direction and start from the root and work towards the bottom level. We could also start from any middle level and work towards both directions of the tree. However, we must fix the aspect ratio of the initial two rectangles only in the level we start. This is to avoid having unused spaces in the treemap.

Here, we will show the behavior of our treemap algorithms on a large unbalanced height-4 tree with 220 leaf nodes.
We use the Flare data.
Flare \cite{Heer2007Flare} is an ActionScript library for making web-based interactive Flash visualizations.
It is the spiritual successor to the Prefuse \cite{Heer2005Prefuse} Java library and an early predecessor of D3\cite{bostock2011d3}.
The nodes in our tree are the ActionScript 3 classes in the library, arranged by the class hierarchy.
Each node has a weight for the size of the class.
This data is widely available as part of many online visualizations,%
\foothrefhttps{observablehq.com/@d3/hierarchical-edge-bundling\#file-readme-flare-imports-json}\textsuperscript{,}%
\foothrefhttps{vega.github.io/editor/\#/examples/vega/tree-layout}\textsuperscript{,}%
\foothrefhttps{observablehq.com/@mahog/vega-tutorial-2-working-with-trees-solution}\textsuperscript{,}%
\foothrefhttps{vega.github.io/editor/\#/examples/vega/radial-tree-layout}\textsuperscript{,}%
\foothrefhttps{vega.github.io/editor/\#/examples/vega/circle-packing}\textsuperscript{,}%
\foothrefhttps{vega.github.io/editor/\#/examples/vega/sunburst}\textsuperscript{,}%
\foothrefhttps{observablehq.com/@mahog/vega-tutorial-2-working-with-trees-solution}
and in particular treemaps generated with various algorithms including \modelSliceDice and \modelSquarified.%
\foothrefhttps{observablehq.com/@d3/treemap?collection=@d3/d3-hierarchy}\textsuperscript{,}%
\foothrefhttps{vega.github.io/editor/\#/examples/vega/treemap}%

In \cref{fig:HierarchicalTreemaps} we show treemaps of the Flare data generated by our algorithms, \modelSquarified, and \modelDC.
See the figure caption for a comparative discussion.
We exclude \modelOptimization due to long run time.
For spiral algorithms we illustrate the results for $\rho_s=2$. Table \ref{tab:FlareComputationalResults} shows the comparison of metrics between the algorithms on this large data set. Our \modelDynamicProg still provides the best overall performance on this data set.

\begin{table*}[htp]
    \caption{Comparison of the metrics between the algorithms for the large unbalanced height-4 weighted tree of Flare \cite{Heer2007Flare} visualization library class hierarchy containing 220 leaf nodes. The \bestName and \rnupName value in each column is shown using color. We set $c=2$ in our modified divided and conquer approach. The three spiral algorithms are considered for two cases, when $\rho_s=2$ and $\rho_s=\phi=1.618$. }
    \label{tab:FlareComputationalResults}
    \centering
    \begin{tabular}{lrrrrrr}
        \textbf{Approach}   & \textbf{Perimeter}& \textbf{maxAR}& \textbf{avgAR}& \textbf{AWAR} & \textbf{maxHD}& \textbf{avgHD}\\
        \hline  
     \modelSquarified \cite{bruls2000squarified} & \rnup{53.2894} & 11.1962 & \rnup{1.5218} & \rnup{1.4025} & 0.9941 & \rnup{0.3669} \\
     \modelDC \cite{liang2015divide}  & 53.7136 &  \best{5.4250} & 1.6539 & 1.5832 & \best{0.7004} & \best{0.3068} \\
     \modelDynamicProg     &  \best{52.8362} & 7.7917 & \best{1.3788} & \best{1.3449}  & 1.2388 & 0.5172 \\
     \modelModifiedDC   & 53.7366 & \rnup{5.6814} & 1.6167 & 1.6278 & 1.0451 & 0.4449 \\
      &       &      &	    &       &       &  \\
    \modelSymmetricSpiral $(\rho_s=2)$   & 75.1244 & 274.9278 & 15.9498 & 6.696 & \rnup{0.9191} &  0.4248\\
    \modelSquareBundle $(\rho_s=2)$ & 54.4496 & 7.8461 & 1.7155 & 1.7913 & 1.1493 & 0.5066 \\
    \modelStripBundle $(\rho_s=2)$ & 55.4316 & 21.381 & 1.8899 & 2.0744 & 1.1757 & 0.5305 \\
    	 &       &      &	    &       &       &  \\
        \modelSymmetricSpiral  $(\rho_s=1.618)$  & 74.2520 & 274.9278 & 15.8101 & 6.4117 &  0.9838 & 0.4527 \\
    \modelSquareBundle $(\rho_s=1.618)$ & 58.1248 & 39.5781 & 3.9537 & 2.1383 & 1.3837 & 0.5508 \\
     \modelStripBundle $(\rho_s=1.618)$  & 56.9472 & 16.2207 & 3.2577 & 1.9992 & 1.1325 & 0.4437 \\
    \end{tabular}
\end{table*}

\begin{figure*}[tbp]%
    \centering%
        \begin{subfigure}[b]{.4\textwidth}%
        \centering
        \includegraphics[sqrtofarea=.7\textwidth]{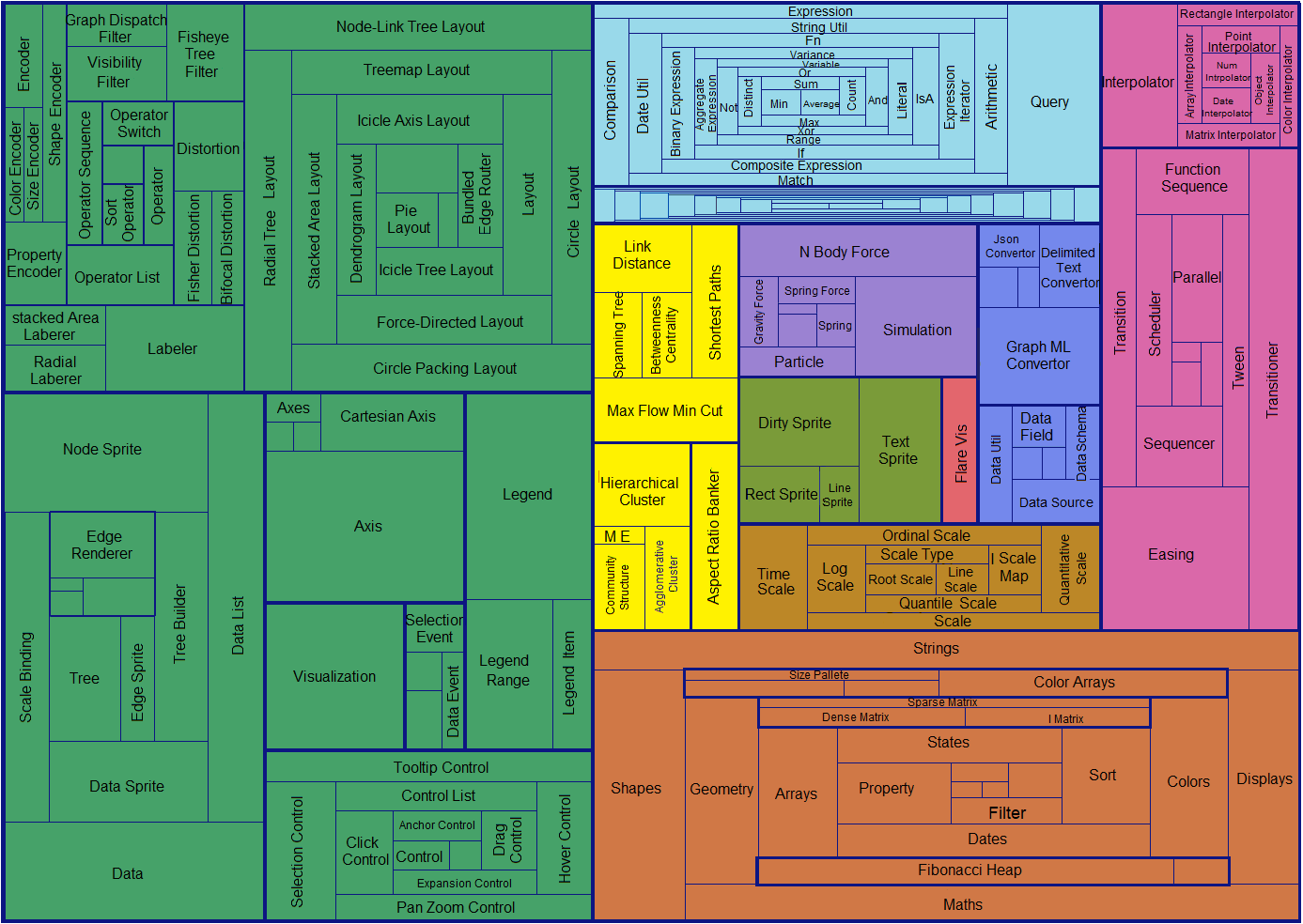}%
        \caption*{\modelSymmetricSpiral}%
        \label{fig:Hierarchical_Spiral_Fibonacci}%
    \end{subfigure}%
    \hfill
    \begin{subfigure}[b]{.3\textwidth}%
        \centering%
        \includegraphics[sqrtofarea=.85\textwidth]{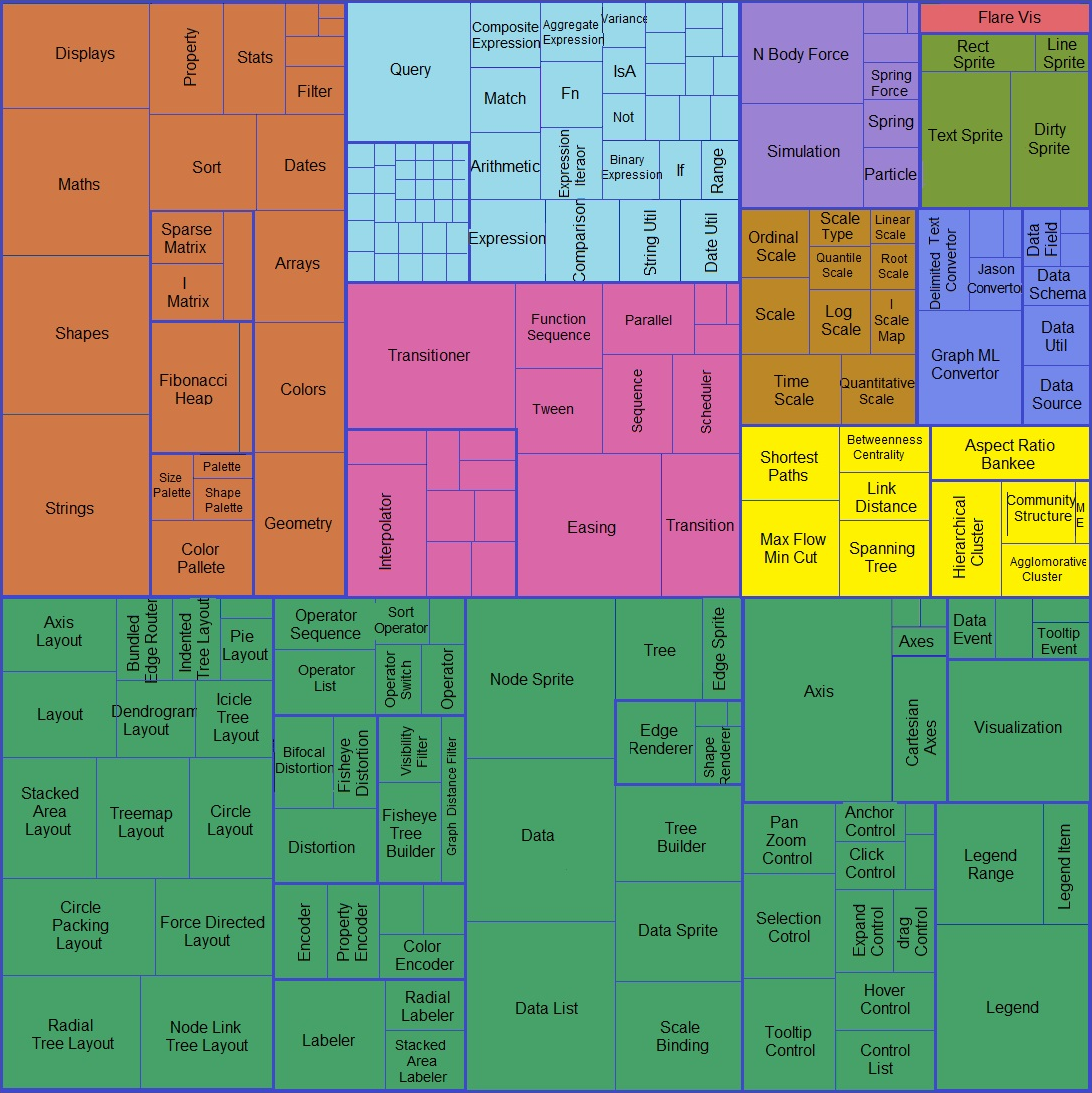}%
        \caption*{\modelSquarified}%
        \label{fig:Hierarchical_Squarified}%
    \end{subfigure}%
    \hfill
     \begin{subfigure}[b]{.3\textwidth}%
        \centering%
        \includegraphics[sqrtofarea=.85\textwidth]{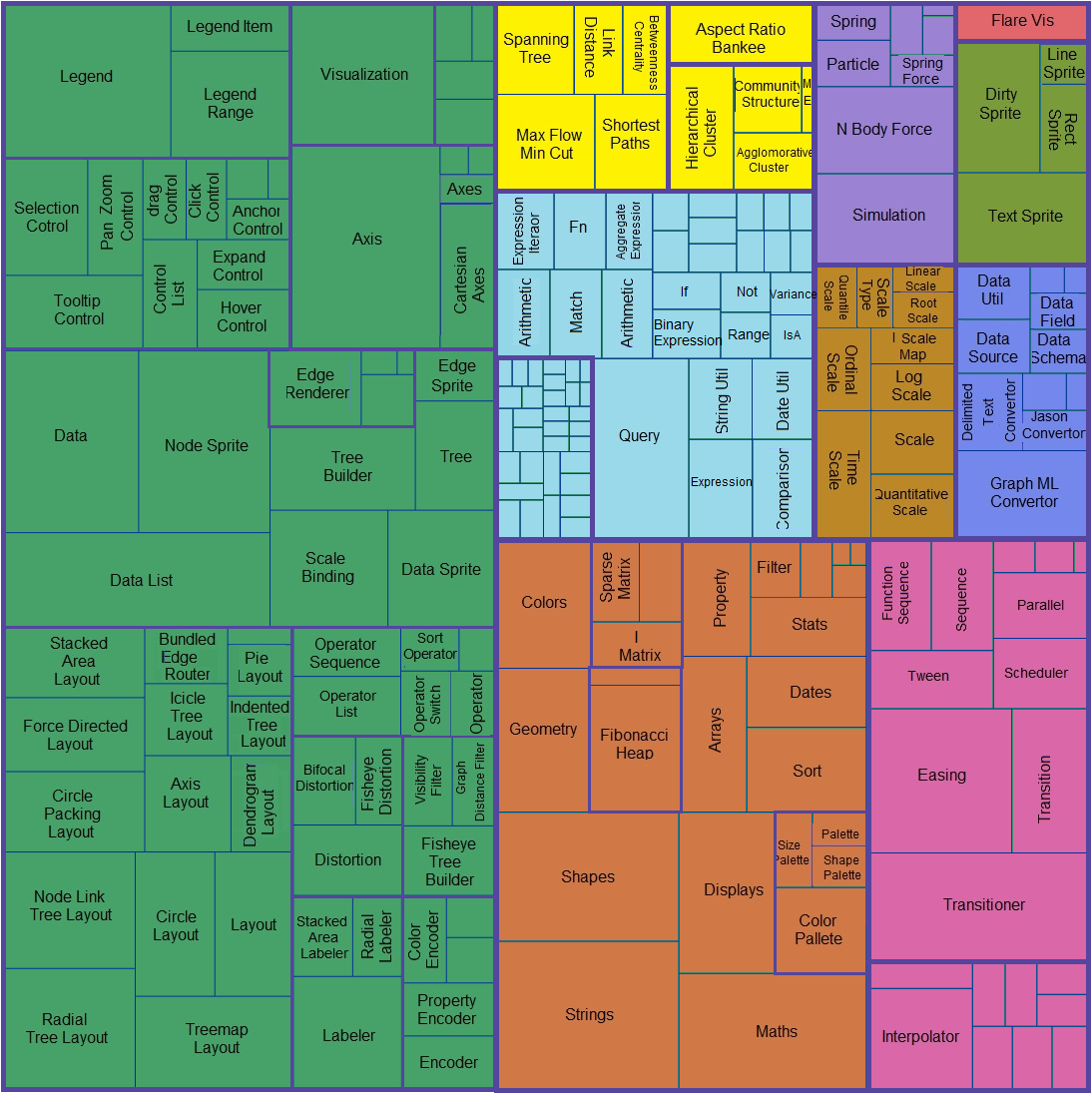}%
        \caption*{\modelDC}%
        \label{fig:Hierarchical_DCinLit}%
    \end{subfigure}%
    \\
        \begin{subfigure}[b]{.4\textwidth}%
        \centering%
        \includegraphics[sqrtofarea=.7\textwidth]{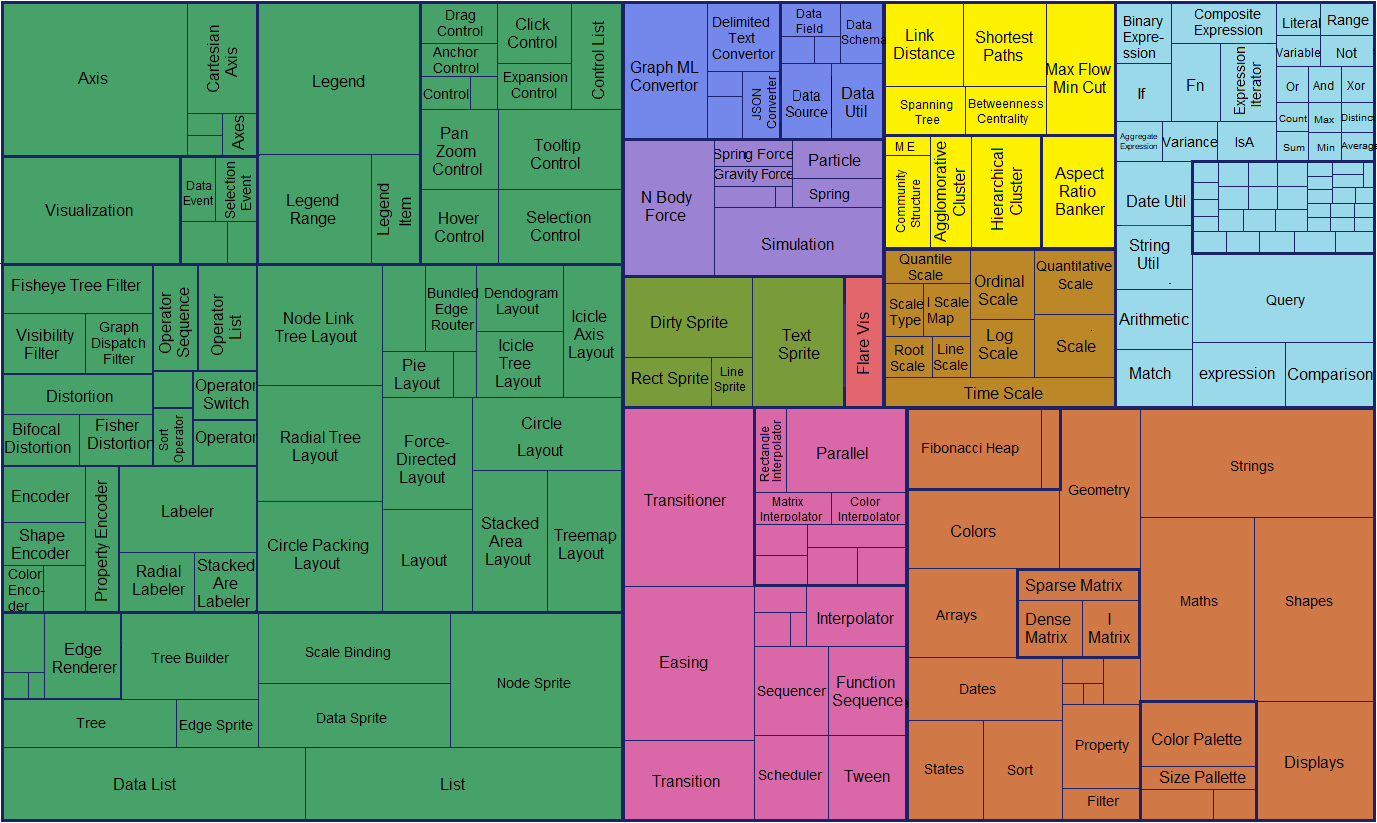}%
        \caption*{\modelSquareBundle}%
        \label{fig:Hierarchical_Spiral_SquareBundled}%
    \end{subfigure}%
    \hfill
     \begin{subfigure}[b]{.3\textwidth}%
        \centering%
     \includegraphics[sqrtofarea=.85\textwidth]{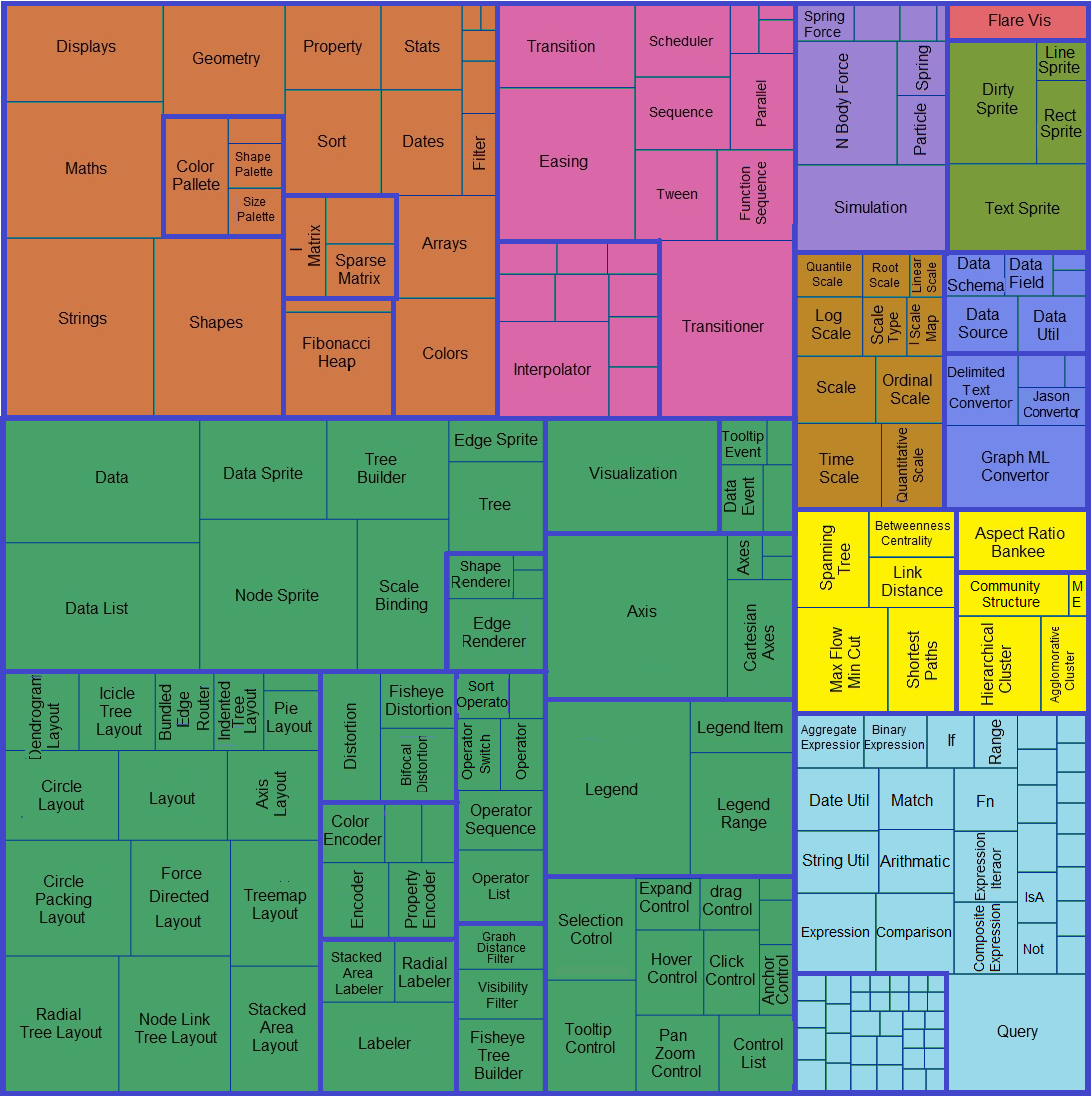}%
        \caption*{\modelDynamicProg}%
         \label{fig:Hierarchical_DynamicProg}%
    \end{subfigure}%
    \hfill
     \begin{subfigure}[b]{.3\textwidth}%
        \centering%
     \includegraphics[sqrtofarea=.85\textwidth]{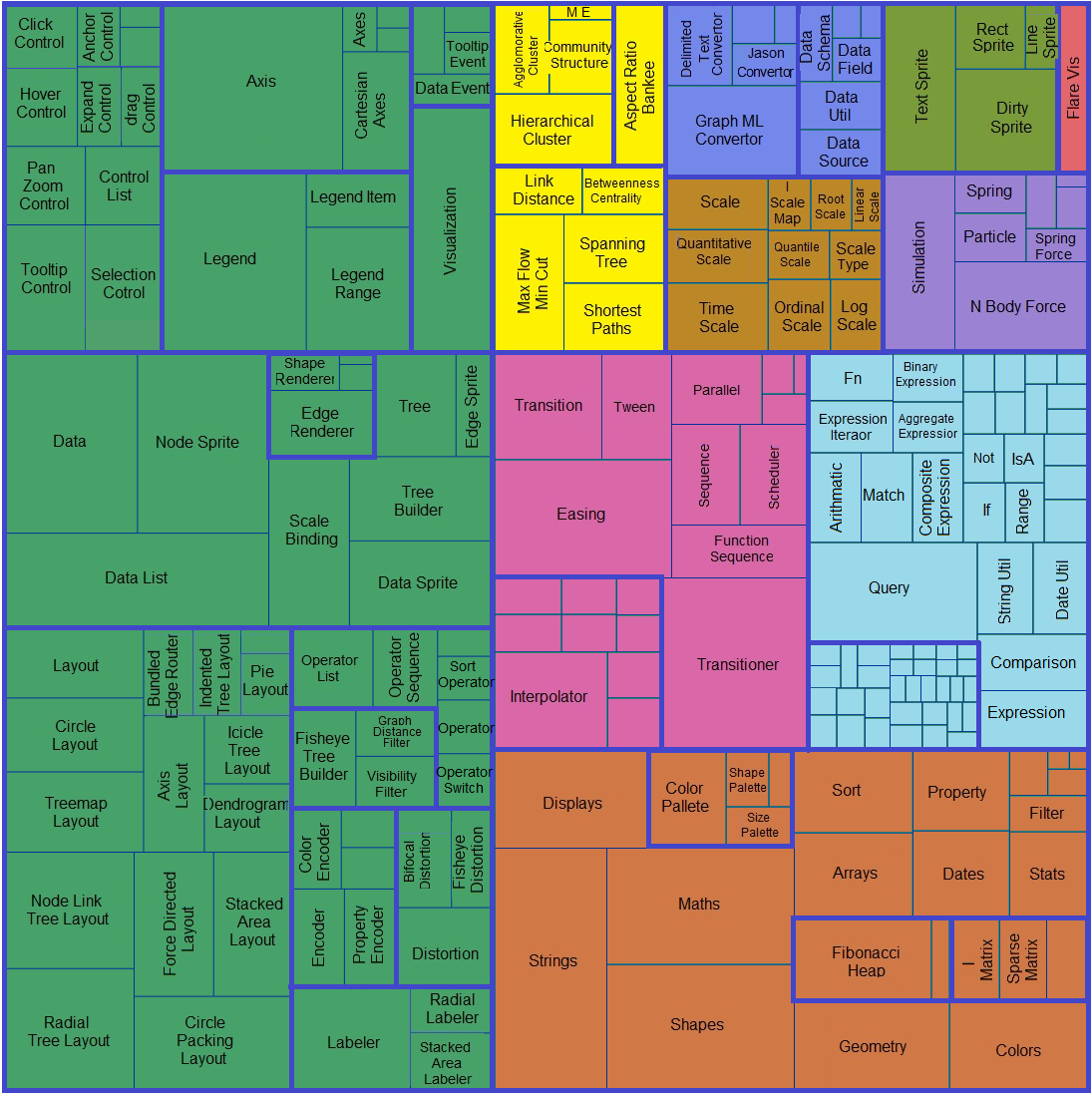}%
        \caption*{\modelModifiedDC}%
         \label{fig:Hierarchical_ModifiedDC}%
    \end{subfigure}%
    \\
    \begin{subfigure}[b]{\textwidth}%
        \centering%
        \includegraphics[sqrtofarea=.38\textwidth]{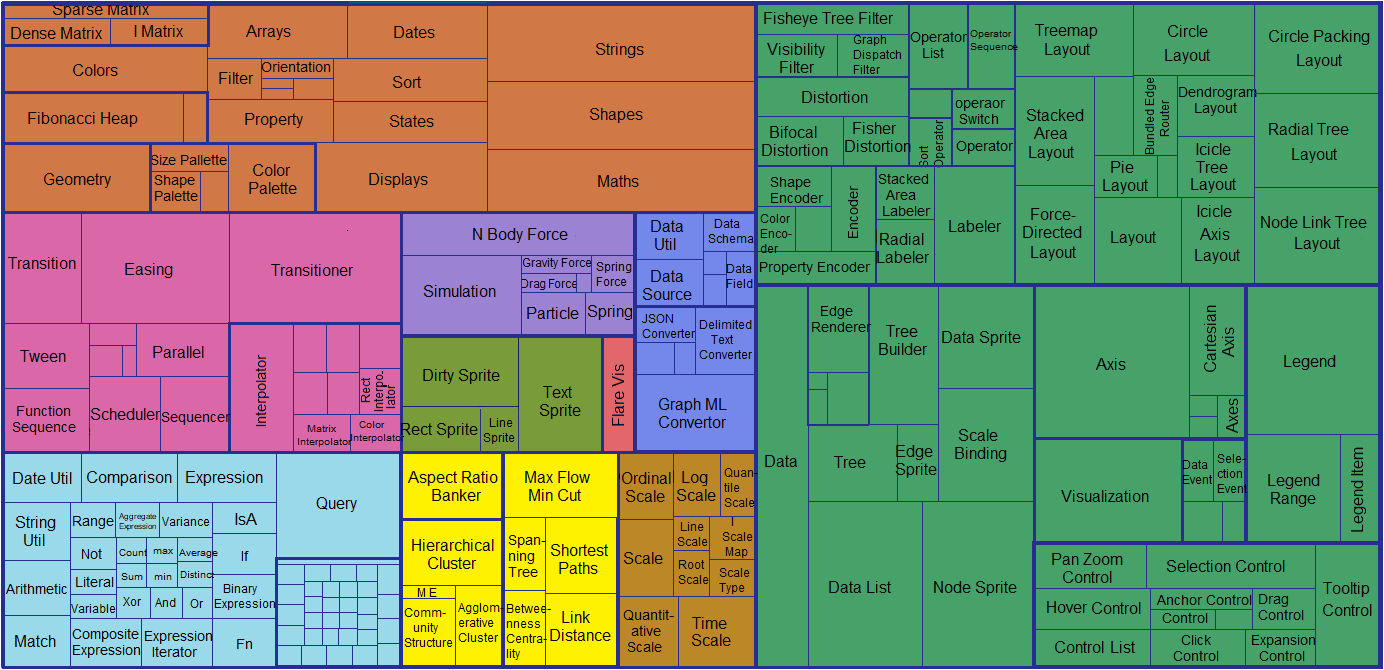}%
        \caption*{\modelStripBundle}%
        \label{fig:Hierarchical_Spiral_StripBundled}%
    \end{subfigure}%
    \subfigsCaption{\protect%
        Comparison of our five treemap algorithms, \modelSquarified, and \modelDC, each showing the Flare \cite{Heer2007Flare} visualization library class hierarchy.
        The data is an unbalanced ``height-4'' weighted tree, with ``220 leaf nodes'' representing classes with weights as the class size.
        We can see that \modelSymmetricSpiral made some unfortunate choices with regards to aspect ratio, in particular for the many skinny children of the \texttt{query$\rightarrow$methods} class (shown in the color \treemapColor{99D9EA}, top).
        However, the Symmetric Spiral is much more apparent than the spirals in the other spiral treemaps, where rectangles are predominantly fat.
        The spiral nature becomes more readily visible when there are many similar-weight siblings, e.g., the children and grandchildren of \texttt{vis} \treemapColor{46A269} (left in \modelSquareBundle, right in \modelStripBundle).
        All of our approaches stand in stark contrast with \modelSquarified, which pushes the smaller rectangles to the top-right, e.g., the children of \texttt{util} (\treemapColor{D07846}).
    }%
    \label{fig:HierarchicalTreemaps}%
\end{figure*}

\section{Conclusion}
\label{sec:conclusion}

We present an \modelOptimization and five new treemapping algorithms.
We believe that \modelOptimization, which minimizes the total sub-rectangle perimeter as the objective function, is the first optimization approach to treemapping.
Besides only having the total perimeter as the objective function it performs the best on all of our aspect ratio measures. Our computational experiments show superior results vs. \modelSquarified \cite{bruls2000squarified} and \modelDC \cite{liang2015divide}, on four metrics for aspect ratio. It also outperformed our five new algorithms.
As such, we recommend \modelOptimization be used as a benchmark for comparing future algorithms on aspect ratio related measures.

Our new \modelDynamicProg and \modelModifiedDC significantly improve on \modelDC \cite{liang2015divide} for almost all measures, albeit with additional computational complexity for \modelDynamicProg, despite avoiding the extreme instances for which \modelDC is particularly vulnerable.

The three proposed spiral algorithms had weaker performance on aspect ratio measures but very good performance on stability. They were also able to find solutions even better than our \modelOptimization on aspect ratio metrics for a few of instances. This is due to the lack of restriction to input region and suggest that performance of our spiral algorithms is problem dependent. 
They also provide an arrangement that mimics the Fibonacci/Golden spiral which we believe is more appealing and could be easier to follow. However, user studies are necessary to validate this claim.

In this paper we assumed that $\AR=1$ is the best for sub-rectangles, although we pursued other aspect ratios for blocks of sub-rectangles in our spiral algorithms to make the overall layout more appealing. This is a uniformly accepted goal to make sub-rectangles as square as possible \cite{vernier2020quantitativeJ}. However, as suggested by \cite{kong2010perceptual,lu2017golden} and many other works in other fields, it could be more desirable to have aspect ratios close to 3/2 or the golden ratio $\phi=\frac{1+\sqrt{5}}{2}\simeq 1.618$ for the sub-rectangles as well. One future research direction is to consider the impact of this change in the objective. 

Here we only evaluated treemapping algorithms on 6 measures of aspect ratio and stability. 
To further explore the strengths and weaknesses of these algorithms, we encourage future researchers to use additional quality metrics and more datasets. Furthermore, one could tweak different parts of the proposed spiral algorithms, e.g., the sorting part, to investigate the sensitivity of the output quality to such changes.

To our knowledge, all divide and conquer treemapping approaches, including ours, initially sort the input areas and as such they do not preserve input order.
An avenue for future research would be algorithmically making a tradeoff between aspect ratio, stability, and preserving order. A machine learning approach similar to that of \cite{bethge2017improving} could also help to achieve the right trade-off. 

Another interesting direction for future research is to consider equal areas being laid out as convex polygonal subregions inside an input region that is not necessarily a rectangle. Clearly, this is very easy when $R$ is a rectangle. However, this becomes quite complex when $R$ is a general convex region. An approach similar to that of \cite{carlsson2016geometric} in partitioning a convex polygon into equal area sub-regions could be adopted to incorporate the additional objective of minimizing total perimeter of all sub-regions.

\section*{Acknowledgments}
The authors gratefully acknowledge support from a Tier-1 grant from Northeastern University. The authors also thank Ben Shneiderman for his valuable comments and Zachary Danziger for his efficient MATLAB function on calculating Hausdorff distance.

\bibliographystyle{IEEEtran}
\bibliography{IEEEabrv,treemap-partitioning.bib}

\eat
{
\appendices
}
\end{document}